\documentclass[twocolumn,superscriptaddress,amsmath,amssymb,aps]{revtex4-1}

\usepackage[pdftex]{graphicx}
\usepackage{dcolumn}
\usepackage{bm}
\usepackage{float}
\usepackage{hyperref}
\usepackage{epstopdf}

\newcommand{\ket}[1]{\left| #1 \right>}
\newcommand{\bra}[1]{\left< #1 \right|}

\newcommand{\Fig}[1]{Fig.\,\ref{#1}}
\newcommand{\Eq}[1]{Eq.\,\eqref{#1}}

\hypersetup{
            colorlinks=true,
            linkcolor=blue,
            anchorcolor=blue,
            citecolor=blue}
\begin{document}

\title{Quantum Theory of Nonlinear Thermal Response}
\author{YuanDong Wang}
\affiliation{School of Electronic, Electrical and Communication Engineering, University of Chinese Academy of Sciences, Beijing 100049, China}
\affiliation{Theoretical Condensed Matter Physics and Computational Materials Physics Laboratory, College of Physical Sciences, University of Chinese Academy of Sciences, Beijing 100049, China}
\author{ZhenGang Zhu}
\email{zgzhu@ucas.ac.cn}
\affiliation{School of Electronic, Electrical and Communication Engineering, University of Chinese Academy of Sciences, Beijing 100049, China}
\affiliation{Theoretical Condensed Matter Physics and Computational Materials Physics Laboratory, College of Physical Sciences, University of Chinese Academy of Sciences, Beijing 100049, China}
\affiliation{CAS Center for Excellence in Topological Quantum Computation, University of Chinese Academy of Sciences, Beijing 100190, China}
\author{Gang Su}
\email{gsu@ucas.ac.cn}
\affiliation{Theoretical Condensed Matter Physics and Computational Materials Physics Laboratory, College of Physical Sciences, University of Chinese Academy of Sciences, Beijing 100049, China}
\affiliation{CAS Center for Excellence in Topological Quantum Computation, University of Chinese Academy of Sciences, Beijing 100190, China}
\affiliation{Kavli Institute of Theoretical Sciences, University of Chinese Academy of Sciences, Beijing 100049, China}
\begin{abstract}
The Linear behavior of thermal transport has been widely explored, both theoretically and experimentally. On the other hand, the nonlinear thermal response has not been fully discussed. In light of the thermal vector potential theory [Phys. Rev. Lett. \textbf{114}, 196601 (2015)],
we develop a general formulation to calculate the linear and nonlinear dynamic thermal responses. In the DC limit, we recover the well-known Mott relation and the Wiedemann-Franz (WF) law at the linear order response, which link the thermoelectric conductivity $\eta$, thermal conductivity $\kappa$ and electric conductivity $\sigma$ together. To be specific, the linear Mott relation describes the linear $\eta$ is proportional to the first derivative of $\sigma$ with respect to Fermi energy (for brevity we call the first derivative, the others are similar); and the linear WF law shows the linear $\kappa$ is proportional to the zero derivative (i.e. the $\sigma$ itself).  We found there are higher-order Mott relation and WF law which follow an order-dependent  relation. At the second order, the Mott relation indicates that the second order $\sigma$ is proportional to the zero derivative of the second order $\eta$; but the second WF law shows that the second $\sigma$ is proportional to the first derivative of $\kappa$. At the third order, the derivative order increases once.
Although we only did explicit calculate up to the third order response, we can deduce that the  $n$-th order electric conductivity is proportional to the $n$-2-th  derivative of the $n$-th order thermoelectric conductivity for the nonlinear Mott relation;  and the $n$-th order electric conductivity is proportional to the $n$-1-th derivative of the $n$-th order thermal conductivity for the nonlinear WF law. Since the second order Hall effect has been studied in experiment, our theory may be tested by measuring the second order Mott and WF as well. Our theory is presented explicitly for fermion, it can also be applied to bosons. As an example, we calculate the second order thermal conductivity of magnons in a strained collinear antiferromagnet on a honeycomb, in which the linear response disappears.
\end{abstract}
\maketitle

\section{Introduction}
The interaction of temperature gradient with matter encompasses a wide range of phenomena, including the conversion of heat and electricity or spins, which is essential for the engineering of thermoelectric and other energy conversion applications. Significant efforts have been devoted to understanding the thermal response in various materials, but most of them are devoted to linear order. In analogy with the anomalous Hall effect, Berry curvature plays a significant role in thermoelectric transport, known as the anomalous Nernst effect (ANE)  \cite{PhysRevLett.97.026603, PhysRevB.78.174508, PhysRevB.79.245424, Zhu_2013}. Owing to the Onsager's reciprocal relations, the Hall conductivity or Nernst coefficient have to be vanishing in a time-reversal invariant system  \cite{PhysRevLett.105.026805, PhysRevB.92.235447, PhysRevLett.115.216806}. However, with increasing interests on nonlinear properties of topological materials, the nonlinear responses could appear in the presence of time-reversal symmetry but with broken inversion symmetry. Recently, the nonlinear anomalous Nernst effect has been predicted in transition-metal dichalcogenides  \cite{PhysRevB.99.115201,PhysRevB.99.201410,PhysRevB.100.245102}. These nonlinear thermal responses appear with distinctive behaviors and have become promising tools for understanding novel materials with low crystalline symmetry in experiments.

Most transport theories of thermally driven lattice systems are mostly phenomenological and numerical. This is because  temperature gradients are macroscopic quantities  after statistical averaging, and thus is impossible to integrate into the Hamiltonian in a straightforward way. However, Luttinger provided a solution in 1964 \cite{PhysRev.135.A1505}. To describe the effect of temperature gradient, he introduced a fictitious scalar field $\Psi$, which is called the ``gravitational" potential, that couples to energy density $h(\bm{r})$. The Luttinger's Hamiltonian is
\begin{equation}\label{lut}
H_{L}=\int d^3rh(\bm{r})\Psi(\bm{r}).
\end{equation}
The Hamiltonian of the system is then given as $H^{\Psi}=\int d^3rh^{\Psi}(\bm{r})$, with the modified energy density $h^{\Psi}(\bm{r}) = [1+\Psi(\bm{r})]h(\bm{r})$. 
By the constriction of Einstein relation, the potential satisfies $\bm{\nabla} \Psi = \bm{\nabla} T/T$. In this way the dynamical response of the system to the varying field $\Psi$ would be equivalent to the response to a temperature gradient assuming that the latter is slowly varying.
Hence the thermal transport coefficient can be directly calculated by linear response theory with Kubo formula. In the following we call this original proposal  thermal scalar potential (TSP) method.

In the half century  since the proposal of the original idea, Luttinger's method has found several applications in the calculation of the linear thermoelectric response.
Nonetheless, a general nonlinear thermoelectric response theory is still lacking. Another point is that the external field may cause the electron to excite to another band or to  move to a nearby $k$ point on the same band. Hence it needs a unified treatment of the two drift effects due to an external field in crystalline systems. This problem is handled in nonlinear optical response calculations, both in length gauge \cite{PhysRevB.52.14636, PhysRevB.53.10751,PhysRevB.61.5337,PhysRevB.90.245423,PhysRevB.94.045434} and velocity gauge \cite{PhysRevB.97.235446,PhysRevB.99.045121}. Motivated by these developments, we devote to developing a quantum theory for thermal response including generally the linear and nonlinear responses.

However, it is proved that a direct application of the coupling Hamiltonian \Eq{lut} often leads to unphysical divergent results as $T\rightarrow 0$. It is shown that the divergence can be eliminated by introducing the vector potential representation \citep{PhysRevLett.114.196601}. By imposing the continuity equation for energy density $\varepsilon$ and energy current density $\bm{j}_{\varepsilon}$, the Luttinger Hamiltonian \Eq{lut} can be transformed into vector potential form
\begin{equation}\label{tata1}
H_{L}=\int d^3r\bm{j}_{\varepsilon}(\bm{r},t)\cdot \bm{A}_{T}(t),
\end{equation}
in which $\bm{j}_{\varepsilon}$ is the energy current density and $\bm{A}_{T}$ is the thermal vector potential, which satisfies
\begin{equation}
\partial_{t}\bm{A}_{T}(\bm{r},t)=\bm{\nabla} \Psi(\bm{r},t)=\frac{\bm{\nabla} T(t)}{T}.
\end{equation}
The Hamiltonian \Eq{tata1} is equivalent to Luttinger's Hamiltonian. The derivation of \Eq{tata1} in Ref. \citep{PhysRevLett.114.196601} is under the assumption that the temperature gradient is static. In order to make it universal significant, we adopt a time-dependent temperature gradient, and the vector potential Hamiltonian \Eq{tata1} is still valid. For a comparison, we call the introduction of the vector potential representation as thermal vector potential (TVP) method.

For the case of electromagnetic vector potential $\bm{A}$, the charge conservation is guaranteed by the U(1) gauge invariance. However, for TVP $\bm{A}_{T}$, there is no such a gauge symmetry.  In velocity gauge, the minimal coupling  free electron Hamiltonian including the thermal vector potential is given by \cite{PhysRevLett.114.196601}
\begin{equation}\label{tata}
\hat{H}_{A_{T}}=\frac{\hbar^2}{2m}\sum_{\bm{k}}(\bm{k}-\varepsilon_{\bm{k}}A_{T})^2\hat{c}_{\bm{k}}^\dagger \hat{c}_{\bm{k}}.
\end{equation}
For a general multi-band Hamiltonian, the minimal coupling Hamiltonian is  generalized to
\begin{equation}\label{mini}
\hat{H}_{A_{T}}=\hat{H}_{0}(\bm{k}-\hat{H}_{0}\bm{A}_{T}).
\end{equation}
The many-body crystalline Hamiltonian reads
\begin{equation}
\hat{H}_{0}=\sum_{p,\bm{k}}\varepsilon_{p\bm{k}}\hat{c}_{p\bm{k}}^\dagger \hat{c}_{p\bm{k}},
\end{equation}
where the latin index $p$ is the band index.

In the present work, we explicitly derive the {\it{dynamical}} thermal-thermal and thermoelectirc response coefficients by developing a theory based on TVP, and consider their DC limit. The frequency dependence of thermal-thermal response is receiving more and more attention in recent years, as a crucial issue especially for the thermal design of microprocessors in which the clock frequencies work in GHz. It is crucial to cool the Joule heat in such system \cite{PhysRevLett.87.074301}. Shastry \cite{PhysRevB.73.085117,Shastry_2008} and others \cite{doi:10.1063/1.4759366, PhysRevB.91.165311} explored the linear dynamical thermal conductivity and  thermoelectirc response mediated by electrons and phonons via the TVP method; while the nonlinear counterpart has been given less attention which should play an important role when the linear part disappears due to symmetry.
%
We apply a canonical perturbation theory, both in velocity gauge and length gauge, to deal with the thermal nonlinear response with quantum effect fully considered. In this method the nonlinear thermal response fundamentally involves interband processes which are difficult to model semiclassically.

The manuscript is organized as follows: In Sec. \ref{sec2} we introduce the perturbation expansion Hamiltonian in velocity gauge and derive the nonlinear thermal response, including nonlinear Nernst conductivity and nonlinear thermal conductivity. In Sec. \ref{sec3} we present the formula given by length gauge and compare the semiclassical results in static  limit.   As an example of application, we present a calculation of nonlinear magnon Hall effect in a collinear antiferromagnetic system in Sec. \ref{sec4}. The last section is dedicated to a summary of our results.

\section{Perturbation expansion: Diagrammatic approach }\label{sec2}
In analogy with the relation between electric field and electromagnetic vector potential, we can define the corresponding ``thermal field" ($\mathbf{E}_{T}$) to thermal vector potential $\mathbf{A}_{T}$ as
\begin{equation}
\bm{E}_{T}=-\frac{\partial \bm{A}_{T}}{\partial t}=-\frac{\bm{\nabla} T(t)}{T},
\end{equation}
and their Fourier transformation
\begin{equation}
\bm{E}_{T}(\omega)=i\omega \bm{A}_{T}(\omega).
\end{equation}
The spatial variation of the temperature gradient is assumed to be much larger than the material, so that the thermal field has no spatial dependence.
The particle current is expanded in powers of the thermal field
\begin{equation}
\begin{aligned}
\langle \hat{J}^{\alpha}_{N} \rangle (\omega)&=\int d\omega_{1}L_{12}^{\alpha\beta}(\omega;\omega_{1})E_{T}^{\beta}\delta_{\omega_{1},\omega}\\
&+\int d\omega_{1}d\omega_{2}L_{12}^{\alpha\beta\gamma}(\omega;\omega_{1},\omega_{2})E_{T}^{\beta}E_{T}^{\gamma}\delta_{\omega_{1}+\omega_{2},\omega}\\
&+\cdots .
\end{aligned}
\end{equation}
The Greek indices $\mu,\alpha,\beta,\cdots \in \{x,y,z\}$ are the space indexes, and $L_{12}^{\mu\alpha_{1}\cdots  \alpha_{n}}(\omega;\omega_{1}\cdots \omega_{n})$ is defined as the $n$-th order thermoelectric conductivity tensor.  The frequency before the semicolon in the response thermoelectric conductivity tensor $L_{12}^{\mu\alpha_{1}\cdots  \alpha_{n}}(\omega;\omega_{1}\cdots \omega_{n})$ represents the frequency of the
output response, and the frequencies after the semicolon represent the frequencies of the input forces.

Before expanding the minimal coupling Hamiltonian \Eq{mini} in Taylor series, one should deal with the $k$ space derivatives carefully. The important fact is that the Hamiltonian operator is differentiated first and then  its matrix elements are calculated. Owing to this covariance $k$ derivative of operator $\hat{\mathcal{O}}(\bm{k})$ is \cite{PhysRevB.91.235320,PhysRevB.99.045121}
\begin{equation}\label{bigd}
\hat{\bm{D}}_{\bm{k}}[\hat{\mathcal{O}}(\bm{k})]_{pq}\equiv[\nabla_{\bm{k}}\hat{\mathcal{O}}(\bm{k})]_{pq}
=\nabla_{\bm{k}}\mathcal{O}(\bm{k})_{pq}-i[\bm{\mathcal{A}}_{\bm{k}},\hat{\mathcal{O}}(\bm{k})]_{pq}.
\end{equation}
Here the covariant derivative operator is defined by  $\hat{D}^{\mu}$. In  \Eq{bigd} $\bm{\mathcal{A}}_{\bm{k}}$ is the Berry connection, and its component in $\alpha$ direction is $\mathcal{A}^{\alpha}_{pq}(\mathbf{k})=i\bra{u_{p\bm{k}}}\frac{\partial}{\partial k^{\alpha}} \ket{u_{q\bm{k}}}$.

The partition function with thermal field is
written as the path integral
\begin{equation}
\mathcal{Z}=\int d[\bar{c},c] {\rm{exp}}\left(-i\int dt K_{A_{T}}\right).
\end{equation}  
In which  $ K_{A_{T}}=H_{A_{T}}-\mu N=\hat{K}_{0}(\bm{k}-\hat{K}_{0}\bm{A}_{T})$, with $K_{0}=\sum_{p,\bm{k}}\tilde{\varepsilon}_{p\bm{k}}\hat{c}_{p\bm{k}}^\dagger \hat{c}_{p\bm{k}}$,   and $\tilde{\varepsilon}_{p}=\varepsilon_{p}-\mu $ is the energy measured from the Fermi energy.

Different from the direct expansion of Hamiltonian in series of electromagnetic vector potential in calculating the nonlinear electric conductivity, the Hermiticity should be ensured in expanding $\hat{K}_{A_{T}}$ in series of thermal vector potential $A_{T}$.
For example, the first order perturbation of $\hat{K}_{A_{T}}$ is
\begin{equation}
\hat{K}_{A_{T}}\approx \hat{K}_{0}+\frac{1}{2}A_{T}^{\alpha}\left[\hat{K}_{0},\hat{D}^{\alpha}[\hat{K}_{0}]\right]_{+},
\end{equation}
where the sum over space index $\alpha$ is implicit, and $[\cdots]_{+}$ is the anti-commutation operation. To distinguish from the normal bracket, we use $[\cdots]_{-}$ to denote the commutation operation in the following.
The grand-canonical ensemble energy operator $K_{A_{T}}$ can be expanded by Taylor series in terms of thermal vector potential
\begin{widetext}
\begin{equation}\label{hexp}
\hat{K}_{A_{T}}=\hat{K}_{0}+\sum_{n=1}^{\infty}\frac{1}{n!}\prod_{k=1}^{n}\frac{1}{2}{A}_{T}^{\alpha_{1}}
\left[\hat{K}_{0},\hat{D}^{\alpha_{1}}\left[\frac{1}{2}{A}_{T}^{\alpha_{2}}\left[\hat{K}_{0},\cdots,\hat{D}^{\alpha_2}
\left[\frac{1}{2}{A}_{T}^{\alpha_{k}}\left[\hat{K}_{0},\hat{D}^{\alpha_{k}}[\hat{K}_{0}]\right]_{+}\right]\right]_{+}\right]\right]_{+}.
\end{equation}
\end{widetext}
\Eq{bigd} can be used to write the velocity operator of the unperturbed system as
\begin{equation}
\hat{\bm{v}}=\hat{\bm{D}}[\hat{K}_{0}].
\end{equation}
The higher order {\it direct} derivatives of the unperturbed Hamiltonian is written as
\begin{equation}
\hat{h}^{\alpha_{1}\cdots \alpha_{n}}=\hat{D}^{\alpha_{1}}\cdots \hat{D}^{\alpha_{n}}[\hat{K}_{0}].
\end{equation}
We introduce the superoperator $\mathcal{D}^{\alpha}$ which is defined as the {\it Hermitian} derivative
\begin{equation}\label{superd}
\hat{\mathcal{D}}^{\alpha}[\hat{\mathcal{O}}]=\frac{1}{2}\left[\hat{K}_{0},\hat{D}^{\alpha}[\hat{\mathcal{O}}]\right]_{+}.
\end{equation}
It should be noted that the the Hermitian derivative superoperators defined in \Eq{superd} carry an additional dimension $[\rm{energy}]^1$  than that of the direct derivative. Hence the Hermitian derivative of the unperturbed $K_{0}$ is defined as
\begin{equation}
\hat{\mathcal{K}}^{\alpha_{1}\cdots \alpha_{n}}=\hat{\mathcal{D}}^{\alpha_{1}}\cdots \hat{\mathcal{D}}^{\alpha_{n}}[\hat{K}_{0}].
\end{equation}
Again, the dimension of $n$-th order Hermitian derivative of the unperturbed Hamiltonian is $n$-power higher than that of the $n$-th order direct derivative of the unperturbed Hamiltonian.

Through Fourier transformation, the expanded $K_{A_{T}}$ is simplified as
\begin{eqnarray}\label{perth}
\hat{K}_{A_{T}}
&=&\hat{K}_{0}+\sum_{n=1}^{\infty}\frac{1}{n!}\prod_{k=1}^{n}\int d\omega_{k}e^{i\omega_{k}t}\frac{-i}{\hbar\omega_{k}}E^{\alpha_{k}}_{T}(\omega_{k})  \notag\\
&\times& \hat{\mathcal{K}}^{\alpha_{1}\cdots \alpha_{k}}.
\end{eqnarray}
Very recently, a diagrammatic approach has been developed to calculate the optical conductance in velocity gauge \cite{PhysRevB.99.045121, Jo_o_2019}. We generalize it in calculating the dynamical thermal response: the propagation of the temperature gradient is defined as a quasiparticle "thermalon". With the aid of TVP concept, the linear and nonlinear thermoelectric response can be derived and the mutual-relation between heat and charge can be studied at nonlinear level revealing deeper physics beyond the linear response.

The local particle current operator is defined as $\hat{\bm{J}}_{N}\equiv \hat{\bm{v}}_{T}$, here $\hat{\bm{v}}_{T}$ is the velocity operator in the perturbed system depending on the thermal field 
\begin{equation}\label{velo}
\begin{aligned}
\hat{\upsilon}^{\alpha}_{T}(t)=&\hat{D}^{\alpha}[\hat{K}_{A_{T}}]\\
=&\sum_{n=1}^{\infty}\frac{1}{n!}\prod_{k=1}^{n}\int d\omega_{k}e^{i\omega_{k}t}\frac{-i}{\hbar\omega_{k}}E_{T}^{\alpha_{k}}(\omega_{k})\hat{D}^{\alpha}[\hat{\mathcal{K}}^{\alpha_{1}\cdots \alpha_{k}}].
\end{aligned}
\end{equation}
The local heat current operator is defined as $\hat{\bm{J}}_{Q} = \hat{\bm{J}}_{E} - \mu \hat{\bm{J}}_{N}$, with $\mu$ the chemical potential. 
 An exact from of the energy current operator $\hat{\bm{J}}_{E}$ can be derived form the conservation equation using Luttinger's Hamiltonian \cite{PhysRevB.55.2344}
\begin{equation}
\frac{\partial \hat{h}^{\Psi}(\bm{r})}{\partial t}=\frac{1}{i\hbar}[\hat{h}^{\Psi}(\bm{r}),\hat{H}^{\Psi}]=-\bm{\nabla}\cdot \hat{\bm{J}}_{E}(\bm{r}).
\end{equation}
 Using $H^{\Psi}=H_{A_{T}}$, the result is (for the derivation in detail see Appendix.\ref{app11})
\begin{equation}\label{hct}
\hat{J}^{\alpha}_{Q}=\frac{1}{2}(\hat{v}^{\alpha}_{T}\hat{K}_{A_{T}}+\hat{K}_{A_{T}}\hat{v}^{\alpha}_{T}) -\frac{i\hbar}{8}\sum_{\gamma}\nabla_{\gamma} (\hat{v}^{\alpha}_{T}\hat{v}^{\gamma}_{T}-\hat{v}^{\alpha}_{T}\hat{v}^{\gamma}_{T}).
\end{equation}
It has been proved that the last term cancels when calculating the Kubo formula. In this case the heat current operator converts to the usual anticommutator representation  $\hat{\bm{J }}_{Q}=\frac{1}{2}[\hat{K}_{A_{T}},\bm{\hat{v}}_{T}]_{+}$. It is worth noting that the heat current operator defined through the conservation equation is compatible with the definition via the thermodynamics of the entropy flux (see Appendix.\ref{app11}). 
An important issue in thermally driven current transport is the magnetization effect. Owing to the orbital motion of Bloch electrons, the magnetization current should be subtracted from the local current \cite{PhysRevB.55.2344,  PhysRevLett.97.026603, RevModPhys.82.1959}
\begin{equation}\label{Jne-defi}
\bm{J}_{N(E)}^{\rm{tr}}=\bm{J}_{N(E)}-\bm{\nabla} \times \bm{M}_{N(E)}(\bm{r}).
\end{equation}
In which $\bm{J}_{N(E)}^{\rm{tr}}$ is the electric (energy) current for transport,  $\bm{J}_{N(E)}$ is the local charge (energy) current, and $\bm{M}_{N(E)}(\bm{r})$ is the particle (energy) magnetization density.  The transport heat current is evaluated as
\begin{equation}\label{jq1}
\bm{J}_{Q}^{\rm{tr}}=\bm{J}_{E}^{\rm{tr}} -\mu\bm{J}_{N}^{\rm{tr}}.
\end{equation}
Alternatively, the heat magnetization can be introduced through the relation \cite{PhysRevLett.107.236601, Zhang_2016, PhysRevB.102.235161}
\begin{equation}\label{mheat}
\bm{M}_{Q}(\bm{r})\equiv \bm{M}_{E}(\bm{r})-\mu \bm{M}_{N}(\bm{r}).
\end{equation}
The transport heat current is given as
\begin{equation}\label{jq2}
\bm{J}_{Q}^{\rm{tr}}=\bm{J}_{Q}-\bm{\nabla} \times \bm{M}_{Q}(\bm{r}),
\end{equation}
in which the local heat current is given as
\begin{equation}\label{localjq}
\bm{J}_{Q}=\bm{J}_{E} -\mu\bm{J}_{N}.
\end{equation}
Combining \Eq{Jne-defi}, \Eq{mheat} and \Eq{localjq}, one can verify that the two definitions of transport heat current \Eq{jq1} and \Eq{jq2} are equivalent:
\begin{eqnarray}
    \bm{J}_{Q}^{\rm{tr}} &=& \bm{J}_{Q}-\bm{\nabla} \times \bm{M}_{Q} \notag\\
                         &=& \bm{J}_{E}-\mu\bm{J}_{N}-\bm{\nabla}\times (\bm{M}_{E}-\mu\bm{M}_{N}) \notag\\ 
                         &=& (\bm{J}_{E} - \bm{\nabla}\times \bm{M}_{E}) -\mu(\bm{J}_{N}-\bm{\nabla}\times \bm{M}_{E}) \notag \\
                         &=&\bm{J}_{E}^{\rm{tr}}-\mu\bm{J}_{N}^{\rm{tr}}.
    \label{equivalenceEq}
    \end{eqnarray}
 Similar derivation can be found in \cite{PhysRevB.101.235430}. In the rest of this paper, the transport heat current is calculated through \Eq{jq2}. 
The density matrix is written as
\begin{equation}
\hat{\rho}\approx \hat{\rho}_{\rm{leq}} + \hat{\rho}_{1},
\end{equation}
where $\hat{\rho}_{\rm{leq}}$ is the local equilibrium density matrix characterized by the local chemical potential $\mu(\bm{r})$ and local temperature $T(\bm{r})$
\begin{equation}
\hat{\rho}_{\rm{leq}} = \frac{1}{Z}{\rm{exp}}\left[-\int d\bm{r}\frac{\hat{h}(\bm{r})-\mu(\bm{r})\hat{n}(\bm{r})}{k_{B}T(\bm{r})} \right],
\end{equation}
and $\rho_{1}$ is the linear response correction to the local equilibrium density matrix.
Therefore the local current is contributed by two parts
\begin{equation}
\bm{J}_{N(Q)}=\bm{J}^{\rm{Kubo}}_{N(Q)}+\bm{J}^{\rm{leq}}_{N(Q)},
\end{equation}
where $\bm{J}_{N(Q)}^{\rm{Kubo}}$ is the direct response current, which is the direct conjugate variables of magnetic vector potential $\bm{A}$ (which will be noted as $\bm{A}_{B}$ in the following  for clarity) and TVP $\bm{A}_{T}$. $\bm{J}^{\rm{leq}}_{N(Q)}$ is the local equilibrium current, which comes from the inhomogeneous local chemical potential and temperature field.
The local equilibrium current satisfies \cite{PhysRevLett.107.236601}
\begin{equation}\label{leqc}
\bm{J}^{\rm{leq}}_{N}=\bm{\nabla} \times \bm{M}_{N}(\bm{r})-\bm{M}_{N}(\bm{r})\times \bm{E}_{T},
\end{equation}
\begin{equation}\label{leqh}
\bm{J}^{\rm{leq}}_{Q}=\bm{\nabla} \times \bm{M}_{Q}(\bm{r})-\bm{M}_{N}(\bm{r})\times \bm{E}-2\bm{M}_{Q}(\bm{r})\times \bm{E}_{T}.
\end{equation}
The expressions for the local equilibrium current \Eq{leqc} and \Eq{leqh} convert to the bulk magnetization current when considering a finite system \cite{PhysRevB.55.2344}. Noting that for transport current, the magnetization current should be subtracted (see \Eq{Jne-defi}). For electric-electric response, the local equilibrium current exactly cancels the magnetization current, and the transport is uniquely determined by Kubo formula. However, for electric-thermal, thermoelectric and thermal-thermal responses, the terms proportional to external fields do not cancel the magnetization, which leave as the correction to Kubo formula.
Hence the transport currents become
\begin{equation}\label{jtr}
\bm{J}_{N}^{\rm{tr}}=\bm{J}^{\rm{Kubo}}_{N}-\bm{M}_{N}(\bm{r})\times \bm{E}_{T},
\end{equation}
\begin{equation}
\bm{J}_{Q}^{\rm{tr}}=\bm{J}^{\rm{Kubo}}_{Q}-\bm{M}_{N}(\bm{r})\times \bm{E}-2\bm{M}_{Q}(\bm{r})\times \bm{E}_{T}.
\end{equation}
The expectation values of Kubo response currents $\bm{J}_{c(h)}^{\rm{Kubo}}$ are
\begin{equation}
\bm{J}_{N(Q)}^{\rm{Kubo}}=\frac{1}{\mathcal{Z}}\frac{\delta \mathcal{Z}[\bm{A}_{B(T)}]}{\delta \bm{A}_{B(T)}},
\end{equation}
with the path-integral form
\begin{equation}\label{jresc}
\begin{aligned}
\langle \hat{\bm{J}}_{N(Q)}^{\rm{Kubo}} (t)\rangle =&\frac{1}{\mathcal{Z}}{\rm{Tr}}\left[\mathcal{T}\hat{\bm{J}}_{N(Q)}(t)e^{-i\int dt^\prime K_{A_{B(T)}}(t^\prime)}\right]  \\
=&\frac{1}{\mathcal{Z}}\int d[\bar{c},c] \bm{J}_{N(Q)}(t){\rm{exp}}\left[-i\int dt^\prime K_{A_{B(T)}}(t^\prime)\right],
\end{aligned}
\end{equation}
where $d[\bar{c},c]$ denotes the functional measure with $\bar{c},c$ the Grassmann variables constructing the Hamiltonian. 

The zero-field expectation values of the particle magnetization and heat magnetization are
\begin{equation}
\bm{M}_{N(Q)}=-\lim_{\bm{B}_{B(T)}\rightarrow 0}\frac{\delta \Omega[\bm{B}_{B(T)}]}{\delta \bm{B}_{B(T)}},
\end{equation}
where $\Omega=F-TS$ is the grand thermodynamic potential, the Landau free energy can be written as $F=-\frac{1}{\beta}{\rm{log}}(\mathcal{Z})$. It is convenient to introduce the auxiliary particle (heat) magnetization
\begin{equation}
\begin{aligned}
\tilde{\bm{M}}_{N(Q)}=&-\lim_{\bm{B}_{B(T)}\rightarrow 0}\frac{\delta F[\bm{B}_{B(T)}]}{\delta \bm{B}_{B(T)}}\\
\end{aligned}
\end{equation}
which can be  alternatively written in a TVP form by taking the long-wavelength limit \cite{PhysRevLett.99.197202, PhysRevLett.107.236601, PhysRevB.102.235161}
\begin{equation}
\tilde{\bm{M}}_{N(Q)}=\lim_{\bm{l}\rightarrow 0}\bm{\nabla}_{\bm{l}}\times \frac{\delta F[\bm{A}_{B(T)}]}{\delta \bm{A}_{B(T),\bm{l}}}.
\end{equation}
$\tilde{\bm{M}}_{N}$ and $\tilde{\bm{M}}_{N}$ in path-integral formalism are written as
\begin{equation}\label{jm}
\begin{aligned}
\langle\tilde{\bm{M}}_{N}(t)\rangle=&\frac{\beta}{2i}\bm{\nabla}_{\bm{l}}\times \frac{1}{\mathcal{Z}}\int d[\bar{c},c] K_{A_{T},\bm{-l}}(t)\bm{D}[K_{0,\bm{l}}]\\
&\times {\rm{exp}}\left[-i\int dt^\prime K_{A_{T}}(t^\prime)\right], 
\end{aligned}
\end{equation}
\begin{equation}\label{jm2}
\begin{aligned}
\langle\tilde{\bm{M}}_{Q}(t)\rangle=&\frac{\beta}{2i}\bm{\nabla}_{\bm{l}}\times \frac{1}{\mathcal{Z}}\int d[\bar{c},c] K_{A_{T},\bm{-l}}(t)\bm{\mathcal{D}}[K_{0,\bm{l}}]\\
&\times {\rm{exp}}\left[-i\int dt^\prime K_{A_{T}}(t^\prime)\right].
\end{aligned}
\end{equation}
By use of the Maxwell relation
$
\partial S/\partial B=\partial M/\partial T ,
$
the particle (heat) magnetization satisfies \cite{PhysRevLett.99.197202, PhysRevLett.107.236601, PhysRevB.102.235161}
\begin{equation}\label{pb1}
\frac{\partial (\beta\bm{M}_{N})}{\partial \beta}=\tilde{\bm{M}}_{N},
\end{equation}
\begin{equation}\label{pb2}
\frac{\partial (\beta\bm{M}_{Q}-\beta\delta M_{Q})}{\partial \beta}=\tilde{\bm{M}}_{Q}.
\end{equation}

With the notation  $J_{1(2)}\equiv J_{c(h)}$, $E_{1}\equiv E$, $E_{2}\equiv E_{T}$ and $M_{1(2)}\equiv M_{N(Q)}$, we introduce the set  of transport equations at $n$-th order
\begin{equation}
\begin{bmatrix}  J^{(n),\alpha}_{1}(\omega) \\ J^{(n),\alpha}_{2}(\omega) \end{bmatrix}=\begin{bmatrix} L_{11}^{{\rm{tr}},(n)} & L_{12}^{{\rm{tr}},(n)} \\ L_{21}^{{\rm{tr}},(n)} & L_{22}^{{\rm{tr}},(n)} \end{bmatrix} \begin{bmatrix}  \prod_{k=1}^{n} E_{1}^{\alpha_{k}}(\omega_{k}) \\ \prod_{k=1}^{n} E^{\alpha_{k}}_{2}(\omega_{k}) \end{bmatrix},
\end{equation}
with the response functions
\begin{widetext}
\begin{equation}
L_{ij}^{{\rm{tr}},(n)}=\left[\prod_{k=1}^{n} \int d\omega _{k}\right]\left[ L_{ij}^{{\rm{Kubo}},\alpha\alpha_{1}\cdots \alpha_{n}}(\omega;\omega_{1}\cdots \omega_{n})-\epsilon^{\alpha\alpha_{1}\gamma}C_{ij} M_{ij}^{\gamma\alpha_{2}\cdots \alpha_{n}}(\omega;\omega_{1}\cdots \omega_{n})\right]\delta_{\omega,\omega_{1}+\cdots +\omega_{n}},
\end{equation}
where $C_{11}=0$, $C_{12}=C_{21}=1/\beta$, $C_{22}=2/\beta$ and $\epsilon^{\alpha\beta\gamma}$ is the Levi-Civita symbol.
The Kubo responses $L_{ij}^{\rm{Kubo}}$ are given by
\begin{equation}\label{lijk}
L_{ij}^{{\rm{Kubo}},\alpha\alpha_{1}\cdots \alpha_{n}}(\omega;\omega_{1}\cdots\omega_{k}) =\int \frac{dt}{2\pi}e^{i\omega t}\prod_{k=1}^{n}\int  \frac{dt_{k}}{2\pi}e^{i\omega_{k} t_{k}}    \frac{\delta}{\delta E_{j}^{\alpha_{k}}(\omega_{k})}\left\langle \hat{J}^{\alpha}_{i}(t)\right\rangle |_{E_{j}^{\alpha_{k}}(\omega_{k})=0},
\end{equation}
and we define the magnetization response $M_{ij}$
\begin{equation}\label{mij}
M_{ij}^{\gamma\alpha_{2}\cdots \alpha_{n}}(\omega;\omega_{1}\cdots\omega_{k}) =\int \frac{dt}{2\pi}e^{i\omega t}\prod_{k=1}^{n}\int  \frac{dt_{k}}{2\pi}e^{i\omega_{k} t_{k}}    \frac{\delta}{\delta E_{j}^{\alpha_{k}}(\omega_{k})}\left\langle \hat{M}_{i}^{\gamma} (t)E_{j}^{\alpha_{1}}(t)\right\rangle|_{E_{j}^{\alpha_{k}}(\omega_{k})=0}.
\end{equation}
\end{widetext}

Based on the form of  \Eq{jresc} and \Eq{jm}, the Kubo contribution of the charge current is dually expanded in powers of TVP, given that the velocity operator and the exponent depend on TVP, while the magnetization is singly expanded. Hence the $n$th order response is computed by drawing all connected diagrams. One should pay attention to drawing the diagrams  that the out-going vertex which corresponds to the expansion of $v$ and in-coming vertex which corresponds to the expansion of action should be distinguished.


Thus, the n-th order thermoelectric response is calculated using the following rules:

1. For the Kubo contribution $L_{12}^{\rm{Kubo}}$, draw all the connected diagrams including $n$ incoming thermalon lines connected by incoming vertexes (symboled as  $\bullet$) and an outgoing photon line connected by one outgoing vertex (symboled as  $\circ$). All the inner lines are composed of electron propagators.

 For the magnetization $M_{12}$,  a subtle point is that two types of incoming vertices should be distinguished. One of which  (symboled as  {\tiny $\blacksquare$}) connects a thermalon line with the momentum $\bm{l}$, the other is the one identical to that of $L_{12}^{{\rm{Kubo}}}$. The outgoing vertex (symboled as $\odot$) connects only one photon line with the momentum $\bm{-l}$.

2. Integrate over the internal frequencies. The electron propagator is the free fermion Green's function $G_{p}(\omega)=1/(\omega-\varepsilon_{p}+\mu)$. The propagation of thermalon is treated classically, with the propagator being unity.
For the Kubo term $L_{12}^{\rm{Kubo}}$, the value of incoming vertex connecting $n$ thermalon is $\prod_{k=1}^{n}(\frac{i}{\hbar \omega_{k}})\mathcal{K}_{pq}^{\alpha_{1}\cdots\alpha_{n}}$. And the value of outgoing vertex connecting $n$ thermalon is $\prod_{k=1}^{n}(\frac{i}{\hbar \omega_{k}})D^{\alpha_{1}}[\mathcal{K}]_{pq}^{\alpha_{2}\cdots\alpha_{n}}$.

For the magnetization $M_{12}$, the $\bm{l}$ dependent incoming vertex is  $\frac{1}{2}\left[\prod_{k=1}^{n}(\frac{i}{\hbar \omega_{k}})\mathcal{K}_{pq,\bm{k}}^{\alpha_{1}\cdots\alpha_{n}}+\prod_{k=1}^{n}(\frac{i}{\hbar \omega_{k}})\mathcal{K}_{pq,\bm{k}+\bm{l}}^{\alpha_{1}\cdots\alpha_{n}}\right]$. The outgoing vertex is $\frac{1}{2}(h^{\alpha}_{pq,\bm{k}}+h^{\alpha}_{pq,\bm{k}+\bm{l}})$. Then calculate the curl with respect to $\bm{l}$ in the long-wavelength limit $\bm{l}\rightarrow 0$, and integrating the auxiliary  magnetization with respect to $\beta$ by use of the relation \Eq{pb1} and \Eq{pb2} to obtain the magnetization.

3. Multiply the symmetry factor by permuting $\alpha_{k}$ and $\omega_{k}$.

\subsection{Linear thermolectirc response}
\begin{widetext}
\begin{table*}
\caption{Values of vertices for the Kubo contribution of electric-electric thermal-electric, electric-thermal and thermal-thermal response.}\label{tab1}
\begin{ruledtabular}
\begin{tabular}{l c r}
 & Incoming vertex {\large $\bullet$} & Outgoing vertex {\large $\circ \quad \quad \quad$} \\ \hline
$L_{11}^{\rm{Kubo}}$ & $\prod_{k=1}^{n}(\frac{ie}{\hbar \omega_{k}})h_{pq}^{\alpha_{1}\cdots\alpha_{n}}$ & $e\prod_{k=1}^{n}(\frac{i}{\hbar \omega_{k}})h_{pq}^{\mu\alpha_{1}\cdots\alpha_{n}}$\\
$L_{12}^{\rm{Kubo}}$ & $\prod_{k=1}^{n}(\frac{i}{\hbar \omega_{k}})\mathcal{K}_{pq}^{\alpha_{1}\cdots\alpha_{n}}$ & $e\prod_{k=1}^{n}(\frac{i}{\hbar \omega_{k}})D^{\mu}[\mathcal{K}]_{pq}^{\alpha_{1}\cdots\alpha_{n}}$\\
$L_{21}^{\rm{Kubo}}$ & $\prod_{k=1}^{n}(\frac{ie}{\hbar \omega_{k}})h_{pq}^{\alpha_{1}\cdots\alpha_{n}}$ & $\prod_{k=1}^{n}\sum_{a=0}^{n}(\frac{i}{\hbar \omega_{k}})\frac{1}{2}[\mathcal{K}^{\alpha_{1}\cdots \alpha_{a}}, e^{n-a}h^{\mu\alpha_{a}\cdots\alpha_{n-a}}]_{pq}$\\
$L_{22}^{\rm{Kubo}}$ & $\prod_{k=1}^{n}(\frac{i}{\hbar \omega_{k}})\mathcal{K}_{pq}^{\alpha_{1}\cdots\alpha_{n}}$ & $\prod_{k=1}^{n}\sum_{a=0}^{n}(\frac{i}{\hbar \omega_{k}})\frac{1}{2}[\mathcal{K}^{\alpha_{1}\cdots \alpha_{a}}, \mathcal{K}^{\mu\alpha_{a}\cdots\alpha_{n-a}}]_{pq}$
\end{tabular}
\end{ruledtabular}
\end{table*}
\begin{table*}
\caption{Values of the momentum dependent vertices for the particle and heat magnetization.}\label{tab2}
\begin{ruledtabular}
\begin{tabular}{l c r}
 & Incoming vertex {\tiny $\blacksquare$} & Outgoing vertex {\small $\odot \quad \quad \quad$} \\ \hline
 $M_{11}$ & $\frac{1}{2}\left[\prod_{k=1}^{n}(\frac{i}{\hbar \omega_{k}})h_{pq,\bm{k}}^{\alpha_{1}\cdots\alpha_{n}}+\prod_{k=1}^{n}(\frac{i}{\hbar \omega_{k}})h_{pq,\bm{k}+\bm{l}}^{\alpha_{1}\cdots\alpha_{n}}\right]$ & $\frac{1}{2}(h^{\alpha}_{pq,\bm{k}}+h^{\alpha}_{pq,\bm{k}+\bm{l}})$\\
$M_{12}$ & $\frac{1}{2}\left[\prod_{k=1}^{n}(\frac{i}{\hbar \omega_{k}})\mathcal{k}_{pq,\bm{k}}^{\alpha_{1}\cdots\alpha_{n}}+\prod_{k=1}^{n}(\frac{i}{\hbar \omega_{k}})\mathcal{K}_{pq,\bm{k}+\bm{l}}^{\alpha_{1}\cdots\alpha_{n}}\right]$ & $\frac{1}{4}\left[(\tilde{\varepsilon}_{p,\bm{k}}+\tilde{\varepsilon}_{q,\bm{k}}) h^{\alpha}_{pq,\bm{k}}+(\tilde{\varepsilon}_{p,\bm{k}+\bm{l}}+\tilde{\varepsilon}_{q,\bm{k}+\bm{l}})h^{\alpha}_{pq,\bm{k}+\bm{l}}\right]$\\
$M_{21}$ & $\frac{1}{2}\left[\prod_{k=1}^{n}(\frac{i}{\hbar \omega_{k}})h_{pq,\bm{k}}^{\alpha_{1}\cdots\alpha_{n}}+\prod_{k=1}^{n}(\frac{i}{\hbar \omega_{k}})h_{pq,\bm{k}+\bm{l}}^{\alpha_{1}\cdots\alpha_{n}}\right]$ & $\frac{1}{2}(h^{\alpha}_{pq,\bm{k}}+h^{\alpha}_{pq,\bm{k}+\bm{l}})$\\
$M_{22}$ & $\frac{1}{2}\left[\prod_{k=1}^{n}(\frac{i}{\hbar \omega_{k}})\mathcal{K}_{pq,\bm{k}}^{\alpha_{1}\cdots\alpha_{n}}+\prod_{k=1}^{n}(\frac{i}{\hbar \omega_{k}})\mathcal{K}_{pq,\bm{k}+\bm{l}}^{\alpha_{1}\cdots\alpha_{n}}\right]$ & $\frac{1}{4}\left[(\tilde{\varepsilon}_{p,\bm{k}}+\tilde{\varepsilon}_{q,\bm{k}}) h^{\alpha}_{pq,\bm{k}}+(\tilde{\varepsilon}_{p,\bm{k}+\bm{l}}+\tilde{\varepsilon}_{q,\bm{k}+\bm{l}})h^{\alpha}_{pq,\bm{k}+\bm{l}}\right]$
\end{tabular}
\end{ruledtabular}
\end{table*}
\end{widetext}
The linear thermoelectric response is given by $L_{12}^{{\rm{Kubo}},\alpha\beta}(\omega;\omega_{1})-\epsilon^{\alpha\beta\gamma} M_{N}^{\gamma}|_{E_{T}=0}$.
Following these rules, $L_{12}^{{\rm{Kubo}},\alpha\beta}(\omega;\omega_{1})$ is found to be
\begin{equation}
\begin{aligned}
L_{12}^{{\rm{Kubo}},\alpha\beta}(\omega;\omega_{1}) &=\frac{i}{\hbar  \omega_{1}}\sum_{p,q}\int_{\bm{k}}\int d\omega^\prime \left\{ \mathcal{K}_{pq}^{\beta}G_{q}(\omega^\prime+\omega)\right. \\
&\left. \times  h_{pq}^{\alpha}G_{p}(\omega^\prime)+ D^{\alpha}[\mathcal{K}^{\beta}]_{pp}G_{p}(\omega^\prime)\right\}.
\end{aligned}
\end{equation}
\begin{figure}[htb]
\centering
\includegraphics [width=2.8in]{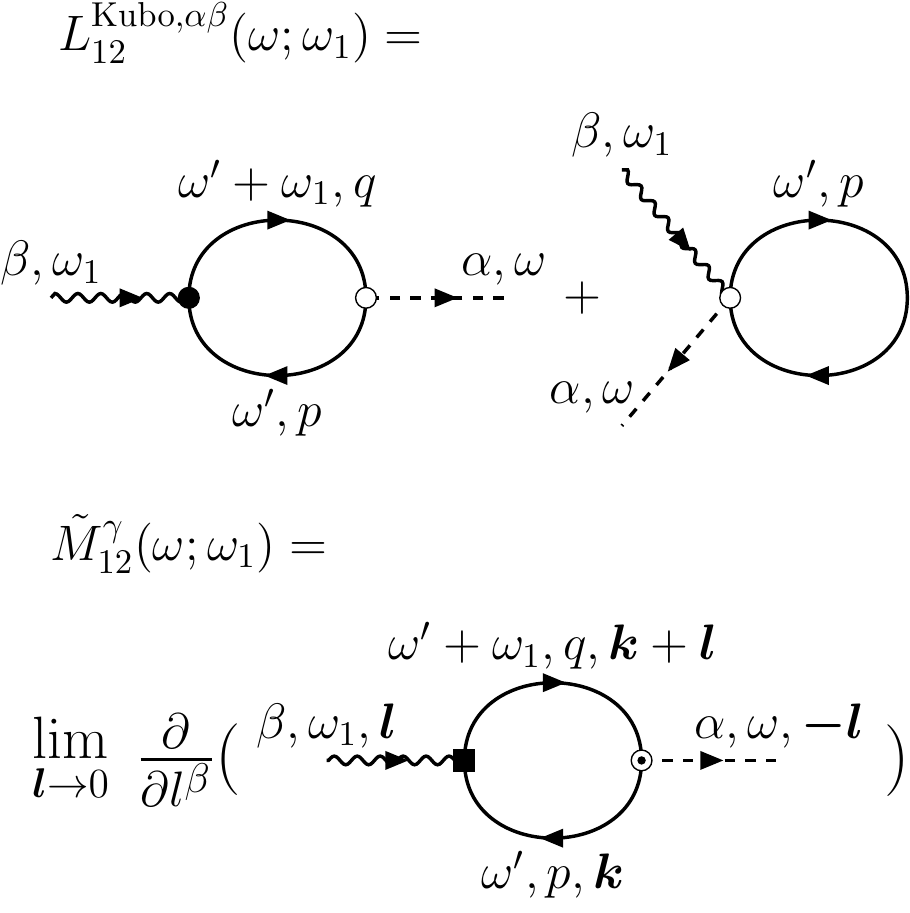}
\caption{Diagrammatic representation of $L_{12}^{{\rm{Kubo}},\alpha\beta}$ and $\tilde{M}_{12}^{\gamma}$. The dashed line connects to a current operator, and the wavy lines are thermalons describing the couplings to thermal field. The momentum of the electron propagators in $L_{12}^{{\rm{Kubo}},\alpha\beta}$ is suppressed.}\label{fig1}
\end{figure}
The integration is over the first Brillouin zone (FBZ), with $\int_{\bm{k}}=\int_{\text{FBZ}}d^{3}k/(2\pi)^{3}$.
And the corresponding diagrams are shown in \Fig{fig1}. This expansion closely resembles that of \cite{PhysRevB.99.045121} but has several differences due to the structure of the minimal coupling thermally perturbed Hamiltonian \Eq{tata}.
The first order Hermitian derivative  is expanded as
\begin{equation}\label{mh1}
\begin{aligned}
\hat{\mathcal{K}}^{\alpha}=\frac{1}{2}(\hat{K}_{0}\hat{h}^{\alpha}+\hat{h}^{\alpha}\hat{K}_{0}),
\end{aligned}
\end{equation}
and
\begin{equation}
\begin{aligned}
D^{\alpha}[\hat{\mathcal{K}}^{\beta}]=&\frac{1}{2}D^{\alpha}\left[\hat{K}_{0}D^{\beta}[\hat{K}_{0}]+D^{\beta}[\hat{K}_{0}]\hat{K}_{0}\right]\\
=&\frac{1}{2}\left(\hat{h}^{\alpha}\hat{h}^{\beta}+\hat{K}_{0}\hat{h}^{\alpha\beta}+\hat{h}^{\alpha\beta}\hat{K}_{0}+\hat{h}^{\beta}\hat{h}^{\alpha}\right).
\end{aligned}
\end{equation}
Noting that the $\hat{K}_{0}$ is a diagonal matrix, the Nernst coefficient becomes [$\alpha^{\alpha\beta}(\omega;\omega_{1})$ reduces to $\alpha^{\alpha\beta}(\omega)$ due to the conservation of energy]
\begin{equation}\label{123}
\begin{aligned}
L_{12}^{{\rm{Kubo}},\alpha\beta}&(\omega)=\frac{i}{\hbar  \omega_{1}}\sum_{p,q}\int_{\bm{k}}\int  d\omega^\prime \left\{ \frac{1}{2}\left[ \tilde{\varepsilon}_{p} h_{pq}^{\alpha}G_{q}(\omega^\prime+\omega)\right. \right. \\
& \left. \times h_{qp}^{\alpha}G_{p}(\omega^\prime) + h_{pq}^{\beta}G_{q}(\omega^\prime+\omega)\tilde{\varepsilon}_{q}h_{qp}^{\alpha}G_{p}(\omega^\prime)\right]\\
&  \left. + \left(\tilde{\varepsilon}_{p} h_{pp}^{\alpha\beta}+\frac{1}{2}h_{pq}^{\alpha}h_{qp}^{\beta}+\frac{1}{2}h_{pq}^{\beta}h_{qp}^{\alpha}\right)G_{p}(\omega^\prime)\right\}.
\end{aligned}
\end{equation}
And $\tilde{\varepsilon}$ should be read as $\varepsilon - \mu$ for simplicity. According to \Eq{bigd}, the 2nd order covariant derivative of $\hat{K}_{0}$ is
\begin{equation}
h^{\alpha\beta}_{pq}=D^{\alpha}\left[h^{\beta}\right]_{pq}=\partial^{\beta}h^{\alpha}_{pq}-i\left[\mathcal{A}^{\beta},h^{\alpha}\right]_{pq}.
\end{equation}
Together with the relation $\mathcal{A}_{pq}^{\alpha}=v_{pq}^{\alpha}/i\varepsilon_{pq}(p\neq q)$ (originating from the relation $v_{pq}^{\alpha}=\partial^{\alpha}\varepsilon_{p}\delta_{pq}-i\left[\mathcal{A}^{\alpha},H_{0}\right]_{pq}$), the linear thermoelectric response is given by
\begin{eqnarray}
L_{12}^{{\rm{Kubo}},\alpha\beta}(\omega)&=&\frac{i}{\hbar   \omega_{1}}\sum_{p,q}\int_{\bm{k}}\left[\tilde{\varepsilon}_{p}\partial^{\beta} f_{p}v_{p}^{\alpha}+\frac{1}{2}(\tilde{\varepsilon}_{p}+\tilde{\varepsilon}_{q})v_{pq}^{\beta}\right.  \notag\\
&\times& v_{qp}^{\alpha} \frac{f_{pq}}{\omega+\varepsilon_{pq}}-\tilde{\varepsilon}_{p}\left(\frac{v_{pq}^{\beta}v_{qp}^{\alpha}}{\varepsilon_{pq}}
-\frac{v_{pq}^{\alpha}v_{qp}^{\beta}}{\varepsilon_{qp}}\right)f_{p}   \notag\\
&+& \left. \frac{1}{2}\left(v_{pq}^{\beta}v_{qp}^{\alpha}+v_{pq}^{\alpha}v_{qp}^{\beta}\right)f_{p}\right],
\notag
\end{eqnarray}
where $f_{pq}=f_{p}-f_{q}$ and $\varepsilon_{pq}=\varepsilon_{p}-\varepsilon_{q}$, and the sum over band indices is only performed over the indices appearing in each term. After some simple algebra, we obtain
\begin{equation}\label{l12k}
\begin{aligned}
L_{12}^{{\rm{Kubo}},\alpha\beta}(\omega)&=\frac{i}{\hbar \omega_{1}}\sum_{p,q}\int_{\bm{k}}\left[\tilde{\varepsilon}_{p}\partial^{\beta} f_{p}v_{p}^{\alpha}+\frac{1}{2}(\tilde{\varepsilon}_{p}+\tilde{\varepsilon}_{q})\right. \\
& \left. \times v_{pq}^{\beta}v_{qp}^{\alpha}\left(\frac{f_{pq}}{\omega+\varepsilon_{pq}}-\frac{f_{pq}}{\varepsilon_{pq}}\right)\right],
\end{aligned}
\end{equation}
where the identity $v_{pq}^{\alpha}=h_{pq}^{\alpha}$ is used. The first term corresponds to the intra-band contribution with normal derivative, playing the role of the Drude weight in the dynamical thermoelectric response. And the later terms are the inter-band contributions, as we demonstrate below, they manifest themselves as the Berry curvature in the static state limit.

Now we give the derivation of $M_{12}^{\gamma}$. Referring to \Eq{jm},  we firstly derive $\tilde{M}_{12}^{\gamma}$ as
\begin{equation}
\begin{aligned}
\tilde{M}_{12}^{\gamma}&=\frac{i}{4\hbar }\sum_{p,q}\int_{\bm{k}} \int d\omega^{\prime} \frac{\partial}{i\partial l^{\beta}}\left[(\tilde{\varepsilon}_{p,\bm{k}}+\tilde{\varepsilon}_{q,\bm{k}+\bm{l}}) \right. \\
&\left. \times G_{p,\bm{k}}(\omega^\prime+\omega) (h_{pq,\bm{k}}^{\alpha}+h_{pq,\bm{k}+\bm{l}}^{\alpha})G_{q,\bm{k}+\bm{l}}(\omega)\right].
\end{aligned}
\end{equation}
Performing the frequency integral, it becomes
\begin{equation}
\begin{aligned}
\tilde{M}_{12}^{\gamma}&=\frac{i}{4\hbar }\sum_{p,q}\int_{\bm{k}} \int d\omega^{\prime} \frac{\partial}{i\partial l^{\beta}}\left[\vphantom{\frac{1}{2}} (\tilde{\varepsilon}_{p,\bm{k}}+\tilde{\varepsilon}_{q,\bm{k}+\bm{l}})\right.\\
&\left. \times(v_{pq,\bm{k}}^{\alpha}+v_{pq,\bm{k}+\bm{l}}^{\alpha})\frac{f_{p,\bm{k}}-f_{q,\bm{k}+\bm{l}}}{\omega-(\varepsilon_{p,\bm{k}}-\varepsilon_{q,\bm{k}+\bm{l}})}   \right].
\end{aligned}
\end{equation}
We first consider the interband contribution for $p\neq q$. In the long-wavelength limit $\bm{l}\rightarrow 0$, we have
\begin{equation}\label{m12pq}
\begin{aligned}
\tilde{M}_{12}^{\gamma,{\rm{inter}}}=&\frac{i}{\hbar }\sum_{p\neq q}\int_{\bm{k}}\frac{1}{2}(\tilde{\varepsilon}_{p}+\tilde{\varepsilon}_{q})\frac{v_{pq}^{\alpha}{v_{qp}^{\beta}}}{(\omega-\varepsilon_{pq})\varepsilon_{pq}}f_{pq}.
\end{aligned}
\end{equation}
The intraband contribution $\tilde{M}_{12}^{\gamma,{\rm{intra}}}$ at $\bm{l} \rightarrow 0$  when $p=q$ is given by
\begin{equation}\label{m12pp}
\begin{aligned}
\tilde{M}_{12}^{\gamma,{\rm{intra}}}=&\frac{i}{2\hbar }\sum_{p,q}\int_{\bm{k}}\left[ -\tilde{\varepsilon}_{p}\frac{(v_{pq}^{\alpha}{v_{qp}^{\beta}}-v_{pq}^{\beta}{v_{qp}^{\alpha}})}{(\omega -\varepsilon_{pq})\varepsilon_{pq}}\frac{\partial f_{p}}{\partial \varepsilon_{p}}\right].
\end{aligned}
\end{equation}
Therefore, we have
\begin{equation}\label{m12}
\begin{aligned}
\tilde{M}_{12}^{\gamma}&=\frac{i}{2\hbar }\sum_{p,q}\int_{\bm{k}}\left[(\tilde{\varepsilon}_{p}+\tilde{\varepsilon}_{q})\frac{v_{pq}^{\alpha}{v_{qp}^{\beta}}}{(\omega-\varepsilon_{pq})\varepsilon_{pq}}f_{pq}\right. \\
&\left. -\tilde{\varepsilon}_{p}\frac{(v_{pq}^{\alpha}{v_{qp}^{\beta}}-v_{pq}^{\beta}{v_{qp}^{\alpha}})}{(\omega -\varepsilon_{pq})\varepsilon_{pq}}\frac{\partial f_{p}}{\partial \varepsilon_{p}}\right].
\end{aligned}
\end{equation}
Integrating \Eq{m12} with respect to $\beta$ from Eq. (\ref{pb1}), we obtain (see Appendix.\ref{appx} for detail)
\begin{equation}\label{m12f}
\begin{aligned}
M_{12}^{\gamma}&=\frac{i}{\hbar }\sum_{p,q}\int_{\bm{k}}\left\{\frac{(v_{pq}^{\alpha}{v_{qp}^{\beta}}-v_{pq}^{\beta}{v_{qp}^{\alpha}})}{(\omega -\varepsilon_{pq})\varepsilon_{pq}} \right. \\
&\left. \times \left[\frac{1}{2}(\tilde{\varepsilon}_{p}-\tilde{\varepsilon}_{q})f_{p}+\frac{1}{\beta}{\rm{ln}}(1+e^{-\beta\varepsilon_{p}})\right]\right\}.
\end{aligned}
\end{equation}
In the DC limit, it becomes
\begin{equation}\label{m12fdc}
\begin{aligned}
M_{12}^{\gamma}=&\sum_{p}\int_{\bm{k}}\left[m_{1,p}^{\gamma}f_{p}+\frac{1}{\beta\hbar}\Omega_{p}^{\gamma}{\rm{ln}}(1+e^{-\beta\tilde{\varepsilon}_{p}})\right].
\end{aligned}
\end{equation}
The first term manifests itself as the particle magnetic moment, which is given as \cite{PhysRevLett.97.026603,PhysRevB.59.14915}
\begin{equation}
m^{\gamma}_{p} = -\frac{1}{\hbar }\epsilon^{\alpha\beta\gamma}{\rm{Im}}\bra{\partial^{\alpha} u_{p}}  (\hat{H}_{0}-\varepsilon_{p}) \ket{\partial^{\beta} u_{p}}.
\label{orbm}
\end{equation}
In Refs. \cite{PhysRevLett.97.026603,PhysRevB.59.14915}, the derivation of \Eq{orbm} starts from a wave packet hypothesis, however, its final expression does not depend on the actual shape and size of the wave packet and only depends on the Bloch functions. Therefore the orbital moment is an intrinsic property of the band. Alternatively, integrating by part, the magnetization \Eq{m12fdc} can be given as
\begin{equation}\label{m12fdcap}
\begin{aligned}
M_{12}^{\gamma}=&\sum_{p}\int_{\bm{k}}\left[m^{\gamma}_{p}f_{p}-\frac{1}{e^2}\int d\varepsilon\sigma^{\gamma}_{p}(\varepsilon)f_{p}\right],
\end{aligned}
\end{equation}
where ${\sigma}_{p}^{\gamma}(\varepsilon)=\frac{e^2}{\hbar}\int [d\bm{k}]\Theta(\varepsilon -\varepsilon_{\bm{k}}){\Omega}^{\gamma}_{p}(\bm{k})$ is the $p$ band contribution to the zero-temperature Hall conductivity  with Fermi energy $\varepsilon$.
Combining \Eq{l12k}  and \Eq{m12f}, we finally obtain the dynamical linear thermoelectric response
\begin{equation}
\begin{aligned}
L_{12}^{\rm{tr},\alpha\beta}&=\frac{i}{\hbar }\sum_{p,q}\int_{\bm{k}}\left[\frac{1}{\omega}\tilde{\varepsilon}_{p}v_{p}^{\beta}v_{p}^{\alpha}\frac{\partial f_{p}}{\partial \varepsilon_{p}} -\frac{(v_{pq}^{\alpha}{v_{qp}^{\beta}}-v_{pq}^{\beta}{v_{qp}^{\alpha}})}{(\omega -\varepsilon_{pq})\varepsilon_{pq}}\right. \\
&\left. \times \left[\tilde{\varepsilon}_{p}f_{p}+k_{B}T\ln \left(1+e^{-\beta\tilde{\varepsilon}_{p}}\right)\right]\vphantom{\frac{1}{2}}\right].
\end{aligned}
\end{equation}
In the DC limit, $\sum_{q}(v_{pq}^{\beta}v_{qp}^{\alpha}-v_{pq}^{\alpha}v_{qp}^{\beta})/\varepsilon_{qp}^{2}$ is recognized as the Berry curvature. Hence we have
\begin{equation}\label{l12length}
\begin{aligned}
L_{DC,12}^{{\rm{tr}},\alpha\beta}(\omega)&=\frac{1}{\hbar}\sum_{p}\int_{\bm{k}}\left\{i\frac{1}{\omega}\tilde{\varepsilon}_{p}v_{p}^{\beta}v_{p}^{\alpha} \frac{\partial f_{p}}{\partial \varepsilon_{p}}+\epsilon^{\alpha\beta\gamma}\Omega_{p}^{\gamma}\right.\\
&\left. \times \left[\tilde{\varepsilon}_{p}f_{p}+k_{B}T\ln \left(1+e^{-\beta\tilde{\varepsilon}_{p}}\right)\right]\right\}.
\end{aligned}
\end{equation}


The first term corresponds to the Drude weight of energy current transport, which diverges in the DC limit. This is because the considered system is a clean one. In real materials the electrons are scattered and have finite lifetime, where the electrons are not accelerated everlastingly. The second term is the topological contribution, which is represented by the Berry curvature. It is seen that the fictitious divergence is eliminated in the TVP method.

The thermoelectric conductivity $\eta$ is related to the thermoelectric response by $\eta^{\alpha\alpha_{1}\cdots\alpha_{n}} = L_{12}^{{\rm{tr}},\alpha\alpha_{1}\cdots\alpha_{n}}/T^{n}$.
The linear anomalous Nernst conductivity is given by $\eta^{\alpha\beta}=L_{12}^{{\rm{tr}},\alpha\beta}/T$. By introducing the entropy density $S_{p}=-f_{p}\ln f_{p}-(1-f_{p})\ln (1-f_{p})$ of $p$ band electrons and neglecting the Drude term, the anomalous Nernst conductivity can be written as
\begin{equation}\label{nnn}
\eta^{\alpha\beta}(\omega)=\frac{ek_{B}}{\hbar}\varepsilon^{\alpha\beta\gamma}\sum_{p}\int_{\bm{k}}\Omega_{p}^{\gamma}S_{p}.
\end{equation}
Referring to \Eq{nnn}, the expression of anomalous Nernst conductivity is consistent with the formula derived by wave packet theory in Ref. \cite{PhysRevLett.97.026603}.





\subsection{Linear thermal-thermal response}
The rules of dynamical thermal conductivity are similar to that of thermoelectric response, but with different vertex functions. The value of outgoing vertex connecting $n$ photon is $\prod_{k=1}^{n}(\frac{i}{\hbar \omega_{k}})\frac{1}{2}[h^{\alpha_{1}\cdots \alpha_{p}}, h^{\alpha_{\mu}\alpha_{p}\cdots\alpha_{n-p}}]_{pq}$, and for incoming vertex it is  $\prod_{k=1}^{n}(\frac{i}{\hbar \omega_{k}})\mathcal{K}_{pq}^{\alpha_{1}\cdots\alpha_{n}}$. Hence the linear thermal-thermal response is given by
\begin{equation}
\begin{aligned}
&L_{22}^{{\rm{Kubo}},\alpha\beta}(\omega)\\
&=\frac{i}{\hbar  \omega_{1}}\sum_{p,q}\int_{\bm{k}}\int d\omega^\prime \mathcal{K}_{pq}^{\beta}G_{q}(\omega^\prime+\omega)\mathcal{K}_{qp}^{\alpha}G_{p}(\omega^\prime)\\
&+\frac{e}{\hbar \omega_{1}}\sum_{p}\int_{\bm{k}}\int d\omega^\prime \left(\mathcal{K}^{\alpha\beta}_{pp}+ \frac{1}{2}[\mathcal{K}^{\beta},h^{\alpha}]_{pp}\right) G_{p}(\omega^\prime),
\end{aligned}
\end{equation}
where the expansion of the $2$nd order Hermitian derivative  $\mathcal{K}^{\mu\alpha}$ is involved (see Appendix.\ref{app1}).
Integrating the  Matsubara frequencies, it yields
\begin{equation}\label{kappa}
\begin{aligned}
L_{22}^{{\rm{Kubo}},\alpha\beta}&(\omega)=\frac{i}{\hbar   \omega_{1}}\sum_{p,q}\int_{\bm{k}}\left[ \tilde{\varepsilon}_{p}^{2}\partial^{\beta} f_{p}v_{p}^{\alpha}+\frac{1}{4}(\tilde{\varepsilon}_{p}+\tilde{\varepsilon}_{q})^{2} \right. \\
&\times v_{pq}^{\beta}v_{qp}^{\alpha}\frac{f_{pq}}{\omega-\varepsilon_{pq}}+\frac{1}{2}\tilde{\varepsilon}_{p}(h_{pq}^{\alpha}h_{qp}^{\beta}+h_{pq}^{\beta}h_{qp}^{\alpha})f_{p}\\
&+\frac{1}{4}(\tilde{\varepsilon}_{p}+\tilde{\varepsilon}_{q})(h_{pq}^{\alpha}h_{qp}^{\beta}+h_{pq}^{\beta}h_{qp}^{\alpha})f_{p}\\
&\left. -\tilde{\varepsilon}_{p}^{2}\left(\frac{h_{pq}^{\beta}h_{qp}^{\alpha}}{\varepsilon_{pq}}-\frac{h_{pq}^{\alpha}h_{qp}^{\beta}}{\varepsilon_{qp}}\right)f_{pq}\right],
\end{aligned}
\end{equation}
which can be written in a compact form
\begin{equation}\label{l22f}
\begin{aligned}
L_{22}^{{\rm{Kubo}},\alpha\beta}(\omega)&=\frac{i}{\hbar \omega}\sum_{p,q}\int_{\bm{k}}\left\{\tilde{\varepsilon}_{p}^{2}v_{p}^{\alpha}\partial^{\beta} f_{p} +\frac{1}{4}\left(\tilde{\varepsilon}_{p}+\tilde{\varepsilon}_{q}\right)^{2}\right.\\
&\times \left. v_{pq}^{\beta}v_{qp}^{\alpha}f_{pq}
\left(\frac{1}{\omega-\varepsilon_{pq}}+\frac{1}{\varepsilon_{pq}}\right)\right\}.
\end{aligned}
\end{equation}

Now we give the derivation of $M_{22}^{\gamma}$. According to \Eq{jm2}, we have
\begin{equation}
\begin{aligned}
\tilde{M}_{22}^{\gamma}=&\frac{i}{4\hbar }\sum_{p,q}\int_{\bm{k}} \int d\omega^{\prime} \frac{\partial}{i\partial l^{\beta}}\left[(\tilde{\varepsilon}_{p,\bm{k}}+\tilde{\varepsilon}_{q,\bm{k}+\bm{l}})G_{p,\bm{k}}(\omega^\prime+\omega)  \right. \\
&\left. \times (\mathcal{K}^{\alpha}_{pq,\bm{k}}+\mathcal{K}^{\alpha}_{pq,\bm{k}+\bm{l}})G_{q,\bm{k}+\bm{l}}(\omega)\right].
\end{aligned}
\end{equation}
Performing the frequency integral, it becomes
\begin{equation}
\begin{aligned}
\tilde{M}_{22}^{\gamma}=&\frac{i}{4\hbar }\sum_{p,q}\int_{\bm{k}} \int d\omega^{\prime} \frac{\partial}{i\partial l^{\beta}}\left[\vphantom{\frac{1}{2}} (\tilde{\varepsilon}_{p,\bm{k}}+\tilde{\varepsilon}_{q,\bm{k}+\bm{l}})\right.\\
&\left. \times(\tilde{\varepsilon}_{p,\bm{k}}v_{pq,\bm{k}}^{\alpha}+v_{pq,\bm{k}+\bm{l}}^{\alpha}\tilde{\varepsilon}_{q,\bm{k}+\bm{l}})\frac{f_{p,\bm{k}}-f_{q,\bm{k}+\bm{l}}}{\omega-(\varepsilon_{p,\bm{k}}-\varepsilon_{q,\bm{k}+\bm{l}})}   \right].
\end{aligned}
\end{equation}
Following the same steps as in the previous section by collecting both the intra- and inter-band contribution, we have
\begin{equation}\label{m22}
\begin{aligned}
\tilde{M}_{22}^{\gamma}=&-\frac{i}{4\hbar }\sum_{p,q}\int_{\bm{k}}\left[2(\tilde{\varepsilon}_{p}+\tilde{\varepsilon}_{q})^{2}\frac{v_{pq}^{\alpha}{v_{qp}^{\beta}}}{(\omega-\varepsilon_{pq})\varepsilon_{pq}}f_{pq}\right. \\
&\left. +(\tilde{\varepsilon}_{p}\varepsilon_{pq}^{2}-4\tilde{\varepsilon_{p}}^{2}\varepsilon_{pq})\frac{(v_{pq}^{\alpha}{v_{qp}^{\beta}}-v_{pq}^{\beta}{v_{qp}^{\alpha}})}{(\omega -\varepsilon_{pq})\varepsilon_{pq}}\frac{\partial f_{p}}{\partial \varepsilon_{p}}\right].
\end{aligned}
\end{equation}
Integrating \Eq{m22} with respect to $\beta$ (via Eq. (\ref{pb2})) from $\beta$ to  $\infty$, we obtain (see Appendix.\ref{appx} for details)
\begin{equation}\label{m22f}
\begin{aligned}
M_{22}^{\gamma}&=-\frac{i}{\hbar }\sum_{p,q}\int_{\bm{k}}\left\{\frac{(v_{pq}^{\alpha}{v_{qp}^{\beta}}-v_{pq}^{\beta}{v_{qp}^{\alpha}})}{(\omega -\varepsilon_{pq})\varepsilon_{pq}} \right. \\
&\left. \times \left[\frac{1}{4}(\tilde{\varepsilon}_{p}+\tilde{\varepsilon}_{q})^{2}f_{p}+\int_{\tilde{\varepsilon}_{p}}^{\infty} d\lambda \lambda^{2} \frac{\partial f_{p}(\lambda)}{\partial \lambda} \right]\right\}.
\end{aligned}
\end{equation}
By use of the identity $\int_{\tilde{\varepsilon}_{p}}^{\infty} d\lambda \lambda^{2} \frac{\partial f_{p}(\lambda)}{\partial \lambda}=-\int_{0}^{f_{p}}(\log\frac{1+t}{t})^{2}dt=c_{2}(f_{p})$ and taking DC limit, we have
\begin{equation}\label{m22fdc}
\begin{aligned}
M_{22}^{\gamma}=-&\frac{1}{\hbar }\sum_{p}\int_{\bm{k}} \left[w_{p}^{\gamma}f_{p} +c_{2}(f_{p})\Omega_{p}^{\gamma}\right],
\end{aligned}
\end{equation}
where the weight function  is 
$c_{2}(f_{p})=(f_{p}-1)\ln^{2}(f_{p}^{-1}-1)+\ln^{2}f_{p}+2{\rm{Li}}_{2}(f_{p})$,
with ${\rm{Li}}_{2}(x)$ being polylogarithm function.
And we introduce the notation
\begin{equation}\label{m2p}
w^{\gamma}_{p} = \frac{1}{\hbar }\varepsilon^{\alpha\beta\gamma}\sum_{p}\frac{1}{2}{\rm{Im}}\bra{\partial^{\alpha} u_{p}}  (\hat{H}_{0}+\varepsilon_{p})^{2} \ket{\partial^{\beta} u_{p}}.
\end{equation}
Combining \Eq{m22f} and \Eq{l22f}, we have
\begin{equation}\label{l22trf}
\begin{aligned}
&L_{22}^{{\rm{tr}},\alpha\beta}(\omega)\\
=&\frac{i}{\hbar }\sum_{p,q}\int_{\bm{k}}\left[\frac{1}{\omega}\tilde{\varepsilon}_{p}^{2}\partial^{\beta} f_{p}v_{p}^{\alpha}+\frac{(v_{pq}^{\alpha}{v_{qp}^{\beta}}-v_{pq}^{\beta}{v_{qp}^{\alpha}})}{(\omega -\varepsilon_{pq})\varepsilon_{pq}}   c_{2}(f_{p})\right].
\end{aligned}
\end{equation}
In the DC limit, it yields
\begin{equation}\label{l22length}
\begin{aligned}
L_{22}^{{\rm{tr}},\alpha\beta}(\omega)=&\frac{1}{\hbar  }\sum_{p}\int_{\bm{k}}\left[i\frac{1}{\omega}\tilde{\varepsilon}_{p}^{2}\partial^{\beta} f_{p}v_{p}^{\alpha}  -c_{2}(f_{p})\Omega_{p}^{\gamma}\right].
\end{aligned}
\end{equation}
The first term is the Drude-type term in heat transport, the second term is the Berry curvature contribution.

Now we present a study of correlations between the thermal conductivity and electric conductivity. Including the transverse transport, the Lorentz number should be generalized into a tensor form
\begin{equation}
\frac{\kappa^{\alpha\beta}}{\sigma^{\alpha\beta}}=L^{\alpha\beta}T,
\end{equation}
where $L^{\alpha\beta}$ is defined as the Lorentz tensor.
Firstly we consider the longitudinal transport. Note that the Drude term in the linear response of $\kappa$ and $\sigma$ corresponds to the contribution of intra-band elements. When $\alpha=\beta$, the topological term vanishes and only the Drude term survives. Hence the Longitudinal response is fully determined by the Drude term. We write the longitudinal electric conductivity as
\begin{equation}
\begin{aligned}
\sigma^{xx}_{L}(\omega;\omega_{1})=\frac{e^2}{\hbar}\sum_{p}\int_{\bm{k}}\frac{\partial^{x} f_{p}v_{p}^{x}}{\omega},
\end{aligned}
\end{equation}
which is written as
\begin{equation}\label{sigmaxx}
\begin{aligned}
\sigma^{xx}_{L}(\omega;\omega_{1})=\frac{e^2}{\hbar \omega}\sum_{p}\int d\varepsilon_{p}\int_{\bm{k}} \frac{\partial f_{p}}{\partial \varepsilon_{p}}\left(\frac{\partial \varepsilon_{p}}{\partial k^{x}}\right)^{2}\delta (\varepsilon_{p}-\varepsilon_{p,\bm{k}}).
\end{aligned}
\end{equation}
The Longitudinal thermal conductivity is given by
\begin{equation}\label{kappaxx}
\begin{aligned}
\kappa^{xx}_{L}(\omega;\omega_{1})&=\frac{1}{ T \hbar \omega}\sum_{p}\int d\varepsilon_{p}\int_{\bm{k}} \varepsilon_{p}^{2} \frac{\partial f_{p}}{\partial \varepsilon_{p}}\left(\frac{\partial \varepsilon_{p}}{\partial k^{x}}\right)^{2}\\
&\times\delta (\varepsilon_{p}-\varepsilon_{p,\bm{k}}).
\end{aligned}
\end{equation}
We can make use of the low-temperature expansion
\begin{equation}\label{somo}
\begin{aligned}
-\frac{\partial f_{p}}{\partial \varepsilon_{p}}&= \delta(\varepsilon_{p}-\mu)+\frac{\pi^2}{6}(k_{B}T)^{2}\frac{\partial^2}{\partial \varepsilon_{p}^{2}}\delta(\varepsilon_{p}-\mu)\\
&+\frac{7\pi^{4}}{360}(k_{B}T)^{4}\frac{\partial^{4}}{\partial \varepsilon_{p}^{4}}\delta(\varepsilon_{p}-\mu)+\cdots.
\end{aligned}
\end{equation}
Inserting \Eq{somo} into \Eq{kappaxx} and \Eq{sigmaxx}, the WF law in Longitudinal direction is obtained
\begin{equation}
\frac{\kappa^{xx}_{L}}{\sigma^{xx}_{L}}=LT,
\end{equation}
with  $L=\frac{1}{3}\left(\frac{k_{B}\pi}{e}\right)^2=2.44\times 10^{-8}$ watt-ohm/K$^{2}$ is the well-known Lorentz number \cite{ashcroft1976solid}.

For transverse transport, according to the expression \Eq{l22length}, the thermal conductivity can be rewritten as
\begin{equation}
\kappa_{T}^{xy}=-\frac{1}{e^{2}T}\int d\epsilon (\epsilon-\mu)^{2} \frac{\partial f(\epsilon)}{\partial \epsilon} \sigma^{xy}(\epsilon).
\end{equation}
where $\sigma^{xy}(\epsilon)=\frac{-e^{2}}{\hbar}\sum_{p}\int_{\bm{k}}\theta
(\epsilon - \varepsilon_{p,\bm{k}})\Omega^{xy}(\bm{k})$ is the intrinsic anomalous Hall conductivity at zero temperature with Fermi energy $\epsilon$.
Given a similar low temperature expansion, the  WF law for transverse transport is verified \cite{Smrcka_1977}, with the  off-diagonal elements of the Lorentz tensor given by $L^{xy}=L$. We conclude that the linear WF law reads
\begin{equation}
\kappa^{\alpha\beta}=LT\sigma^{\alpha\beta},
\label{WFlaw}
\end{equation}
which states that the linear thermal conductivity is proportional to the linear electric conductivity both for the longitudinal and transverse transport. In this work, we call \Eq{WFlaw} as the linear WF law or the 1st order WF law.

\begin{figure}[htb]
\centering
\includegraphics [width=3.0in]{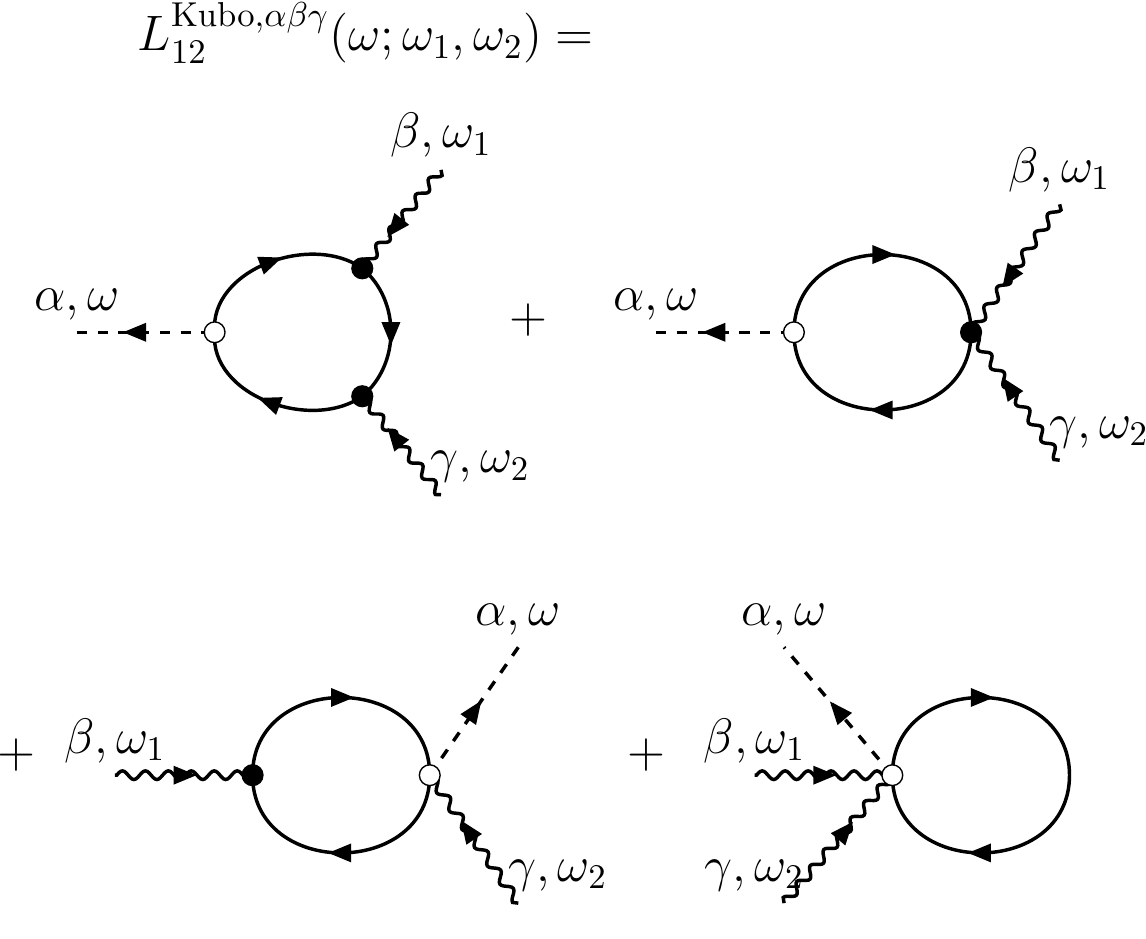}\\
\includegraphics [width=3.0in]{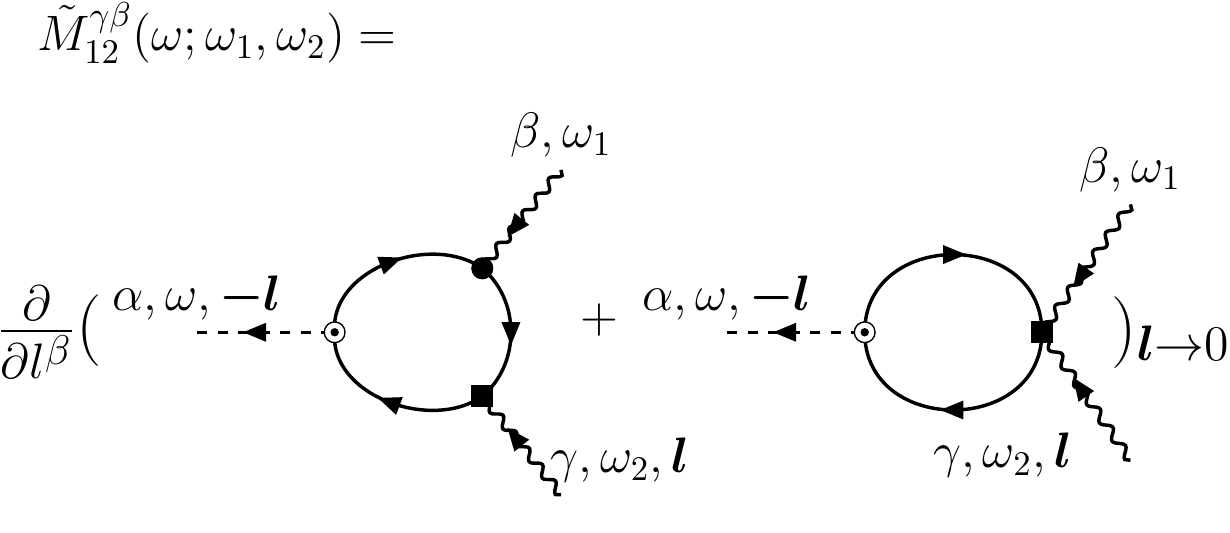}
\caption{Diagrammatic representation of $2$nd-order thermoelectric response, including the $2$nd-order Kubo contribution $L_{12}^{{\rm{Kubo}},\alpha\beta\gamma}$ and the local equilibrium contribution $\tilde{M}_{12}^{\gamma\beta}$.}\label{fig2}
\end{figure}


\subsection{Second-order thermoelectric response}

Now we consider the second-order thermoelectric response $L_{12}^{\alpha\beta\gamma}$. At second order it is composed of four types of diagrams, as shown in \Fig{fig2}.
By using of the Hermitian derivation operator $\mathcal{K}^{\alpha_{1}\cdots \alpha_{k}}$ defined in  Sec. \ref{sec2}, the Kubo contribution to second-order thermoelectric response is given by
\begin{widetext}
\begin{equation}
\begin{aligned}
L_{12}^{{\rm{Kubo}},\alpha\beta\gamma}(\omega;\omega_{1},\omega_{2})&=-\frac{1}{\hbar^{2}\omega_{1}\omega_{2}}\sum_{p,q,r}\int_{\bm{k}}\int  d\omega^\prime \left\{[D^{\alpha}[\mathcal{K}^{\beta\gamma}]]_{pp}G_{p}(\omega^\prime) +2 G_{p}(\omega^\prime)\mathcal{K}_{pq}^{\beta}G_{q}(\omega^\prime+\omega_{1})D^{\alpha}[\mathcal{K}^{\gamma}]_{qp} \right. \\
&\left. + G_{p}(\omega^\prime)\mathcal{K}_{pq}^{\beta\gamma}G_{q}(\omega^\prime+\omega_{12})h_{qp}^{\alpha} + G_{p}(\omega^\prime)\mathcal{K}_{pq}^{\beta}G_{q}(\omega^\prime+\omega_{1}) \mathcal{K}_{qr}^{\gamma} G_{r}(\omega^\prime+\omega_{2})h_{rp}^{\alpha}\right\}\\
&+ (\beta \leftrightarrow \gamma , \omega_{1} \leftrightarrow \omega_{2}),
\end{aligned}
\end{equation}
where $(\beta \leftrightarrow \gamma , \omega_{1} \leftrightarrow \omega_{2})$ denotes symmetrization under simultaneous swap of
the indices $(\beta, \gamma)$ and the frequencies $(\omega_{1}, \omega_{2})$. And the energy conservation is constrained by $\omega=\omega_{12}=\omega_{1}+\omega_{2}$. It can be seen from \Fig{fig2} that for the Kubo contribution, the first diagram describes a process where thermalons interact  sequentially. In contrast, the other three diagrams contain vertices of order greater than one, which is described by instantaneous processes with two or three interaction events.
Performing the integral over Matsubara frequencies, we obtain
\begin{equation} \label{nernst}
\begin{aligned}
L_{12}^{{\rm{Kubo}},\alpha\beta\gamma}(\omega;\omega_{1},\omega_{2})&=-\frac{1}{\hbar^{2}\omega_{1}\omega_{2}}\sum_{p,q,r}\int_{\bm{k}} \left\{ \frac{1}{2}f_{p} \left[D^{\alpha}[\mathcal{K}^{\beta\gamma}]\right]_{pp}  \frac{f_{pq}}{\omega_{1}-\varepsilon_{pq}}\mathcal{K}_{pq}^{\beta}D^{\alpha}[\mathcal{K}^{\gamma}]_{qp}
+\frac{1}{2}\frac{f_{pq}}{\omega_{1}+\omega_{2}-\varepsilon_{pq}}\mathcal{K}_{pq}^{\beta\gamma}h_{qp}^{\alpha}   \right. \\
&\left. +\mathcal{K}_{pq}^{\beta}\mathcal{K}_{qr}^{\gamma}h_{rp}^{\alpha}\frac{(\omega_{1}-\varepsilon_{rq})f_{pq}+(\omega_{1}-\varepsilon_{qp})f_{rq}}
{(\omega_{1}-\varepsilon_{qp})(\omega_{2}-\varepsilon_{rq})(\omega_{1}+\omega_{2}-\varepsilon_{qp})}\right\}.
\end{aligned}
\end{equation}
To keep the shorthand notation, we leave the expansion of the vertices in Appendix.\ref{app1}.
The magnetization response is given by
\begin{equation}
\begin{aligned}
\tilde{M}_{12}^{\gamma\beta}(\omega;\omega_{1},\omega_{2})&=\frac{i}{4\hbar }\sum_{p,q,r}\int_{\bm{k}} \int d\omega^{\prime} \frac{\partial}{i\partial l^{\beta}}\left[   G_{p,\bm{k}}(\omega^\prime) (\mathcal{K}^{\gamma}_{pq,\bm{k}}+\mathcal{K}^{\gamma}_{pq,\bm{k}+\bm{l}})G_{q,\bm{k}+\bm{l}}(\omega^\prime +\omega_{1}+\omega_{2})(h_{qp,\bm{k}}^{\alpha}+h_{qp,\bm{k}+\bm{l}}^{\alpha})\right.\\
&\left.+   G_{p,\bm{k}}(\omega^\prime)(\tilde{\varepsilon}_{p,\bm{k}}+\tilde{\varepsilon}_{q,\bm{k}+\bm{l}}) G_{q,\bm{k}+\bm{l}}(\omega^\prime+\omega_{1}) (h_{qr,\bm{k}}^{\alpha}+h_{qr,\bm{k}+\bm{l}}^{\alpha})G_{r,\bm{k}}(\omega^\prime+\omega_{1}+\omega_{2})\mathcal{K}^{\beta}_{rp,\bm{k}}\right].
\end{aligned}
\end{equation}
After the integral over $\omega'$, we derive
\begin{equation}
\begin{aligned}
\tilde{M}_{12}^{\gamma\beta}(\omega;\omega_{1},\omega_{2})&=\frac{i}{4\hbar }\sum_{p,q,r}\int_{\bm{k}}   \frac{\partial}{i\partial l^{\beta}}\left[\frac{1}{2}(\mathcal{K}^{\gamma}_{pq,\bm{k}}+\mathcal{K}^{\gamma}_{pq,\bm{k}+\bm{l}})    (h_{qp,\bm{k}}^{\alpha}+h_{qp,\bm{k}+\bm{l}}^{\alpha})\frac{f_{pq}}{\omega_{1}+\omega_{2}-\varepsilon_{pq}}\right.\\
&\left.+(\tilde{\varepsilon}_{p,\bm{k}}+\tilde{\varepsilon}_{q,\bm{k}+\bm{l}})   (h_{qr,\bm{k}}^{\alpha}+h_{qr,\bm{k}+\bm{l}}^{\alpha})\mathcal{K}^{\beta}_{qr,\bm{k}}\frac{(\omega_{1}-\varepsilon_{rq})f_{pq}+(\omega_{1}-\varepsilon_{qp})f_{rq}}
{(\omega_{1}-\varepsilon_{qp})(\omega_{2}-\varepsilon_{rq})(\omega_{1}+\omega_{2}-\varepsilon_{qp})} \right].
\end{aligned}
\end{equation}
\end{widetext}
Considering that the partial differential in $\tilde{M}_{12}^{\gamma\beta}$  involves many terms, an analytical treatment of $\tilde{M}_{12}^{\gamma\beta}$ is rather tedious. Instead, it is more convenient to treat it numerically. The same process applies to the
2nd order electric-thermal response $L_{21}^{{\rm{tr}},\alpha\beta\gamma}$ and thermal-thermal response $L_{22}^{{\rm{tr}},\alpha\beta\gamma}$.

Different methods are proposed to include finite relaxation rates into nonlinear responses \cite{PhysRevB.93.085403,PhysRevB.96.035431,Cheng_2014}, both in length gauge and velocity gauge. Referring to \Eq{nernst}, which involves the electron transfer processes between two or more bands leading to different relaxation times, it is more accurate to correct the covariant derivative by relaxation rate $\Gamma_{mn}$ for excited states \cite{PhysRevResearch.2.033100}
\begin{equation}
v_{pq}^{\alpha}=\partial^{\alpha}\varepsilon_{p}\delta_{pq}-i
\varepsilon_{pq}\mathcal{A}_{pq}^{\alpha}-\sum_{r}(\mathcal{A}_{pr}^{\alpha}\Gamma_{rq}-\Gamma_{pr}\mathcal{A}^{\alpha}_{rq}).
\end{equation}
In order to make a direct connection to the semiclassical result, the simple replacement $\omega \rightarrow \omega + i\Gamma$ is adopted.
 Due to the finite lifetime of electrons,  the propagator is  replaced by $1/(\omega +i\Gamma)$,  where $\Gamma$ is the imaginary part of the self-energy and $\tau = 1/\Gamma $ is the electron relaxation time.

The expansion of the vertices might appear pretty verbose, but crucially it allows us a straightforward identification of the physical processes. By taking $\omega \rightarrow 0$
the static limit response can be directly implemented in numerics. However, a direct conversion to the static state results is rather laborious. As we show in the following, it is much easier to do this in length gauge.

\section{Static state results: Length gauge}\label{sec3}
As discussed in Sec. \ref{sec2}, the formalism given in velocity gauge pertains to more apparent physical picture for the resonant structure of interband transition induced by the thermal field. However, in most cases we focus on the  analysis of the steady-state response of temperature gradient and it is easier to do it in length gauge.
Both approaches yield identical results in the clean limit. And
the wave functions between the two gauges are related by a time-dependent unitary transformation \cite{PhysRevB.96.035431, PhysRevB.96.195413, PhysRevResearch.2.033100}. After taking many sum rules the results in velocity gauge are transformed to those of length gauge .

Perturbed by the thermal field, the Hamiltonian in length gauge is given as
\begin{equation}\label{lenh}
\hat{H}_{E_{T}}=\hat{H}_{0}+\frac{1}{2} (\hat{H}_{0}\hat{\bm{r}}+\hat{\bm{r}}\hat{H}_{0})\cdot\bm{E}_{T}.
\end{equation}
In terms of the relation $\hat{\bm{r}}=i\hat{\bm{D}}$ between the covariant derivative and the position operator, $H_{E_{T}}$ is rewritten as
\begin{equation}\label{lenh2}
\hat{H}_{E_{T}}=\hat{H}_{0}+i \hat{\bm{\mathcal{D}}}\cdot \bm{E}_{T},
\end{equation}
where the definition $\hat{\bm{\mathcal{D}}}[\mathcal{O}]=\frac{1}{2}[\hat{H}_{0},\hat{\bm{D}}[\mathcal{O}]]_{+}$ is used.
We adopt the reduced density matrix (RDM) equations of motion approach \cite{PhysRevB.96.035431} to calculate the nonlinear thermal response in length gauge. The RDM in band space is given by the average of the product of a creation and a destruction operator in Bloch states
\begin{equation}
\rho_{\bm{k}pq}(t)\equiv \langle c_{p\bm{k}}^{\dagger}(t)c_{q\bm{k}}(t) \rangle.
\end{equation}
The standard density-matrix formalism is performed by expanding the RDM in powers of the thermal field in calculating the nonlinear thermal response.

In analogy with the optical conductivity $\sigma(\omega)$ which describes the response of the transient charge current to an time-dependent electric field $\bm{E}(t)$, we can define the dynamical Nernst (or thermal Hall) conductivity, as the response of the transient charge (heat) current to a  time-dependent temperature gradient field $\bm{\nabla} T(t)$.

The expectation values of the Kubo contribution of the charge (heat) current are given by
\begin{equation}
J_{N(Q)}^{{\rm{Kubo}},\alpha}(t)={\rm{Tr}}[\hat{J}_{N(Q)}^{\alpha}\rho(t)],
\end{equation}
where $\alpha=x,y,z$, $\hat{J}_{c}^{\alpha}\equiv e \hat{v}^{\alpha}$, and $\hat{J}_{h}^{\alpha}\equiv \frac{1}{2}[\hat{H}_{0}, \hat{v}^{\alpha}]_{+}$.
For simplicity we suppose that the system is only perturbed by the thermal field. According to \Eq{m12fdc} and \Eq{m22fdc}, the particle magnetization   which can be expressed in form of the RDM
\begin{equation}\label{m12fdca}
\begin{aligned}
M_{N}^{\gamma}=&{\rm{Tr}}\left[\int_{\bm{k}}m^{\gamma}\rho-\frac{1}{e^2}\int d\varepsilon\sigma^{\gamma}(\varepsilon)\rho\right],
\end{aligned}
\end{equation}
where the orbital magnetic moment and zero-temperature Hall conductivity are generalized to the matrix form $m^{\gamma}_{pq}= m^{\gamma}_{p}\delta_{pq}$, $\sigma^{\gamma}_{pq}= \sigma^{\gamma}_{p}\delta_{pq}$. And similar for the heat magnetization
\begin{equation}\label{m22fdca}
\begin{aligned}
M_{Q}^{\gamma}=&{\rm{Tr}}\left[\int_{\bm{k}} w^{\gamma}\rho -\frac{1}{e^2}\int d\varepsilon \tilde{\varepsilon}\sigma^{\gamma}(\varepsilon)\rho\right],
\end{aligned}
\end{equation}
with $w^{\gamma}_{pq}= w^{\gamma}_{p}\delta_{pq}$ and $\tilde{\varepsilon}_{pq}=\tilde{\varepsilon}_{p}\delta_{pq}$.

The equation of motion of the RDM is given by
\begin{equation}\label{rdm}
\begin{aligned}
i\hbar \frac{\partial \rho_{\bm{k}pq}(t) }{\partial t}&={\rm{Tr}} \left[i\hbar \frac{\partial \rho(t)}{\partial t}c_{p\bm{k}}^{\dagger}c_{q\bm{k}}\right]\\
&=\left\langle \left[c_{p\bm{k}}^{\dagger}(t)c_{q\bm{k}}(t), H_{E_{T}}(t)\right]_{-} \right\rangle .
\end{aligned}
\end{equation}
Substituting the Hamiltonian \Eq{lenh} into \Eq{rdm}, and expanding RDM in powers of the external field $\rho = \sum_{n}\rho^{(n)}$, the equation of motion can be solved recursively
\begin{equation}
\left(i\hbar\frac{\partial}{\partial t}-\varepsilon_{\bm{k}pq}\right)\rho_{\bm{k}pq}^{(n)}(t) = \bm{E}_{T}\cdot \hat{\bm{\mathcal{D}}}\left[\rho^{(n-1)}(t)\right]_{\bm{k}pq}.
\label{iterativeRhon}
\end{equation}
Therefore, the $n$th-order RDM can be expressed via the zeroth-order RDM by iterating Eq. (\ref{iterativeRhon}), and the zeroth-order RDM is the Fermi-Dirac distribution function times the unit matrix in band space  $\rho^{(0)}_{pq}=f_{p}\delta_{pq}$. To solve the equation, we need to transform it into frequency space. The time derivative in the equations of motion is replaced by a frequency factor that is collected into an energy denominator $d_{\bm{k}pq}(\omega)=1/(\omega-\varepsilon_{\bm{k}pq})$, and the iterative relation is given by
\begin{equation}
\begin{aligned}
\rho_{\bm{k}pq}^{(n)}(\omega)=&i\int \frac{d\omega^{\prime}}{2\pi}E^{\alpha_{1}}_{T}\left[d(\omega) \circ \hat{\mathcal{D}}^{\alpha_{1}}\left[\rho^{(n-1)}(\omega-\omega^\prime)\right]\right]_{\bm{k}pq},
\end{aligned}
\end{equation}
where $\circ$ is the Hadamard product $(A\circ B)_{pq}=A_{pq}B_{pq}$.
The $n$-th order RDM is
\begin{widetext}
\begin{equation}\label{jrhok}
\begin{aligned}
\rho^{(n)}(\omega)_{pq}=(i)^{n}\left[\prod_{i=1}^{n}\int d\omega_{i}E_{T}^{\alpha_{i}}(\omega_{i})  \right]\left[d(\omega)\circ \left[\mathcal{D}^{\alpha_{1}}\left[d(\omega-\omega_{1})\cdots \left[\mathcal{D}^{\alpha_{k}}\left[d(\omega-\omega_{k}) \cdots \circ\left[\mathcal{D}^{\alpha_{n}}\left[\rho^{(0)}\right]\right]\right]\right]\right]\right]\right]\delta(\omega_{[n]}-\omega),
\end{aligned}
\end{equation}
\end{widetext}
where $\omega_{[n]}\equiv \sum_{i}^{n}\omega_{n}$.
The $n$-th order components of  the Kubo particle (heat) current are written as
\begin{equation}\label{jrhok}
\begin{aligned}
J^{{\rm{Kubo}},(n),\alpha}_{i}(\omega)=\int_{\bm{k}}{\rm{Tr}}\left[\hat{J}^{\alpha}_{i}\rho^{(n)}(\omega)\right].
\end{aligned}
\end{equation}
For the $n$-th order current, the magnetization is expanded up to the $(n-1)$-th order of thermal field, which is given by
\begin{equation}\label{m12rdm}
\begin{aligned}
&M_{N}^{(n),\gamma}(\omega)\\
 =&{\rm{Tr}}\left[\int_{\bm{k}}m^{\gamma}\rho^{(n-1)}(\omega)+\frac{1}{e}\int d\varepsilon\sigma^{\gamma}(\varepsilon) \rho^{(n-1)}(\omega)\right],
\end{aligned}
\end{equation}
\begin{equation}\label{m22rdm}
\begin{aligned}
&M_{Q}^{(n),\gamma}(\omega)\\
 =&{\rm{Tr}}\left[\int_{\bm{k}}w^{\gamma}\rho^{(n-1)}(\omega)-\frac{1}{e^2}\int d\varepsilon (\varepsilon -\mu)\sigma^{\gamma}(\varepsilon) \rho^{(n-1)}(\omega)\right],
\end{aligned}
\end{equation}
The higher order derivatives follow from an expansion of the time evolution of the instantaneous eigenstates beyond linear approximation.

\subsection{Linear thermoelectric and thermal-thermal response}
Firstly, we rederive the first-order thermoelectric response coefficient, as a pedagogical demonstration of our method. $L_{12}^{{\rm{Kubo}},\alpha\beta}$ is related to the 1st order  RDM, which is expanded as
\begin{equation}\label{rho1}
\begin{aligned}
\rho^{(1)}_{pq}=&iE_{T}^{\beta}(\omega)\left[d(\omega) \circ \hat{\mathcal{D}}^{\beta}\left[\rho^{(0)}\right]_{-}\right]_{pq}\\
=&iE_{T}^{\beta}(\omega)\left[\frac{\tilde{\varepsilon}_{p}}{\omega}\partial^{\beta}f_{p}\delta_{pq} -i\frac{\tilde{\varepsilon}_{p}+\tilde{\varepsilon}_{q}}{2(\varepsilon_{pq}+\omega)}\mathcal{A}_{pq}^{\beta}f_{pq}\right].
\end{aligned}
\end{equation}
The linear Kubo current is
\begin{equation}
\begin{aligned}
J_{N}^{{\rm{Kubo}},(1),\alpha}(\omega)=\int_{\bm{k}}{\rm{Tr}}\left[\hat{J}_{N}^{\alpha}\rho^{(1)}(\omega)\right],
\end{aligned}
\end{equation}
and the Kubo contribution of transport coefficient is found as (Appendix.\ref{app4})
\begin{equation}\label{L12lres}
\begin{aligned}
&L_{12}^{{\rm{Kubo}},\alpha\beta}(\omega)\\
=&i\sum_{p,q}\int_{\bm{k}}\left[\frac{\tilde{\varepsilon}_{p}}{\omega}v_{p}^{\alpha}\partial^{\beta}f_{p} -i\frac{\tilde{\varepsilon}_{p}+\tilde{\varepsilon}_{q}}{2(\varepsilon_{pq}+\omega)}\mathcal{A}_{pq}^{\beta}v_{qp}^{\alpha}f_{pq}\right].
\end{aligned}
\end{equation}
It is equivalent to the expression \Eq{l12k} derived by diagrammatic approach in velocity gauge.
Using \Eq{m12rdm}, the first order particle magnetization density is written as
\begin{equation}
\begin{aligned}\label{L12lleq}
M_{N}^{(1),\gamma}=\sum_{p}\int_{\bm{k}}\left[m^{\gamma}_{p}f_{p}+k_{B}T\Omega_{p}^{\gamma}{\rm{ln}}(1+e^{-\beta \tilde{\varepsilon}_{p}})\right].
\end{aligned}
\end{equation}
Considering the DC limit by taking $\omega\rightarrow 0$, we obtain the linear thermoelectric response for transport current by collecting the Kubo contribution \Eq{L12lres} and the magnetization correction \Eq{L12lleq}, which is given by
\begin{equation}
\begin{aligned}
L_{DC,12}^{{\rm{tr}},\alpha\beta}(\omega)=L_{12,DC}^{{\rm{Kubo}},\alpha\beta}(\omega)-\epsilon^{\alpha\beta\gamma} M_{N}^{\gamma}.
\end{aligned}
\end{equation}
By separating all the terms with the Berry connection, the transport coefficient can be written as
\begin{equation}
\begin{aligned}
L_{DC,12}^{{\rm{tr}},\alpha\beta}(\omega)=L_{D,12}^{\alpha\beta}(\omega)+L_{A,12}^{\alpha\beta}(\omega),
\end{aligned}
\end{equation}
in which the first term is the usual Drude term
\begin{equation}\label{l12d1}
\begin{aligned}
L_{D,12}^{\alpha\beta}(\omega)=\frac{i}{\hbar }\sum_{p}\int_{\bm{k}}\frac{1}{\omega}v_{p}^{\alpha}v_{p}^{\beta} \frac{\partial f_{p}}{\partial \varepsilon_{p}},
\end{aligned}
\end{equation}
and the second term is the anomalous term contributed by the Berry curvature
\begin{equation}\label{l12a1}
\begin{aligned}
L_{A,12}^{\alpha\beta}(\omega)=\frac{e}{\hbar }\sum_{p}\int_{\bm{k}}\Omega_{p}^{\gamma} \left[\tilde{\varepsilon}_{p}f_{p}+k_{B}T\ln \left(1+e^{-\beta\tilde{\varepsilon}_{p}}\right)\right].
\end{aligned}
\end{equation}
Not surprisingly, \Eq{l12d1} and \Eq{l12a1} recover \Eq{l12length}  obtained in length gauge.

In analogy, the linear thermal-thermal response  $L_{22}^{{\rm{tr}},\alpha\beta}$ is derived in a similar process. The linear Kubo heat current is
\begin{equation}
\begin{aligned}
J_{Q}^{{\rm{Kubo}},(1),\alpha}(\omega)=\int_{\bm{k}}{\rm{Tr}}\left[\hat{J}_{Q}^{\alpha}\rho^{(1)}(\omega)\right],
\end{aligned}
\end{equation}
and the Kubo contribution to transport coefficient is given by
\begin{equation}\label{L22lres}
\begin{aligned}
&L_{22}^{{\rm{Kubo}},\alpha\beta}(\omega)\\
=&i\sum_{p,q}\int_{\bm{k}}\left[\frac{1}{\omega}\tilde{\varepsilon}_{p}^{2}v_{p}^{\alpha}\partial^{\beta}f_{p}-i\frac{(\tilde{\varepsilon}_{p}+\tilde{\varepsilon}_{q})^{2}}{4(\varepsilon_{pq}-\omega)}\mathcal{A}_{pq}^{\alpha}v_{qp}^{\mu}f_{pq}\right].
\end{aligned}
\end{equation}
The heat magnetization is
\begin{equation}
\begin{aligned}\label{L22lleq}
M_{Q}^{(1),\gamma}=&\int_{\bm{k}}{\rm{Tr}}\left\{w^{\gamma}\rho^{(0)}-\Omega^{\gamma}c_{2}\left[\rho^{(0)}\right] \right\}\\
=&\sum_{p}\int_{\bm{k}}\left[w_{p}^{\gamma}f_{p}-\Omega_{p}^{\gamma}c_{2,p}(f_{p})\right].
\end{aligned}
\end{equation}
Combining \Eq{L22lres} and \Eq{L22lleq} and taking the DC limit, we obtain the
response coefficient for transport thermal current
\begin{equation}
\begin{aligned}
L_{22}^{{\rm{tr}},\alpha\beta}(\omega)=&\frac{1}{\hbar}\sum_{p}\int_{\bm{k}}\left[i\frac{1}{\omega}\tilde{\varepsilon}_{p}^{2}\partial^{\beta} f_{p}v_{p}^{\alpha} -c_{2}(f_{p})\Omega_{p}^{\gamma}\right],
\end{aligned}
\end{equation}
which recovers the expression \Eq{l22length}.

\subsection{Second-order thermoelectric conductivity and Mott relation}
The Kubo contribution to the 2nd order thermoelectric response coefficient is related to the 2nd order RDM
\begin{equation}\label{rho2}
\begin{aligned}
\rho^{(2)}&=-\int d\omega_{1}\int d\omega_{2}E_{T}^{\beta}(\omega_{1})E_{T}^{\delta}(\omega_{2}) d(\omega)\\
&\circ \mathcal{D}^{\beta}\left[d(\omega-\omega_{1}) \circ\mathcal{D}^{\delta}[\rho^{(0)}]\right]\delta(\omega_{[2]}-\omega).
\end{aligned}
\end{equation}
 We aim to obtain the expression in the $\omega \rightarrow 0$ limit and then compare with the semiclassical results. The Kubo contribution of the 2nd order particle current is given by
\begin{equation}\label{L12kabc}
\begin{aligned}
J_{N}^{{\rm{Kubo}},(2),\alpha}(\omega)=\int_{\bm{k}}{\rm{Tr}}\left[j_{N}^{\alpha} \rho^{(2)}\right].
\end{aligned}
\end{equation}
Substituting \Eq{rho2} into \Eq{L12kabc}, and using \Eq{lijk}, the 2nd order thermoelectric response is expanded as the summation of four integral kernels
\begin{equation}
\begin{aligned}
&L_{12}^{{\rm{Kubo}},\alpha\beta\delta}=e\int_{\bm{k}}\left[\Pi^{(2),\beta\delta}+\Pi^{(2),\beta}+\Pi^{(2),\delta}+\Pi^{(2)}\right],
\end{aligned}
\end{equation}
where the superscripts $\alpha$, $\beta$, and $\delta$ ($\alpha,\beta,\delta=x,y,z$) of $\Pi$ denote the $k^{\alpha}$, $k^{\beta}$, and $k^{\delta}$ Hermitian derivatives defined in \Eq{superd} and the superscript $(2)$ denotes the 2nd order.
The expressions for the integral kernels are obtained as (detailed derivation is sketched in Appendix. \ref{app4})
\begin{equation}\label{pi1}
\begin{aligned}
&\Pi^{(2),\beta\delta}= \sum_{p}v_{p}^\alpha \frac{1}{\omega}\frac{1}{\omega-\omega_{1}}\tilde{\varepsilon}_{p}\partial^{\beta}(\tilde{\varepsilon}_{p}\partial^{\delta}f_{p}),\\
&\Pi^{(2),\beta}=\sum_{p,q}\frac{-i}{2} v_{pq}^{\alpha}\frac{1}{\omega-\varepsilon_{qp}}\tilde{\varepsilon}_{p}\partial^{\beta}\left[\frac{1}{\omega-\omega_{1}-\varepsilon_{qp}}\right.\\
& \quad\quad\quad \times\left. (\tilde{\varepsilon}_{p}+\tilde{\varepsilon}_{q})\mathcal{A}^{\delta}_{qp}f_{pq}\vphantom{\frac{1}{\omega-\omega_{1}-\varepsilon_{qp}}}\right],\\
&\Pi^{(2),\delta}=\sum_{p,q}\frac{-i}{2}v_{pq}^{\alpha}\frac{1}{(\omega-\varepsilon_{qp})}\frac{1}{(\omega-\omega_{1})}\mathcal{A}^{\beta}_{qp}\\
&\quad\quad\quad \times \left(\tilde{\varepsilon}_{p}\tilde{\varepsilon}_{q}\partial^{\delta} f_{pq}+\tilde{\varepsilon}_{p}^{2}\partial^{\delta}f_{p}-\tilde{\varepsilon}_{q}^{2}\partial^{\delta}f_{q}\right),\\
&\Pi^{(2)}=-\sum_{p,q,r}\frac{1}{4} v_{pq}^{\alpha}\frac{1}{\omega-\varepsilon_{qp}}(\tilde{\varepsilon}_{q}+\tilde{\varepsilon}_{r})\mathcal{A}^{\beta}_{qr}\frac{1}{\omega-\omega_{1}-\varepsilon_{rp}}\\
&\quad \quad \quad \times(\tilde{\varepsilon}_{r}+\tilde{\varepsilon}_{p})\mathcal{A}^{\delta}_{rp}(f_{rp}-f_{qr}).
\end{aligned}
\end{equation}
Here $\Pi^{(2),\beta\delta}$ is the intraband contribution, which is the generalized 2nd order Drude term. The others are the interband contributions which contain the Berry connection.

Now we consider the static state. The dominating terms are distinguished by the $\omega$ dependent denominators of the integral kernels. For $\Pi^{(2),\beta\delta}$, it is proportional to $1/(\omega\omega_{2})$ (considering $\omega_{2}=\omega-\omega_{1}$), which diverges at DC limit (as $\omega_{1}$, $\omega_{2}$ approaching zero). For $\Pi^{(2),\delta}$, it is proportional to $1/\omega_{2}$, which also diverges at DC limit.
%
While for $\Pi^{(2),\beta}$ and $\Pi^{(2)}$, there is no divergent dominator and can be safely omitted.
Therefore the dominating terms are from $\Pi^{(2),\beta\delta}$ and $\Pi^{(2),\delta}$ in the DC limit, and the 2nd order thermoelectric conductivity is given by
\begin{equation}\label{tef}
\begin{aligned}
L_{DC,12}^{{\rm{Kubo}},\alpha\beta\delta}&(\omega ;\omega_{1},\omega_{2})\\
=-&\sum_{p,q} \int_{\bm{k}}\left[\frac{1}{\omega\omega_{2}} v_{p}^{\alpha}\tilde{\varepsilon}_{p}\partial^{\beta}(\tilde{\varepsilon}_{p}\partial^{\delta}f_{p})+\frac{i}{2\omega_{2}}\right.\\
\times & \left.\varepsilon_{p}(\varepsilon_{p}+\varepsilon_{q})\frac{v_{pq}^{\alpha}v_{qp}^{\beta}-v_{pq}^{\beta}v_{qp}^{\alpha}}
{\varepsilon_{pq}^{2}}\partial^{\delta}f_{p}\vphantom{\frac{1}{2}}\right].
\end{aligned}
\end{equation}
By use of the identity \Eq{orbm}, the DC 2nd order thermoelectric response can be written into the following more suggestive form
\begin{equation}
\begin{aligned}\label{l12dc_2}
&L_{DC,12}^{{\rm{Kubo}},\alpha\beta\delta}(\omega ;\omega_{1},\omega_{2})\\
=&-\sum_{p}\int_{\bm{k}}\left[\frac{1}{\omega\omega_{2}}v_{p}^{\alpha}\tilde{\varepsilon}_{p}\partial^{\beta}(\tilde{\varepsilon}_{p}\partial^{\delta}f_{p})
+\frac{i}{\omega_{2}}\right.\\
&\left.\times(\tilde{\varepsilon}_{p} m_{p}^{\gamma}  +\tilde{\varepsilon}_{p}^{2}\Omega_{p}^{\gamma})\partial^{\delta} f_{p} \right].
\end{aligned}
\end{equation}
Now we derive the 2nd order particle magnetization density, which is related to the 1st order RDM
\begin{equation}
\begin{aligned}\label{L12lleq2}
M_{N}^{(2),\gamma}(\omega)={\rm{Tr}}\left[\int_{\bm{k}}\rho^{(1)}(\omega) m^{\gamma}+\frac{1}{e}\int d\varepsilon \sigma^{\gamma}(\varepsilon)\rho^{(1)}(\omega) \right].
\end{aligned}
\end{equation}
Referring to  \Eq{rho1}, the second term of $\rho^{(1)}$ is omitted because it is subleading in the DC limit. Hence the 2nd order thermoelectric magnetization response is given as
\begin{equation}
\begin{aligned}\label{m12dc_2}
M_{12,DC}^{\gamma\delta}(\omega)=&i\sum_{p}\int_{\bm{k}}\frac{1}{\omega}\tilde{\varepsilon}_{p} m^{\gamma}_{p}\partial^{\delta}f_{p}\\
&+i\frac{1}{e}\sum_{p}\int d\varepsilon \frac{1}{\omega}\tilde{\varepsilon}_{p} \sigma^{\gamma}(\varepsilon)\partial^{\delta}f_{p}(\varepsilon) .
\end{aligned}
\end{equation}
Combining \Eq{l12dc_2} and \Eq{m12dc_2}, we obtain the 2nd thermoelectric conductivity in the DC limit
\begin{equation}\label{l122final}
\begin{aligned}
L_{12,DC}^{{\rm{tr}},\alpha\beta\delta}(\omega;\omega_{1},\omega_{2})=L_{12,D}^{\alpha\beta\delta}(\omega;\omega_{1},\omega_{2})+L_{12,A}^{\alpha\beta\delta}(\omega;\omega_{1},\omega_{2}).
\end{aligned}
\end{equation}
For the Drude term:
\begin{equation}\label{l12d}
\begin{aligned}
L_{12,D}^{\alpha\beta\delta}=-\sum_{p}\int_{\bm{k}}\frac{1}{\omega\omega_{2}} v_{p}^{\alpha}\tilde{\varepsilon}_{p}\partial^{\beta}(\tilde{\varepsilon}_{p}\partial^{\delta}f_{p}),
\end{aligned}
\end{equation}
and the anomalous term is given as
\begin{equation}\label{l12b}
\begin{aligned}
L_{12,A}^{\alpha\beta\delta}=i\epsilon^{\alpha\beta\gamma}\sum_{p}\int d\varepsilon_{p} \frac{1}{\omega_2}v_{p}^{\delta}\sigma^{\gamma}(\varepsilon_{p})\left[2\tilde{\varepsilon}_{p}\frac{\partial f_{p}}{\partial \varepsilon_{p}}+ \tilde{\varepsilon}_{p}^2 \frac{\partial^2 f_{p}}{\partial \varepsilon_{p}^{2}}\right].
\end{aligned}
\end{equation}
Noting that for the system with time-reversal symmetry, the Drude term vanishes and only the anomalous term survives.

Next we study how the thermoelectric conductivity is related to the electric conductivity at the 2nd order.
The  2nd order electric-electric response is written as
\begin{equation}
\begin{aligned}
J_{N}^{{\rm{tr}},(2),\alpha}(\omega)=&\int_{\bm{k}}{\rm{Tr}}\left[j_{N}^{\alpha} \rho^{(2)}(\omega)\right].
\end{aligned}
\end{equation}
The 2nd order RDM with an electric field perturbation is given as
\begin{equation}
\begin{aligned}
\rho^{(2)}=&-\int d\omega_{1}\int d\omega_{2}E^{\beta}(\omega_{1})E^{\delta}(\omega_{2}) d(\omega)\\
&\circ D^{\beta}\left[d(\omega-\omega_{1}) \circ D^{\delta}[\rho^{(0)}]\right]\delta(\omega_{[2]}-\omega).
\end{aligned}
\end{equation}
Expanding $\rho^{(1)}$, the 2nd order electric-electric response becomes \cite{PhysRevB.96.035431}
\begin{equation}
\begin{aligned}
&L_{11,DC}^{{\rm{tr}},\alpha\beta\delta}(\omega ;\omega_{1},\omega_{2})=L_{11,D}^{\alpha\beta\delta}(\omega ;\omega_{1},\omega_{2})+L_{11,A}^{\alpha\beta\delta}(\omega ;\omega_{1},\omega_{2}),
\end{aligned}
\end{equation}
with
\begin{eqnarray}
L_{11,D}^{\alpha\beta\delta}&=&-\sum_{p}\int_{\bm{k}}\frac{1}{\omega\omega_{2}} \partial^{\beta}v_{p}^{\alpha}\partial^{\delta}f_{p},\\
L_{11,A}^{\alpha\beta\delta}&=&-\epsilon^{\alpha\beta\gamma}\sum_{p}\int_{\bm{k}}i\frac{1}{\omega_{2}}\Omega^{\gamma}_{p} \partial^{\delta}f_{p}.
\end{eqnarray}
Firstly we focus on the anomalous term $L_{11,A}^{\alpha\beta\delta}$. Integrating by part, it can be rewritten as
\begin{equation}
L^{\alpha\beta\delta}_{11,A}= -\frac{i}{\hbar^2}\epsilon^{\alpha\beta\gamma}\sum_{p} \int d\varepsilon_{p} \frac{1}{\omega_{2}} A^{\delta}(\varepsilon_{p})\sigma^{\gamma}(\varepsilon_{p}),
\end{equation}
in which we define $A^{\delta}(\varepsilon_{p})\equiv \frac{\partial f_{p}}{\partial \varepsilon_p}\frac{\partial v^{\delta}(\varepsilon_{p})}{\partial \varepsilon_p}+\frac{\partial^2 f_{p}}{\partial \varepsilon_{p}^2}v^{\delta}(\varepsilon_{p})$.
Using the identity
\begin{equation}\label{id1}
\begin{aligned}
\frac{\partial f_{0}}{\partial \varepsilon_{\bm{k}}}\frac{\partial v^{\alpha}_{\bm{k}}}{\partial k_{\beta}}+\frac{\partial^2 f_{0}}{\partial \varepsilon_{\bm{k}}^2}v^{\alpha}_{\bm{k}}v^{\beta}_{\bm{k}}=v_{\bm{k}}^{\beta}\frac{\partial f_{0}}{\partial \varepsilon_{\bm{k}}}\frac{\partial v^{\alpha}_{\bm{k}}}{\partial \varepsilon_{\bm{k}}}+\frac{\partial^2 f_{0}}{\partial \varepsilon_{\bm{k}}^2}v^{\alpha}_{\bm{k}}v^{\beta}_{\bm{k}},
\end{aligned}
\end{equation}
we have
\begin{equation}
v_{\bm{k}}^{\beta}\frac{\partial v^{\alpha}_{\bm{k}}}{\partial \varepsilon_{\bm{k}}}=\frac{\partial v^{\alpha}_{\bm{k}}}{\partial k_{\beta}}=\frac{\partial^2 \varepsilon_{\bm{k}}}{\partial k_{\alpha}\partial k_{\beta}}=\frac{1}{m^*_{\alpha\beta}}.
\end{equation}
Here $m^{*}_{\alpha\beta}$ is the effective mass of the Bloch electrons. When we consider a limit case that $\bm{v}_{\bm{k}}$ is independent of energy, namely,
$\partial \bm{v}_{\bm{k}}/ \partial \varepsilon =0$ which indicates  a large effective mass. The 2nd order anomalous Hall conductivity is approximated as
\begin{equation}\label{l11b}
L^{\alpha\beta\delta}_{11,A}\approx -\frac{i}{\hbar^2}\epsilon^{\alpha\beta\gamma} \sum_{p} \int d\varepsilon_p \frac{1}{\omega_{2}} \frac{\partial^2 f_{p}}{\partial \varepsilon_{p}^2}v^{\delta}(\varepsilon_{p}) \sigma^{\gamma}(\varepsilon_{p}) .
\end{equation}
By inserting the low-temperature expansion formula \Eq{somo} into \Eq{l11b} and \Eq{l12b}, and considering that the electric conductivity $\sigma_{DC}^{\alpha\beta\delta}$ and thermoelectric conductivity $\eta_{A}^{\alpha\beta\delta}$ satisfy $\sigma_{A}^{\alpha\beta\delta}=L_{A,11}^{{\rm{tr}},\alpha\beta\delta}$ and $\eta_{A}^{\alpha\beta\delta}=L_{12,A}^{{\rm{tr}},\alpha\beta\delta}/T^{2}$, we obtain
\begin{equation}\label{mott2}
\eta^{\alpha\beta\delta}_{A}=\frac{1}{3}\frac{\pi^{2} k_{B}^{2}}{e^{2}}\sigma^{\alpha\beta\delta}_{A}=L\sigma^{\alpha\beta\delta}_{A}.
\end{equation}
It indicates that when the dispersion is weakly dependent on the velocity, the 2nd order thermoelectric conductivity (the 2nd order Nernst coefficient) is proportional to the 2nd order electric conductivity (the 2nd order particle Hall conductivity) at low temperatures, which is different from the Mott relation for the linear order. The linear Mott relation tells us that the linear Nernst coefficient is proportional to the derivative of linear Hall conductivity to the Fermi energy, which is $\eta^{\alpha\beta}_{A}=\frac{\pi^{2}}{3}\frac{k_{B}^{2}T}{e}\frac{\partial \sigma_{A}^{\alpha\beta}(\mu)}{\partial \mu}$ \cite{PhysRevLett.97.026603}. This proportionality between 2nd Nernst and Hall conductivity results from that the 2nd order thermoelectric conductivity has a power of $\varepsilon^{2}/T^{2}$, and the non-zero contribution of the low-temperature \Eq{somo} comes form the second order.

Now we demonstrate that the 2nd order Mott relation \Eq{mott2} applies to the Drude contribution. Integrating by part, the Drude contribution of the 2nd thermoelectric conductivity \Eq{l12d} can be rewritten as
\begin{equation}\label{l12b2                                                                                                                                                                                                                                                                                                                                                                                                                                                                                                                                                                                                                                                                                                                }
\begin{aligned}
L_{12,D}^{\alpha\beta\delta}=-\sum_{p}\int d\varepsilon_{p}\frac{1}{\omega\omega_{2}} & \left[2\tilde{\varepsilon}_{p}\frac{\partial f_{p}}{\partial \varepsilon_{p}}+ \tilde{\varepsilon}_{p}^2 \frac{\partial^2 f_{p}}{\partial \varepsilon_{p}^{2}}\right]\int_{\bm{k}}v_{p}^{\alpha}v_{p}^{\beta}\\ \times &\delta(\varepsilon_{p}-\varepsilon_{p,\bm{k}}).
\end{aligned}
\end{equation}
Using \Eq{id1} and considering the large effective mass limit, the Drude contribution of the 2nd electric conductivity  is given as
\begin{equation}
\begin{aligned}
L_{11,D}^{\alpha\beta\delta}\approx -\sum_{p}\int d\varepsilon_{p}\frac{1}{\omega\omega_{2}} \tilde{\varepsilon}_{p}^2 \frac{\partial^2 f_{p}}{\partial \varepsilon_{p}^{2}}\int_{\bm{k}}v_{p}^{\alpha}v_{p}^{\beta}\delta(\varepsilon_{p}-\varepsilon_{p,\bm{k}}).
\end{aligned}
\end{equation}
By use of the Sommerfeld expansion \Eq{somo}, the 2nd order Mott relation \Eq{mott2} for the Drude term is directly testified.

\subsection{Second-order thermal conductivity and  Wiedemann-Franz law}
According to \Eq{jrhok},  the Kubo contribution to the 2nd order heat current is given by
\begin{equation}
\begin{aligned}
J_{Q}^{{\rm{Kubo}},(2),\alpha}(\omega)=&\int_{\bm{k}}{\rm{Tr}}\left[ \hat{j}^{\alpha}_{Q}\rho^{(2)} \right].
\end{aligned}
\end{equation}
By use of the expansion of the 2nd RDM, the 2nd order thermoelectric response is expressed in form of four  integral kernels
\begin{equation}
\begin{aligned}
&L_{22}^{{\rm{Kubo}},\alpha\beta\delta}=\int_{\bm{k}}\left[\Xi^{(2),\beta\delta}+\Xi^{(2),\beta}+\Xi^{(2),\delta}+\Xi^{(2)}\right],
\end{aligned}
\end{equation}
where $\Xi^{(2),\beta\delta}$, $\Xi^{(2),\beta}$, $\Xi^{(2),\delta}$ and $\Xi^{(2)}$ are given by (see Appendix. \ref{app4} for details)
\begin{equation}\label{xi1}
\begin{aligned}
&\Xi^{(2),\beta\delta}=\sum_{p}v_{p}^\alpha \frac{1}{\omega}\tilde{\varepsilon}_{p}^{2}\partial^{\beta}\left[\frac{1}{\omega-\omega_{1}}\tilde{\varepsilon}_{p}\partial^{\delta}f_{p}\right],\\
&\Xi^{(2),\beta}=\sum_{p,q}-i\frac{1}{2}v_{pq}^{\alpha}(\tilde{\varepsilon}_{p}+\tilde{\varepsilon}_{q})\frac{1}{\omega}\tilde{\varepsilon}_{p}\partial^{\beta}\left[\frac{1}{\omega-\omega_{1}-\varepsilon_{qp}} \right. \\
& \left. \quad\quad\quad \times(\tilde{\varepsilon}_{p}+\tilde{\varepsilon}_{q})\mathcal{A}^{\delta}_{qp}f_{pq}\vphantom{\frac{1}{\omega-\omega_{1}-\varepsilon_{qp}}}\right],\\
&\Xi^{(2),\delta}=\sum_{p,q}-i\frac{1}{4}(\tilde{\varepsilon}_{p}+\tilde{\varepsilon}_{q})v_{pq}^{\alpha}\frac{1}{(\omega-\varepsilon_{qp})}\frac{1}{(\omega-\omega_{1})}\\
&\quad\quad\quad \times \mathcal{A}^{\beta}_{qp}(\tilde{\varepsilon}_{p}\tilde{\varepsilon}_{q}\partial^{\delta} f_{pq}+\tilde{\varepsilon}_{p}^{2}\partial^{\delta}f_{p}-\tilde{\varepsilon}_{q}^{2}\partial^{\delta}f_{q}),\\
&\Xi^{(2)}=\sum_{p,q,r}-\frac{1}{8} v_{pq}^{\alpha}(\tilde{\varepsilon}_{p}+\tilde{\varepsilon}_{q})\frac{1}{\omega-\varepsilon_{qp}}(\tilde{\varepsilon}_{q}+\tilde{\varepsilon}_{r})\mathcal{A}^{\beta}_{qr}\\
&\quad\quad\quad \times\frac{1}{\omega-\omega_{1}-\varepsilon_{rp}}(\tilde{\varepsilon}_{r}+\tilde{\varepsilon}_{p})\mathcal{A}^{\delta}_{rp}(f_{rp}-f_{qr}).
\end{aligned}
\end{equation}
It can be seen that the poles $\Xi^{(2),...}$ is identical to that of $\Pi^{(2),...}$, with the leading term contributed by $\Xi^{(2),\beta\delta}$ and $\Xi^{(2),\delta}$. Hence the Kubo contribution in DC limit is found as
\begin{equation}\label{2l22dc}
\begin{aligned}
&L_{22,DC}^{{\rm{Kubo}},\alpha\beta\delta}(\omega ;\omega_{1},\omega_{2})\\
=-&\sum_{p,q}\int_{\bm{k}}\left[ \frac{1}{\omega\omega_{2}}v_{p}^\alpha \tilde{\varepsilon}_{p}^{2}\partial^{\beta}(\tilde{\varepsilon}_{p}\partial^{\delta}f_{p})  +i\frac{1}{4\omega_{2}}\tilde{\varepsilon}_{p}(\tilde{\varepsilon}_{p}+\tilde{\varepsilon}_{q})^{2}\right. \\
&\left. \times\frac{v_{pq}^{\alpha}v_{qp}^{\beta}-v_{pq}^{\beta}v_{qp}^{\alpha}}{\varepsilon_{pq}^2}\partial^{\delta}f_{p}\right].
\end{aligned}
\end{equation}
By use of the quantity $w_{p}^{\gamma}$ introduced in \Eq{m2p}, it can be rewritten as
\begin{equation}\label{2l22dcf}
\begin{aligned}
&L_{22,DC}^{{\rm{Kubo}},\alpha\beta\delta}(\omega ;\omega_{1},\omega_{2})\\
=-&\sum_{p,q}\int_{\bm{k}}\left[ \frac{1}{\omega\omega_{2}}v_{p}^\alpha \tilde{\varepsilon}_{p}^{2}\partial^{\beta}(\tilde{\varepsilon}_{q}\partial^{\delta}f_{p})  +i\frac{1}{\omega_{2}}\tilde{\varepsilon}_{p}w^{\gamma}_{p}\partial^{\delta}f_{p}\right].
\end{aligned}
\end{equation}
The 2nd order heat magnetization is written as
\begin{equation}\label{L22lleq2}
\begin{aligned}
M_{Q}^{(2),\gamma}(\omega)={\rm{Tr}}\left[\int_{\bm{k}}w^{\gamma}\rho^{(1)}(\omega)-\frac{1}{e^2}\int d\varepsilon\tilde{\varepsilon} \sigma^{\gamma}(\varepsilon)\rho^{(1)}(\omega)\right].
\end{aligned}
\end{equation}
Hence we obtain the 2nd order thermal-thermal magnetization response
\begin{equation}
\begin{aligned}\label{m22dc_2}
M_{22,DC}^{\gamma\delta}(\omega)=&i\sum_{p}\int_{\bm{k}}\frac{1}{\omega}\tilde{\varepsilon}_{p} w^{\gamma}_{p}\partial^{\delta}f_{p}\\
&-i\frac{1}{e^2}\sum_{p}\int d\varepsilon \frac{1}{\omega}\tilde{\varepsilon}^{2}_{p} \sigma^{\gamma}(\varepsilon)\partial^{\delta}f_{p}(\varepsilon) .
\end{aligned}
\end{equation}
From \Eq{2l22dc} and \Eq{m22dc_2}, we obtain the 2nd order thermal-thermal response
\begin{equation}\label{l122final}
\begin{aligned}
L_{22,DC}^{{\rm{tr}},\alpha\beta\delta}(\omega;\omega_{1},\omega_{2})=L_{22,D}^{\alpha\beta\delta}(\omega;\omega_{1},\omega_{2})+L_{22,A}^{\alpha\beta\delta}(\omega;\omega_{1},\omega_{2})
\end{aligned}
\end{equation}
with
\begin{equation}
\begin{aligned}
&L_{22,D}^{\alpha\beta\delta}=-\sum_{p}\int_{\bm{k}}\frac{1}{\omega\omega_{2}} v_{p}^{\alpha}\tilde{\varepsilon}_{p}^{2}\partial^{\beta}(\tilde{\varepsilon}_{p}\partial^{\delta}f_{p}),\\
&L_{22,A}^{\alpha\beta\delta}=i\epsilon^{\alpha\beta\gamma}\sum_{p}\int d\varepsilon_{p}  \frac{1}{\omega_2}v_{p}^{\delta}\sigma^{\gamma}(\varepsilon_p)\left[2\tilde{\varepsilon}_{p}^{2}\frac{\partial f_{p}}{\partial \varepsilon_{p}}+ \tilde{\varepsilon}_{p}^3 \frac{\partial^2 f_{p}}{\partial \varepsilon_{p}^{2}}\right].
\end{aligned}
\end{equation}
By use of the Sommerfeld expansion  \Eq{somo} and the identity $\kappa^{\alpha\beta\delta}=L_{22}^{{\rm{tr}},\alpha\beta\delta}/T^{2}$, it  yields
\begin{equation}\label{2ndWF}
\begin{aligned}
\sigma^{\alpha\beta\delta}=-\frac{e}{2L}\frac{\partial \kappa^{\alpha\beta\delta}(\mu)}{\partial \mu}.
\end{aligned}
\end{equation}
We call \Eq{2ndWF} as the second order WF law. We see that the relation between the 2nd order thermal conductivity $\kappa^{\alpha\beta\delta}$ and the 2nd order electric conductivity $\sigma^{\alpha\beta\delta}(\mu)$ does not obey the linear WF law in \Eq{WFlaw}, which is $\kappa^{\alpha\beta}=LT\sigma^{\alpha\beta}$. In the second order response, the 2nd order electric conductivity $\sigma^{\alpha\beta\delta}$ is proportional to the first derivative of the 2nd order  thermal conductivity $\kappa^{\alpha\beta\delta}(\mu)$ to the chemical potential, rather than to $\kappa^{\alpha\beta\delta}(\mu)$ itself.

\subsection{Third-order thermal response}
The Kubo contribution of the 3rd order electric current is written as
\begin{widetext}
\begin{equation}
\begin{aligned}
J_{N}^{{\rm{Kubo}},(3),\alpha}(\omega)
=&\int_{\bm{k}}{\rm{Tr}}\left[j_{N}^{\alpha} \rho^{(3)}\right],
\end{aligned}
\end{equation}
where  the 3rd order RDM is given by
\begin{equation}
\rho^{(3)}(\omega)=i^{3}\int d\omega_{1}\int d\omega_{2}\int d\omega_{3}E_{T}^{\alpha_{1}}(\omega_{1})E_{T}^{\alpha_{2}}(\omega_{2})E_{T}^{\alpha_{3}}(\omega_{3})\left(d(\omega)\circ \left[\mathcal{D}^{\beta}\left[d(\omega-\omega_{1}) \circ\left[\mathcal{D}^{\delta}\left[d(\omega - \omega_{[2]}) \circ [\mathcal{D}^{\zeta} [\rho^{(0)}]\right]\right]\right]\right]\right),
\end{equation}
in which the expansion of the 3rd order RDM results in eight terms, hence the 3rd Kubo thermoelectric response can be rewritten as (for details see Appendix.\ref{app6})
\begin{equation}
\begin{aligned}
L_{12}^{{\rm{Kubo}},\alpha\beta\delta\zeta}(\omega ;\omega_{1},\omega_{2},\omega_{3})=\int_{\bm{k}}\left[\Pi^{(3),\beta\delta\zeta}+\Pi^{(3),\beta\delta}+\Pi^{(3),\beta\zeta}+\Pi^{(3),\delta\zeta}+\Pi^{(3),\beta}+\Pi^{(3),\delta}+\Pi^{(3),\zeta}+\Pi^{(3)}\right].
\end{aligned}
\end{equation}
The expressions of the $\Pi^{(3),...}$s are shown in the Appendix.\ref{app6}. The derivation of the 3rd order thermoelectric conductivity in the DC limit can be done by calculating the poles of the denominator of $\Pi^{(3),...}$. The divergent terms are $\Pi^{(3),\beta\delta\zeta}$ (with poles of $0$, $\omega_{1}$ and $\omega_{1}+\omega_{2}$), $\Pi^{(3),\beta\zeta}$ (with poles of $\omega_{1}+\omega_{2}$) and $\Pi^{(3),\alpha\delta\zeta}$ (with poles of $\omega_{1}$ and $\omega_{1}+\omega_{2}$). Reserving the leading terms of $O(\omega^{-3})$ ($\Pi^{(3),\beta\delta\zeta}$) and $O(\omega^{-2})$ ($\Pi^{(3),\delta\zeta}$), we obtain the 3rd order thermoelectric conductivity in the DC limit as
\begin{equation}
\begin{aligned}
L_{12,DC}^{{\rm{Kubo}},\alpha\beta\delta\zeta}(\omega ;\omega_{1},\omega_{2},\omega_{3})
&=\sum_{p}\int_{\bm{k}}\left\{-i\frac{1}{\omega\omega_{[2]}\omega_{3}}v_{p}^{\alpha}\tilde{\varepsilon}_{p}\partial^{\beta}
\left[\tilde{\varepsilon}_{p}\partial^{\delta}(\tilde{\varepsilon}_{p}\partial^{\zeta}f_{p})\right]-\frac{1}{2\omega_{[2]}\omega_{3}}
\frac{v_{pq}^{\alpha}v^{\beta}_{qp}}{\varepsilon_{pq}^{2}}\left[\tilde{\varepsilon}_{q}\tilde{\varepsilon}_{p}\partial^{\delta}(\tilde{\varepsilon}_{p}\partial^{\zeta} f_{p})\right.\right.\\
&\left.\left.-\tilde{\varepsilon}_{q}\tilde{\varepsilon}_{p}\partial^{\delta}(\tilde{\varepsilon}_{q}\partial^{\zeta} f_{q}) +\tilde{\varepsilon}_{p}^{2}\partial^{\delta}(\tilde{\varepsilon}_{p}\partial^{\zeta} f_{p})-\tilde{\varepsilon}_{q}^{2}\partial^{\delta}(\tilde{\varepsilon}_{q}\partial^{\zeta} f_{q})\right]\right\},
\end{aligned}
\end{equation}
which can be written in a more compact form
\begin{equation}\label{L12lk3}
\begin{aligned}
L_{12,DC}^{{\rm{Kubo}},\alpha\beta\delta\zeta}(\omega ;\omega_{1},\omega_{2},\omega_{3})
=&\sum_{p}\int_{\bm{k}}\left\{-i\frac{1}{\omega\omega_{[2]}\omega_{3}}v_{p}^{\alpha}\tilde{\varepsilon}_{p}\partial^{\beta}\left[\tilde{\varepsilon}_{p}\partial^{\delta}(\tilde{\varepsilon}_{p}\partial^{\zeta}f_{p})\right]-\frac{1}{\omega_{[2]}\omega_{3}}(\tilde{\varepsilon}_{p}m^{\gamma}_{p}+\tilde{\varepsilon}_{p}^{2}\Omega^{\gamma}_{p})\partial^{\delta}(\tilde{\varepsilon}_{p}\partial^{\zeta} f_{p})\right\}.
\end{aligned}
\end{equation}
The 3rd order particle magnetization is given as
\begin{equation}
\begin{aligned}\label{L12lleq3}
M_{N}^{(3),\gamma}(\omega)={\rm{Tr}}\left[\int_{\bm{k}}\rho^{(2)}(\omega) m^{\gamma}+\frac{1}{e}\int d\varepsilon \sigma^{\gamma}(\varepsilon)\rho^{(2)}(\omega) \right].
\end{aligned}
\end{equation}
Noting that only the terms  up to $O(\omega^{-2})$ are retained. By use of the expansion of $\rho^{(2)}$ (see Appendix.\ref{app6} for details), the leading term is proportional to $\Pi^{(2),\beta\delta}$. Hence we obtain the $3$-rd order thermoelectric magnetization response
\begin{equation}
\begin{aligned}\label{L12lleq3p}
M_{12}^{\gamma\delta\zeta}(\omega;\omega_{1},\omega_{2},\omega_{3})=\sum_{p}\int_{\bm{k}} \frac{1}{\omega\omega_{2}}\tilde{\varepsilon}_{p}m^{\gamma}\partial^{\delta}(\tilde{\varepsilon}_{p}\partial^{\zeta}f_{p})+\frac{1}{e}\int d\varepsilon \frac{1}{\omega\omega_{2}}\tilde{\varepsilon}_{p}\sigma^{\gamma}_{p}(\varepsilon)\partial^{\delta}(\tilde{\varepsilon}_{p}\partial^{\zeta}f_{p}).
\end{aligned}
\end{equation}
Combining \Eq{L12lleq3p} and \Eq{L12lk3}, we finally obtain the 3rd order thermoelectric response
\begin{equation}
\begin{aligned}\label{L12ltr3}
L_{12,DC}^{{\rm{tr}},\alpha\beta\delta\zeta}(\omega ;\omega_{1},\omega_{2},\omega_{3})
=L_{12,D}^{\alpha\beta\delta\zeta}(\omega ;\omega_{1},\omega_{2},\omega_{3})+L_{12,A}^{\alpha\beta\delta\zeta}(\omega ;\omega_{1},\omega_{2},\omega_{3}).
\end{aligned}
\end{equation}
with
\begin{eqnarray}
L_{12,D}^{\alpha\beta\delta\zeta}(\omega ;\omega_{1},\omega_{2},\omega_{3})&=&\sum_{p}\int_{\bm{k}}-i\frac{1}{\omega\omega_{[2]}\omega_{3}}v_{p}^{\alpha}\tilde{\varepsilon}_{p}\partial^{\beta}\left[\tilde{\varepsilon}_{p}\partial^{\delta}(\tilde{\varepsilon}_{p}\partial^{\zeta}f_{p})\right],\\
L_{12,A}^{\alpha\beta\delta\zeta}(\omega ;\omega_{1},\omega_{2},\omega_{3})&=&\epsilon^{\alpha\beta\gamma}\sum_{p}\int d\varepsilon_p \frac{1}{\omega\omega_{2}} \left[6\tilde{\varepsilon}_{p}\frac{\partial f_{p}}{\partial \varepsilon_p}+6\tilde{\varepsilon}_{p}^{2}\frac{\partial^2 f_p}{\partial \varepsilon^2_{p}}+ \tilde{\varepsilon}_{p}^3 \frac{\partial^3 f}{\partial \varepsilon^3_{p}}\right]  v^{\delta}(\varepsilon_p)v^{\zeta}(\varepsilon_p)\sigma^{\gamma}(\varepsilon_p).
\end{eqnarray}

\renewcommand\arraystretch{1.9}
\begin{table*}
\caption{The high order of thermal to electric conductivity, and thermal to thermal conductivity, i.e. the higher order Mott relation and values of WF law are summarized up to the third order. $L=\frac{1}{3}\left(\frac{k_{B}\pi}{e}\right)^2=2.44\times 10^{-8} W\Omega/K^{2}$ is the well-known first order Lorentz number. }\label{highorderMottWF}
\begin{tabular*}{10 cm}{l l r}
\hline
 Order     & Thermal-electric (Mott) & Thermal-thermal (Wiedemann-Franz)  \\
\hline
1st        &  $\sigma^{\alpha\beta}=\frac{1}{eLT}\int_{-\infty}^{\mu}d\varepsilon\eta^{\alpha\beta}(\varepsilon)$   &   $\sigma^{\alpha\beta}=\frac{1}{LT}\kappa^{\alpha\beta}$    \\
\hline
2nd        &     $\sigma^{\alpha\beta\delta}=\frac{1}{L}\eta^{\alpha\beta\delta}$        &   $\sigma^{\alpha\beta\delta}=-\frac{e}{2L}\frac{\partial \kappa^{\alpha\beta\delta}(\mu)}{\partial \mu}$     \\
\hline
3rd &  $\sigma^{\alpha\beta\delta\zeta} = \frac{e}{9T{L}}\frac{\partial \eta^{\alpha\beta\delta\zeta}(\mu)}{\partial \mu}$  &  $\sigma^{\alpha\beta\delta\zeta} = \frac{e^2}{42TL}\frac{\partial^2 \kappa^{\alpha\beta\delta\zeta}(\mu)}{\partial \mu^2} $ \\
\hline
\end{tabular*}
\end{table*}
\end{widetext}

Following the similar process, the Drude part and the anomalous part of the 3rd order electric conductivity is given as
\begin{equation}
\begin{aligned}\label{L11ltr3}
&L_{11,D}^{\alpha\beta\delta\zeta}(\omega ;\omega_{1},\omega_{2},\omega_{3})
=\sum_{p}\int_{\bm{k}}\frac{-i}{\omega\omega_{[2]}\omega_{3}}v_{p}^{\alpha}\partial^{\beta}
\left[\partial^{\delta}(\partial^{\zeta}f_{p})\right],\\
&L_{11,A}^{\alpha\beta\delta\zeta}(\omega ;\omega_{1},\omega_{2},\omega_{3})
=\epsilon^{\alpha\beta\gamma}\sum_{p}\int_{\bm{k}}\frac{1}{\omega_{[2]}\omega_{3}}\Omega^{\gamma}_{p}\partial^{\delta}(\partial^{\zeta} f_{p}).
\end{aligned}
\end{equation}
In the limit of large effective mass, the anomalous part is approximated as
\begin{equation}
\begin{aligned}
&L_{11,A}^{\alpha\beta\delta\zeta}(\omega ;\omega_{1},\omega_{2},\omega_{3})\\
\approx  & - \epsilon^{\alpha\beta\gamma}\sum_{p}\int d\varepsilon_{p}  \frac{1}{\omega_{[2]}\omega_{3}} \frac{\partial^3 f_{p}}{\partial \varepsilon^{3}_{p}} v^{\delta}_{p}v^{\xi}_{p} \sigma^{\beta}(\varepsilon_{p}).
\end{aligned}
\end{equation}
By use of the Sommerfeld expansion \Eq{somo}  and considering that $\sigma^{\alpha\beta\delta\zeta}=L_{DC,11}^{{\rm{tr}},\alpha\beta\delta\zeta}$, $\eta^{\alpha\beta\delta\zeta}=L_{DC,12}^{{\rm{tr}},\alpha\beta\delta\zeta}/T^3$, we obtain
\begin{equation}
\begin{aligned}
\sigma^{\alpha\beta\delta\xi}=\frac{e}{9TL}\frac{\partial\eta^{\alpha\beta\delta\xi}(\mu )}{\partial \mu}.
\end{aligned}
\end{equation}
After a similar derivation for the 3rd order thermal conductivity (see Appendix. \ref{app7}), we obtain
\begin{equation}
\begin{aligned}
\sigma^{\alpha\beta\delta\xi}=\frac{e^2}{42TL} \frac{\partial^2 \kappa^{\alpha\beta\delta\xi}(\mu)}{\partial \mu^2}.
\end{aligned}
\end{equation}
Interestingly, it is found that at 3rd order the electric conductivity is proportional to the first derivative of the 3rd order thermoelectric conductivity. Analogously, the 3rd order  electric conductivity is proportional to the second derivative of the 3rd order thermal conductivity.

 According to the expression of the thermally expanded Hamiltonian \Eq{hexp}, it is seen that expanding one more order of $\bm{E}_{T}$ is accompanied by one more order of the band energy.  Given the fact that the order of band energy in response functions determines the leading terms in low-temperature expansion, hence we reach the conclusion that for the nonlinear Mott relation, the $n$-th order electric conductivity is proportional to the $n-2$-th order  derivative of the $n$-th order thermoelectric conductivity  with respect to the chemical potential. For the nonlinear WF law,  the $n$-th order electric conductivity is proportional to the $n-1$-th  derivative of the $n$-th order thermal conductivity with respect to the chemical potential  (see Table. \ref{highorderMottWF}).

\section{Semiclassical Approach}\label{app7}
In this section we carefully give the derivation of the nonlinear thermal response through the semiclassical approach. We start with the semiclassical Boltzmann equation, then show that it matches the results from the quantum approach in previous sections. In the last we discuss the symmetries of the nonlinear currents.

The local particle or heat current is contributed by two parts: one is from the motion of the wave-packet center, the other is from the self-rotation of the wave-packet, which can be written as
\begin{equation}
\begin{aligned}
\bm{J}_{N}=&\int_{\bm{k}}f(\varepsilon_{\bm{k}})\dot{\bm{r}}+\bm{\nabla}_{\bm{r}}\times\int_{\bm{k}}f(\varepsilon_{\bm{k}})\bm{m}(\bm{k}),  \\
\bm{J}_{Q}=&\int_{\bm{k}}(\varepsilon -\mu)f(\varepsilon_{\bm{k}})\dot{\bm{r}}+\bm{\nabla}\times\int_{\bm{k}}f(\varepsilon_{\bm{k}})\bm{m}^{Q}(\bm{k}).
\end{aligned}
\end{equation}
In which we introduce  the energy and thermal magnetic moment
\begin{equation}
\bm{m}^{E}(\bm{k})=\varepsilon_{\bm{k}}\bm{m}(\bm{k}), \quad \bm{m}^{Q}(\bm{k})=\bm{m}^{E}(\bm{k})-\mu \bm{m}(\bm{k}).
\end{equation}
We write the formula of the transport currents again
\begin{equation}
\bm{J}^{\rm{tr}}_{N(Q)}=\bm{J}_{N(Q)}-\bm{\nabla}\times \bm{M}_{N(Q)}.
\end{equation}
The total particle magnetization can be derived based on the wave-packed theory using a confining potential \cite{RevModPhys.82.1959}
\begin{equation}\label{msn}
\bm{M}_{N}=\int_{\bm{k}}f(\varepsilon_{\bm{k}})\bm{m}(\bm{k})-\frac{1}{e^2}\int d\varepsilon f(\varepsilon)\bm{\sigma}(\varepsilon).
\end{equation}
In which $\bm{\sigma}(\varepsilon)=\frac{e^2}{\hbar}\int_{\bm{k}}\Theta(\varepsilon -\varepsilon_{\bm{k}})\bm{\Omega}(\bm{k})$ is the zero-temperature Hall conductivity  with Fermi energy $\varepsilon$.
The thermal magnetization is written as \cite{Zhang_2016}
\begin{equation}\label{msq}
\bm{M}^{Q}=\int_{\bm{k}}f(\varepsilon_{\bm{k}})\bm{m}^{Q}(\bm{k})-\frac{1}{e^2}\int d\varepsilon (\varepsilon -\mu) f(\varepsilon)\bm{\sigma}(\varepsilon).
\end{equation}
 Note that the first term is from the self rotation of the wave-packet, while the second term is contributed by the edge, as it vanishes in the bulk for a uniform system. Using \Eq{msn} and \Eq{msq}, the transport current is found as
\begin{equation}
\bm{J}^{\rm{tr}}_{N(Q)} = \bm{J}^{D}_{N(Q)} + \bm{J}_{N(Q)}^{A},
\end{equation}
where the first term is the Drude contribution
\begin{eqnarray}
\bm{J}^{D}_{N} &=& \int_{\bm{k}}f(\varepsilon_{\bm{k}})\bm{v}_{\bm{k}}, \\
\bm{J}^{D}_{Q} &=& \int_{\bm{k}}(\varepsilon_{\bm{k}}-\mu)f(\varepsilon_{\bm{k}})\bm{v}_{\bm{k}}.
\end{eqnarray}
The second term is from the anomalous term, manifesting itself as the anomalous Nernst (thermal Hall) effect.
\begin{eqnarray}
\bm{J}^{A}_{N} &=&  - \frac{1}{e^2}\bm{\nabla}\times \int d\varepsilon f(\varepsilon)\bm{\sigma}(\varepsilon), \label{ja1ch} \\
\bm{J}^{A}_{Q} &=&  - \frac{1}{e^2}\bm{\nabla}\times \int d\varepsilon (\varepsilon -\mu) f(\varepsilon)\bm{\sigma}(\varepsilon), \label{ja1ch2}
\end{eqnarray}
and the anomalous Hall effect
\begin{equation}
\begin{aligned}
\bm{J}^{A}_{N} = \frac{e}{\hbar}\bm{E}\times \int_{\bm{k}} f(\varepsilon_{\bm{k}})\bm{\Omega}(\bm{k}).
\end{aligned}
\end{equation}
It is worth noting that the contribution from the particle magnetic moment $\bm{m}(\bm{k})$ cancels out, since it is localized and does not contribute to transport.

The Boltzmann equation is given as
\begin{equation}
(\partial t + \dot{\bm{r}}\cdot \bm{\nabla }_{\bm{r}}+\dot{\bm{k}}\cdot \bm{\nabla }_{\bm{k}})f(\bm{r},\bm{k},t)=\mathcal{I}_{\rm{coll}}[f(\bm{r},\bm{k},t)].
\end{equation}
where the collision integral $\mathcal{I}_{\rm{coll}}[f(\bm{r},\bm{k},t)]$ captures the effect of scattering. In the absence of the magnetic field, the equations of motion are given by
\begin{equation}
\begin{aligned}
&\dot{\bm{r}}=\frac{\partial \varepsilon_{\bm{k}}}{\hbar \partial \bm{k}}-\dot{\bm{k}}\times \bm{\Omega}(\bm{k}),\\
&\hbar\dot{\bm{k}}=-e\bm{E}
\end{aligned}
\end{equation}
By expanding the distribution function $f=\sum_{n=0}^{\infty}f_{n}$ by order of temperature gradient $\bm{\nabla} T$ or electric field $\bm{E}$,  the Hall current at each order is obtained by replacing the distribution function by $f_{n}$.   Since we are interested in the steady-state solution, the  $t$ dependence of $f(\bm{r},\bm{k},t)$ is dropped.

Perturbed by homogeneous electric field, the Boltzmann equations is
\begin{equation}
-\frac{e}{\hbar}\bm{E}\cdot\bm{\nabla}_{\bm{k}}f(\bm{k})=\frac{f_{0}-f(\bm{k})}{\tau},
\end{equation}
where $\tau$ is the relaxation time. The iteration relation is found as
\begin{equation}
f^{E}_{n} = \frac{e}{\hbar}\tau \bm{E}\cdot \bm{\nabla}_{\bm{k}}f^{E}_{n-1}.
\end{equation}
The first two order distribution functions are directly obtained as
\begin{equation}
\begin{aligned}
&f^{E}_{1}=\frac{e\tau}{\hbar }\frac{\partial f_{0}}{\partial \varepsilon_{\bm{k}}}v_{\bm{k}}^{\alpha} E^{\alpha},\\
&f^{E}_{2}=\frac{e^2\tau^2}{\hbar^2 }\left(\frac{\partial f_{0}}{\partial \varepsilon_{\bm{k}}}\frac{\partial v^{\alpha}_{\bm{k}}}{\partial k_{\beta}}+\frac{\partial^2 f_{0}}{\partial \varepsilon_{\bm{k}}^2}v^{\alpha}_{\bm{k}}v^{\beta}_{\bm{k}}\right)E^{\alpha}E^{\beta}.
\end{aligned}
\end{equation}
Following the same procedure, the  Boltzmann equations in the presence of  temperature gradient is
\begin{equation}
\dot{\bm{r}} \cdot\bm{\nabla}_{\bm{r}}f(\bm{r},\bm{k})=\frac{f_{0}-f(\bm{r},\bm{k})}{\tau},
\end{equation}
and the iteration relation is found as
\begin{equation}
f_{n}^{T}=-\tau \bm{v}_{\bm{k}}\cdot \bm{\nabla}_{\bm{r}}f^{T}_{n-1}=(-\tau \bm{v}_{\bm{k}}\cdot \bm{\nabla}_{\bm{r}})^{n}f_{0}.
\end{equation}
The first two order distribution functions are written as
\begin{equation}\label{ff1}
\begin{aligned}
f^{T}_{1}=&\frac{\tau}{\hbar T}\mathcal{F}_{1}(\varepsilon_{\bm{k}})v^{\alpha}_{\bm{k}}\nabla^{\alpha}T, \\
f^{T}_{2}=&  \frac{\tau^2}{\hbar T^2}\mathcal{F}_{2}(\varepsilon_{\bm{k}})v^{\alpha}_{\bm{k}}v^{\beta}_{\bm{k}}\nabla^{\alpha}T\nabla^{\beta}T,
\end{aligned}
\end{equation}
where we define
\begin{equation}
\begin{aligned}
\mathcal{F}_{1}^{T}(\varepsilon_{\bm{k}})=&(\varepsilon_{\bm{k}}-\mu)\frac{\partial f_{0}}{\partial \varepsilon_{\bm{k}}},\\
\mathcal{F}_{2}^{T}(\varepsilon_{\bm{k}})=&\left[2\mathcal{F}_{1}^{T}(\varepsilon_{\bm{k}}) + (\varepsilon_{\bm{k}}-\mu)^2 \frac{\partial^2 f_{0}}{\partial \varepsilon_{\bm{k}}^2}\right].
\end{aligned}
\end{equation}
By use of the relation
\begin{equation}
\bm{\nabla}_{\bm{r}}f_{0} = -\frac{1}{T}(\varepsilon_{\bm{k}}-\mu)\frac{\partial f_{0}}{\partial \varepsilon_{\bm{k}}}\bm{\nabla}T
\end{equation}
and substituting the formula of $f^{T}_{n}$ of \Eq{ff1} into \Eq{ja1ch} and \Eq{ja1ch2}, one  obtains the 2nd order anomalous Nernst (thermal Hall) conductivity
\begin{equation}\label{ji2}
\begin{aligned}
\eta^{\gamma\alpha\delta}=&-\frac{e\tau}{\hbar^{2}T^2}\epsilon^{\alpha\beta\gamma} \int d\varepsilon \mathcal{F}_{2}^{T}(\varepsilon)v^{\delta}(\varepsilon) \sigma^{\beta}(\varepsilon), \\
\kappa^{\gamma\alpha\delta}=&\frac{\tau}{\hbar^{2}T^2}\epsilon^{\alpha\beta\gamma} \int d\varepsilon (\varepsilon-\mu)\mathcal{F}_{2}^{T}(\varepsilon)v^{\delta}(\varepsilon) \sigma^{\beta}(\varepsilon).
\end{aligned}
\end{equation}
In which we assume the temperature is slowly varying in space, and
omit the terms that are of nonlinear temperature gradient.
\Eq{ji2} reproduces the formulas derived from the quantum approach in Sec. \ref{sec3}.

Now we investigate how the large effective mass limit changes the thermal transport coefficient. According to \Eq{ja1ch}, the $n$-th order anomalous currents are given as
\begin{equation}
\begin{aligned}
\bm{J}^{A,(n)}_{N}=&\frac{1}{\hbar}\bm{\nabla}\times \int d\varepsilon F^{(n-1)}(\varepsilon) \int_{\bm{k}}\delta(\varepsilon -\varepsilon_{\bm{k}})\bm{\Omega}(\bm{k}), \\
\bm{J}^{A,(n)}_{Q}=&\frac{1}{\hbar}\bm{\nabla}\times \int d\varepsilon G^{(n-1)}(\varepsilon) \int_{\bm{k}}\delta(\varepsilon -\varepsilon_{\bm{k}})\bm{\Omega}(\bm{k}),
\end{aligned}
\end{equation}
where $F^{(n)}$ and $G^{(n)}$ are the primitive functions of $f^{T}_{1}$ and $(\varepsilon -\mu)f^{T}_{1}$:
\begin{equation}
\begin{aligned}
F^{(n)}=\int_{-\infty}^{\varepsilon}f^{(n)}(\varepsilon^\prime)d\varepsilon^\prime, \quad G^{(n)}=\int_{-\infty}^{\varepsilon}(\varepsilon^\prime -\mu)f^{(n)}(\varepsilon^\prime)d\varepsilon^\prime .
\end{aligned}
\end{equation}
Under the large effective mass limit, $F^{(1)}$ and $G^{(1)}$ are found as
\begin{equation}
\begin{aligned}
F^{(1)} \approx & \frac{\tau}{T}S(f_{0})\bm{v}_{\bm{k}}\cdot \bm{\nabla}T,\\
G^{(1)} \approx & \frac{\tau}{T}C(f_{0})\bm{v}_{\bm{k}} \cdot \bm{\nabla}T,
\end{aligned}
\end{equation}
where we define
\begin{equation}
\begin{aligned}
S(f_{0})=&f_{0}\ln f_{0} + (1-f_{0})\ln (1-f_{0}),\\
C(f_{0})=&(f_{0} -1)\ln^2 (f_{0}^{-1}-1) + \ln^2 f_{0} + 2{\rm{Li}}_{2}f_{0}.
\end{aligned}
\end{equation}
Therefore we have
\begin{eqnarray}
\eta^{\alpha\beta\delta}&\approx & -\frac{e\tau}{\hbar^{2}T^2}\epsilon^{\alpha\beta\gamma} \int d\varepsilon (\varepsilon-\mu)^2 \frac{\partial f_{0}}{\partial \varepsilon} \nonumber \\
&\times& \int_{\bm{k}}\delta(\varepsilon-\varepsilon_{\bm{k}}) v_{\bm{k}}^{\delta}\Omega(\bm{k})^{\gamma}, \label{etas2}\\
\kappa^{\alpha\beta\delta}&\approx & \frac{\tau}{\hbar^{2}T^2}\epsilon^{\alpha\beta\gamma} \int d\varepsilon (\varepsilon-\mu)^3 \frac{\partial f_{0}}{\partial \varepsilon} \nonumber \\
&\times& \int_{\bm{k}}\delta(\varepsilon-\varepsilon_{\bm{k}}) v_{\bm{k}}^{\delta}\Omega(\bm{k})^{\gamma},\label{kappas2}
\end{eqnarray}
which recovers the results in Ref. \cite{PhysRevB.99.201410}.

As it is shown above, a group velocity term and a topological term together constitute the conductivity in the DC limit. Let us consider the transformation of these two terms under time-reversal symmetry $\mathcal{T}$ and inversion symmetry $\mathcal{I}$. For the group velocity term, it is composed of the group velocity or its higher order derivatives. With the definition $v_{p}^{\alpha_{1}\cdots\alpha_{n}}(\bm{k})\equiv \left[\prod_{i=1}^{n}\partial k^{\alpha_{i}}\right]\varepsilon_{p}(\bm{k})$, the time reversal $\mathcal{T}$ or the inversion $I$ give
\begin{equation}
v_{p}^{\alpha_{1}\cdots\alpha_{n}}(\bm{k})=(-)^{n}v_{p}^{\alpha_{1}\cdots\alpha_{n}}(\bm{-k}).
\end{equation}
Therefore the group velocity term in odd-order conductivity is even, leaving the momentum integral vanishes. For example, the group velocity term in the 2nd thermoelectric conductivity is $\Pi^{(2),\beta\delta}$ given by \Eq{pi1}, which is expanded as
\begin{equation}\label{pi11}
\begin{aligned}
\Pi^{(2),\beta\delta}=& \sum_{p}v_{p}^\alpha \frac{1}{\omega}\frac{1}{\omega-\omega_{1}}\left(\tilde{\varepsilon}_{p}v_{p}^{\beta}v_{p}^{\delta}\frac{\partial f_{p}}{\partial \varepsilon_{p}} \right.\\
&+\left.\tilde{\varepsilon}_{p}^{2}v_{p}^{\beta}v_{p}^{\delta}\frac{\partial^{2} f_{p}}{\partial \varepsilon_{p}^{2}}+\tilde{\varepsilon}_{p}^{2}v^{\beta\delta}_{p}\frac{\partial f_{p}}{\partial \varepsilon_{p}}\right).
\end{aligned}
\end{equation}
Referring to \Eq{pi11}, it is easy to see that $\Pi^{(2),\beta\delta}$ is odd. The topological terms are functions of $\bm{\Omega}_{p}$, $\bm{m}_{p}$ and $\bm{w}_{p}$. The time-reversal $\mathcal{T}$ gives
\begin{equation}
\bm{\Omega}_{p}(\bm{k})=-\bm{\Omega}_{p}(\bm{-k}),
\end{equation}
\begin{equation}
\bm{m}_{p}(\bm{k})=-\bm{m}_{p}(\bm{-k}),
\end{equation}
\begin{equation}
\bm{w}_{p}(\bm{k})=-\bm{w}_{p}(\bm{-k}),
\end{equation}
and the inversion $\mathcal{I}$ gives
\begin{equation}
\bm{\Omega}_{p}(\bm{k})=\bm{\Omega}_{p}(\bm{-k}),
\end{equation}
\begin{equation}
\bm{m}_{p}(\bm{k})=\bm{m}_{p}(\bm{-k}),
\end{equation}
\begin{equation}
\bm{w}_{p}(\bm{k})=\bm{w}_{p}(\bm{-k}).
\end{equation}
Therefore the topological term in odd-order conductivities is odd (even) under $\mathcal{T}$ ($\mathcal{I}$), while this term in even-order conductivities is even (odd) under $\mathcal{T}$ ($\mathcal{I}$).

\section{Nonlinear thermal response of magnons}\label{sec4}
Based on the analytical formula of nonlinear thermal conductivity, we attempt to find out a system in which the nonlinear response dominates over the linear effect. Note that although we start from a fermionic Hamiltonian to derive the thermal response, the formulas  are general and can be directly extended to bosonic or other systems.

We consider the magnon transport driven by temperature gradient in a collinear antiferromagnet on a honeycomb lattice. The Hamiltonian is
\begin{equation}\label{HS}
H=J\sum_{\langle ij \rangle}\bm{S}_{i}\cdot \bm{S}_{j}+g_{J}\mu_{B}\sum_{i}S_{i}\cdot \bm{B}+K\sum_{i}S_{iz}^2,
\end{equation}
where $J>0$ is the nearest neighbour antiferromagnetic exchange interaction. The second term  is the Zeeman coupling to the external magnetic field applied parallel to the magnetic ordering direction, in which $g_{J}$ is the Lande's g-factor and $\mu_{B}$ is the Bohr magneton. The third term ($K<0$) is the easy-axis anisotropy which ensures the N\'{e}el vector in the $z$ direction.

As the ground state of \Eq{HS} is a fully aligned antiferromagnetic order, we describe the underlying magnetic excitations by the Holstein-Primakoff transformation,
\begin{equation}
S_{iA}^{+} \approx \sqrt{2S}a_{i}, \quad S_{iA}^{-} \approx \sqrt{2S}a_{i}^\dagger, \quad S_{iA}^{z}=S-a_{i}^\dagger a_{i},
\end{equation}
\begin{equation}
S_{iB}^{+} \approx \sqrt{2S}b_{i}^\dagger , \quad S_{iB}^{-}\approx \sqrt{2S}b_{i}, \quad S_{iB}^{z}=b_{i}^\dagger b_{i}-S.
\end{equation}
Performing a Fourier transformation, the bosonic Bogoliubov-de Gennes (BdG) Hamiltonian defined in the $2\times 2$ form with a vector $\Psi_{\bm{k}}=(a_{\bm{k}}, b_{\bm{k}}^{\dagger})^{T}$ as
\begin{equation}
H_{0}(\bm{k})=S\begin{bmatrix} 3J-K+g_{J}\mu_{B}B & \gamma^{*}(\bm{k}) \\ \gamma(\bm{k}) & 3J-K-g_{J}\mu_{B}B \end{bmatrix}.
\end{equation}
We define $\gamma(\bm{k})=\sum_{i}e^{i\bm{k}\cdot \bm{\delta}_{i}}$, $\bm{\delta}_{1}=(0,1)l$, $\bm{\delta}_{2}=(\frac{\sqrt{3}}{2},-\frac{1}{2})l$ and $\bm{\delta}_{2}=(-\frac{\sqrt{3}}{2},-\frac{1}{2})l$ are the vectors connecting the nearest neighbours. For simplicity, we set $l=\frac{1}{\sqrt{3}}$.

As the next step, the Bogoliubov transformation $c_{\bm{k}}=u_{\bm{k}}a_{\bm{k}}-v_{\bm{k}}b_{\bm{k}}^{\dagger}$ and  $d_{\bm{k}}=u_{\bm{k}}b_{\bm{k}}-v_{\bm{k}}a_{\bm{k}}^{\dagger}$ is used to diagonalize $H_{0}(\bm{k})$. We need to solve the  eigenvalue equation
\begin{equation}
\begin{aligned}
&H_{0}(\bm{k})\bm{t}_{\pm}(\bm{k})=\sigma_{z}\varepsilon_{\pm}(\bm{k})\bm{t}_{\pm}(\bm{k}),\\
&H_{0}(\bm{k})\bm{t}_{\pm}(\bm{k})\sigma_{z}=\varepsilon_{\pm}(\bm{k})\bm{t}_{\pm}(\bm{k}).
\end{aligned}
\end{equation}
We only keep the particle branch (positive excitation), and the dispersions of the two branch magnons of the unstrained Hamiltonian  are given by
\begin{equation}\label{spectrum}
\varepsilon_{p=\uparrow,\downarrow}=S\sqrt{(3J-K)^2-\vert J\gamma(\bm{k}) \vert^2}\pm g_{J}\mu_{B}B.
\end{equation}
In which $\uparrow$ ($\downarrow$) denotes $z$-direction spin angular momentum carried by the magnons.
In the absence of Dzyaloshinskii-Moriya interaction (DMI), the two branches of magnons are degenerate.
The linear spin Nernst coefficient of magnons is given by
\begin{equation}\label{mshe}
\begin{aligned}
\eta^{\alpha\beta}=\sum_{p=\uparrow,\downarrow}\frac{ek_{B}}{\hbar}\epsilon^{\alpha\beta\gamma}\int_{\bm{k}}S(g_{p})\Omega_{p}^{\gamma},
\end{aligned}
\end{equation}
Distinguished from that of electrons, here $g_{p}$ is the Bose-Einstein distribution and $S(g_{p})=g_{p}\ln g_{p}-(1+g_{p})\ln (1+g_{p})$ is the entropy density of $p$ band magnons.
The thermal Hall conductivity is given as \cite{PhysRevLett.106.197202, PhysRevB.84.184406}
\begin{equation}
\kappa^{\alpha\beta}=-\sum_{p=\uparrow,\downarrow}\frac{k_{B}^{2}T}{\hbar}\epsilon^{\alpha\beta\gamma}\int_{\bm{k}}c_{2}(g_{p})\Omega^{\gamma}_{p},
\end{equation}
where the bosonic $c_{2}$ function is $c_{2}(g_{p})=(1+g_{p})(\ln{\frac{1+g_{p}}{g_{p}}})^{2}-(\ln{g_{p}})^{2}-2{\rm{Li}}_{2}(-g_{p})$ \cite{PhysRevLett.106.197202}.

In the absence of DMI, it is demonstrated in Ref. \cite{PhysRevLett.117.217202} the quadratic order expanded Hamiltonian of \Eq{HS} is invariant under combined symmetry of time-reversal ($\mathcal{T}$) and a $180^{\circ}$ rotation around the $x$ axis in the spin space ($c_{x}$). Under $\mathcal{T}c_{x}$, $\varepsilon_{p}(\bm{k})=\varepsilon_{p}(-\bm{k})$ and $\Omega_{p}(\bm{k})=-\Omega_{p}(-\bm{k})$, hence the integrand in \Eq{mshe} is odd and indicates a zero linear spin Nernst coefficient (i.e. $\eta^{\alpha\beta}_{DC}=0$). We shall emphasize that the spin Nernst effect of magnon does not exist at any order if there is no the DMI.
If the DMI is introduced, it breaks $\mathcal{T}c_{x}$ symmetry and changes the dispersion, leaving a nonzero linear spin Nernst coefficient as the leading order \cite{PhysRevLett.117.217202}.
Since we focus on zero DMI case, we will not discuss the spin Nernst effect of magnon in the following.

For the magnon thermal Hall effect (MTHE), things are different. Although the linear MTHE disappears for both zero and nonzero DMI (the two branches of magnons with opposite spin angular momentum flow in opposite transverse directions), the second-order nonlinear MTHE should exist (even for zero DMI) giving rise to a leading order contribution to the MTHE.
%
Assuming that the temperature gradient is applied along $y$ direction,  according to \Eq{kappas2} the 2nd order magnon thermal Hall conductivity is
\begin{equation}\label{kappa3}
\begin{aligned}
\kappa^{xyy}\cong \frac{\tau}{T^{2}}\sum_{p}\int_{\bm{k}}\varepsilon_{p}^{3}\partial^{y}g_{p}\Omega_{p}^{z}(\bm{k}).
\end{aligned}
\end{equation}
In deriving \Eq{kappa3}, the relaxation-time approximation for steady state $\lim_{\omega\rightarrow 0}-i/(\omega + i\Gamma) \cong \tau$ is indicated and the negligible external magnetic field is adopted. It should be noted that $\kappa^{xyy}_{DC}$ becomes zero when $T$ approaches zero \cite{ref1}.

It has been shown that the largest symmetry of a 2D crystal that allows for nonvanishing Berry curvature dipole is a mirror symmetry \cite{PhysRevLett.115.216806}.
The mirror symmetry $M_{y}$ is perpendicular to the mirror line, and the mirror symmetry $M_{y}$ requires
$
\Omega_{p}^{z}(k_{x},k_{y})=-\Omega_{p}^{z}(k_{x},-k_{y}).
$
Together with $\mathcal{T}c_{x}$, we get
$
\Omega_{p}^{z}(k_{x},k_{y})=\Omega_{p}^{z}(-k_{x},k_{y}).
$
The mirror symmetry $M_{y}$ leads to
$
\varepsilon_{p}(k_{x},k_{y})=\varepsilon_{p}(k_{x},-k_{y}),
$
When combining  $\mathcal{T}c_{x}$ and $M_{y}$, it requires
$
\varepsilon_{p}(k_{x},k_{y})=\varepsilon_{p}(-k_{x},k_{y}).
$
Therefore, the partial derivative of Bose function distribution $\partial^{x}g_{p}$ and $\partial^{y}g_{p}$ is both an odd function.

To reduce the $c_{3v}$ space group symmetry of Hamiltonian \Eq{HS} to the single Mirror symmetry $M_{y}$, we apply a uniaxial tensile strain along the  $y$ direction. Hence only the interaction along the $y$-axis changes, without lattice deformation. Hence antiferromagnetic coupling on the $d_{1}$ bonds is changed to $J(1+\delta)$, and the correction to the Hamiltonian is
\begin{equation}
H_{s}(\bm{k})=\begin{bmatrix} \delta J & \delta J \exp(i\bm{k}\cdot \bm{\delta}_{1}) \\ \delta J \exp(-i\bm{k}\cdot \bm{\delta}_{1}) & \delta J \end{bmatrix}.
\end{equation}
The total Hamiltonian is $H=H_{0}+H_{s}$, and the magnon dispersion is given by
\begin{figure}[htp]
\centering
\includegraphics [width=3.4in]{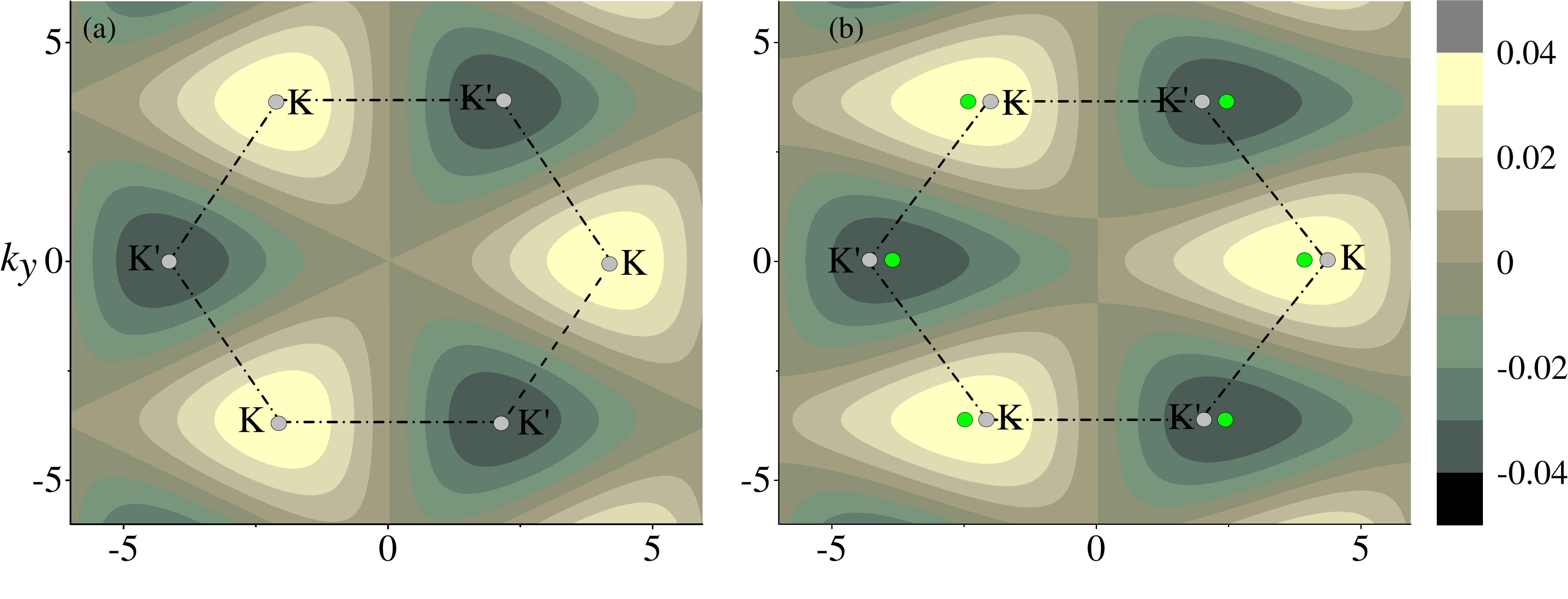}
\includegraphics [width=3.4in]{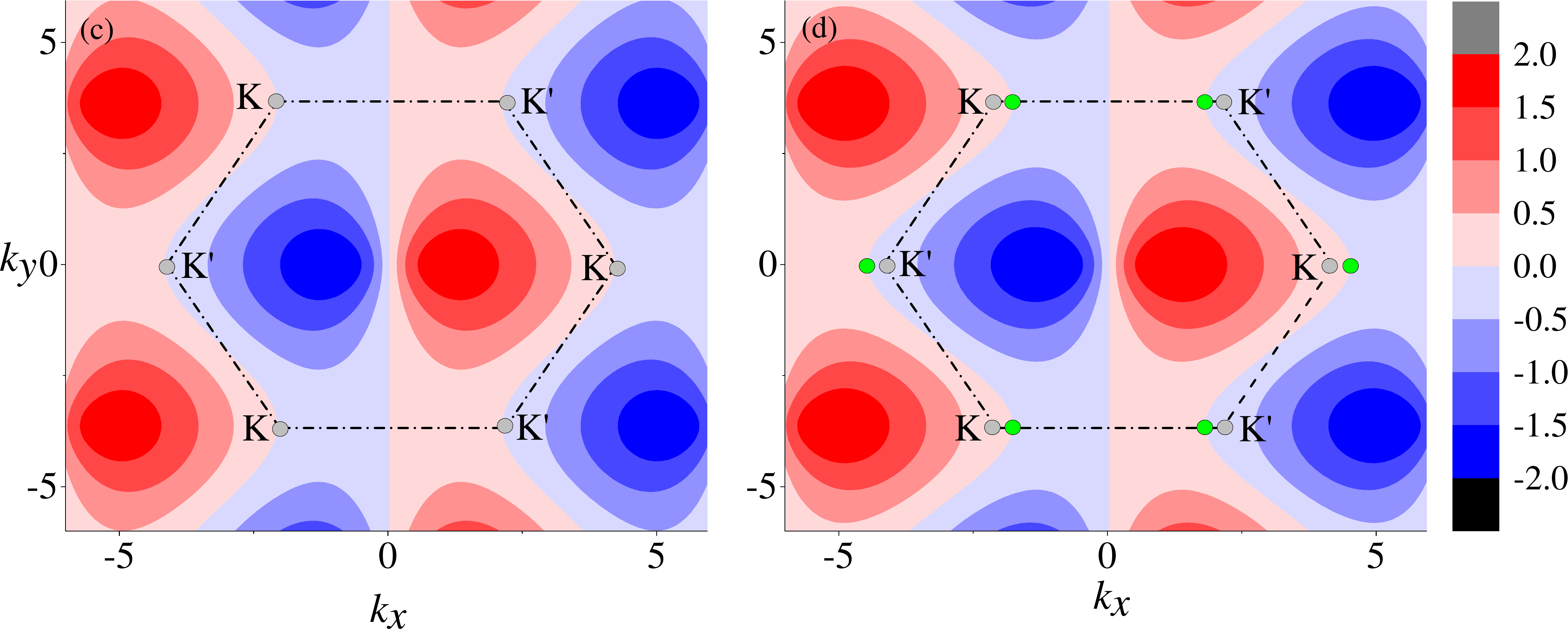}
\caption{(a)-(b). Berry curvature $\Omega_{\uparrow}^{z}(\bm{k})$ of the spin-up magnon mode without strain $\delta J=0$ (a) and with uniform uniaxial strain $\delta J=0.5$ (b).  The gray circles denote the locations of maximum value for the un-strained $\Omega_{\uparrow}^{z}(\bm{k})$, which correspond to $K$ and $K^{\prime}$. The yellow circle denotes locations of maximum value for strained $\Omega_{\uparrow}^{z}(\bm{k})$. (c)-(d). $\partial g_{\uparrow}/\partial k_{y}$ without strain $\delta J=0$ in (c) and with strain $\delta J=0.5$ in (d). The gray (yellow) circles denote the locations of maximum value for the unstrained (strained) $\partial g_{\uparrow}/\partial k_{y}$. Parameters are $J=2$, $K=-0.2$, $k_{B}T=0.5$ and $g_{J}\mu_{B}B=0.01$.  Numbers are in unit of meV. }\label{fig3}
\end{figure}

\begin{equation}\label{disp}
\begin{aligned}
\varepsilon_{p=\uparrow,\downarrow}&=S\sqrt{(3J+\delta J-K)^2-\vert J\gamma(\bm{k})+\delta
 Je^{i\bm{k}\cdot\bm{\delta}_{1}} \vert^2} \\
&\pm g_{J}\mu_{B}B.
\end{aligned}
\end{equation}

\Fig{fig3} shows the unstrained (strained) Berry curvature of spin-up magnon and the associated $\partial g_{\uparrow}/\partial k_{y}$ distribution. Considering that the integral in \Eq{kappa3} is mostly contributed from the region around $K$ and $K^\prime$. In the absence of strain (see \Fig{fig3}(a)), the maximum values of Berry curvature $\Omega_{\uparrow}^{z}(\bm{k})$ locate at $K$ and $K^{\prime}$. Meanwhile, the zero points of  $\partial g_{\uparrow}/\partial k_{y}$ also locate at $K$ and $K^{\prime}$ (see \Fig{fig3}(c)), resulting in the cancellation of the integral around each $K$ and $K^\prime$ and zero $\kappa^{xyy}$. However, when applying the uniaxial strain along $y$ direction , the maximum values of Berry curvature $\Omega_{\uparrow}^{z}(\bm{k})$ are shifted from the original $K$  ($K^\prime$) towards $-k_{x}$  ($k_{x}$) direction (see \Fig{fig3}(b)). And the zero points of $\partial g_{\uparrow}/\partial k_{y}$ are also shifted from the original $K$ ($K^\prime$) towards $k_{x}$ ($-k_{x}$) direction (see \Fig{fig3}(d)). Therefore the integral around each $K$ and $K^\prime$ can not be cancelled, leading to a finite 2nd order magnon thermal Hall conductivity $\kappa^{xyy}$.

\begin{figure}[H]
\centering
\includegraphics [width=3.3in]{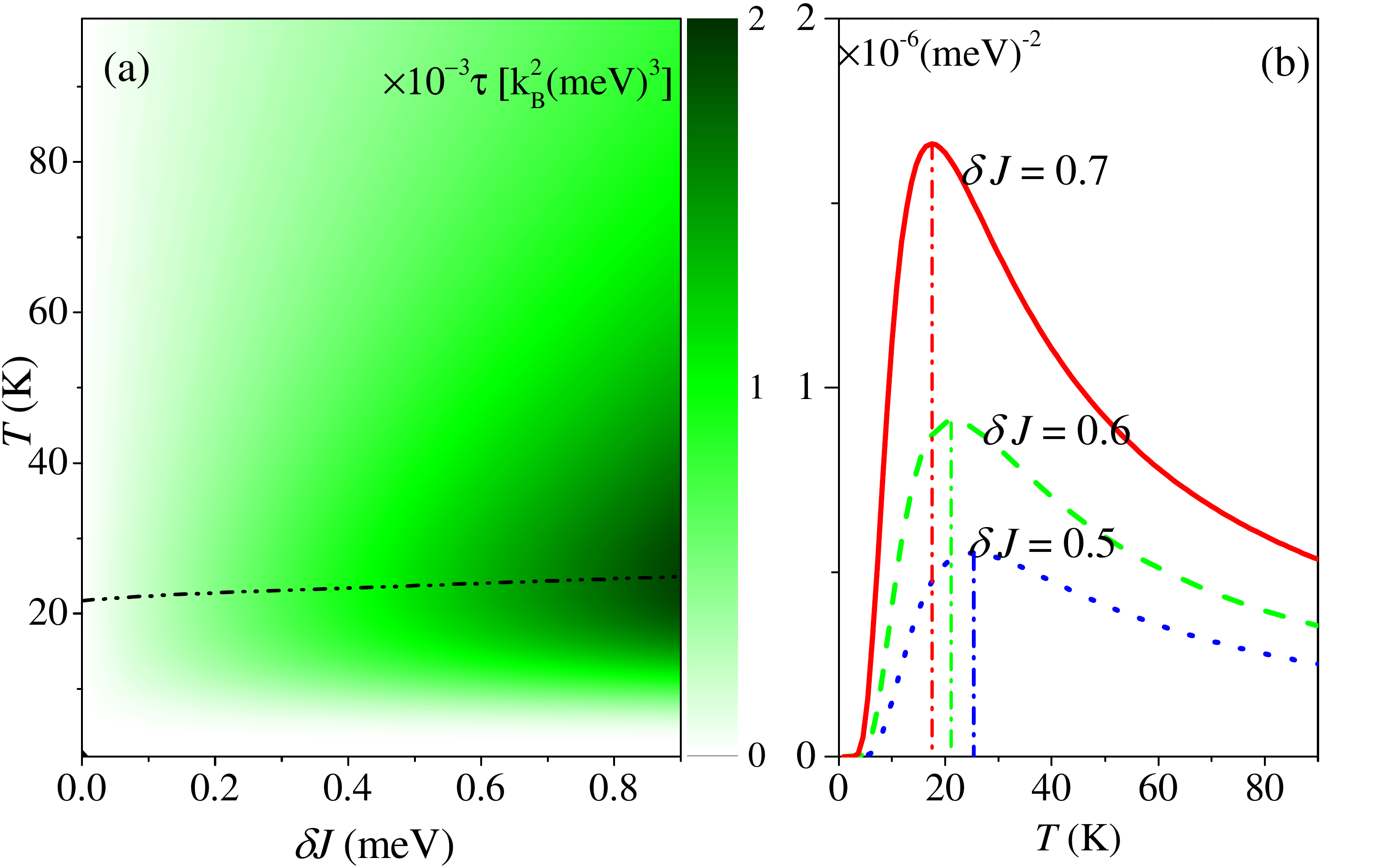}
\caption{(a). The  magnon thermal Hall conductivity up to the 2nd order (i.e. $\kappa^{xyy}$, and 1st order disappears) as a function of strain-induced coupling $\delta J$ and temperature of a collinear antiferromagnets. $J=2$, $K=-0.2$ and $g_{J}\mu_{B}B=0.01$. (b). The $T$-dependent factor $\mathcal{F}$ for different $\delta J$. $\varepsilon_{p}$ is taken to be -0.2. Numbers are in unit of meV. }\label{fig4}
\end{figure}

To further illustrate the above picture, we show the dependence  of $\kappa^{xyy}$ on the temperature and the strain-induced coupling $\delta J$, which is plotted in \Fig{fig4}(a). Notice that $\kappa^{xyy}$ approaches zero as $T$ approaches zero. For fixed $T$,  $\kappa^{xyy}$  increases monotonically with increasing $\delta J$, suggesting the appearance of the nonlinear MTHE induced by the strain. It should be noted that our analysis based on the linear spin wave theory is only valid in the temperature range much lower than the N\'{e}el temperature, which is estimated to be around $200$ K in MnPS$_{3}$ \cite{PhysRevB.91.235425}. However,  $\kappa^{xyy}$ is not monotonic in $T$. For fixed $\delta J$, $\kappa^{xyy}$ increases at first and then decreases with a maximum around $22$ K. To understand this nonmonotonicity, we extract the temperature dependence of \Eq{kappa3}.
The $T$-dependent factor of \Eq{kappa3} is expressed as $\mathcal{F}=\frac{1}{(k_{B}T)^{2}}\frac{\exp{(\beta \varepsilon_{p})}}{(\exp{(\beta \varepsilon_{p})}-1)^{2}}$. In \Fig{fig4}(b) we depict the $T$-dependent factor $\mathcal{F}$ as a function of $T$. For different $\delta J$, $\mathcal{F}$ is maximized around $22$ K, hence we conclude that the temperature-nonmonotonicity of $\kappa^{xyy}$ is determined by $\mathcal{F}$. As shown in \Fig{fig4}(b), the temperature position $T_{\rm{max}}$ of the maximum of $\mathcal{F}$ decreases from $25$ K to $18$ K as $\delta J$ increases from $0.5$ meV to $0.7$ meV. In \Fig{fig4}.(a) we indicate the position $T_{\rm{max}}$ of the maximum of $\kappa^{xyy}$ by the dash-dot line. As a contrast, $T_{\rm{max}}$ increases slightly with the increment $\delta J$. This is because the $T$-dependent $\mathcal{F}$ indicates that all momentum $\bm{k}$ is weighted equally for fixed $T$. However, according to \Eq{kappa3}, the final temperature dependence of $\kappa^{xyy}$ should be weighted by $\varepsilon^{3}_{p}\Omega_{p}^{z}$ additionally.

\section{Concluding Remarks and Discussions }
In summary,
a nonlinear thermal response theory is developed through perturbed expansion approach in favour of thermal vector potential. Based on the diagram rules and values of vertices connecting the propagator of temperature gradient, the general expressions of the dynamical thermoelectric and thermal conductivity are obtained. 
In the DC limit, the results for the linear order and the second order thermoelectric response explicitly reproduce the known formula obtained through wave packet theory or Boltzmann equation.

The choice of the gauge depends on convenience. It is easy to give a cleaner resonance structure and is easier to implement numerically in velocity gauge. For the DC limit and semiclassical limits, it is better to apply the length gauge. By providing the DC limit formula in length gauge, we demonstrate the relations among the thermal response coefficients beyond the linear order. For linear transport, the Mott relation and WF law tell us that the thermoelectric (thermal) conductivity is proportional to the first (zero) derivative of the linear electric conductivity to the Fermi energy.
Beyond the linear order, it is found that there exist higher order Mott relation and WF law.
The 2nd order Mott relation and WF law say that the 2nd order electric  conductivity is proportional to zero (the first) derivative of the thermoelectric (thermal) conductivity with respect to the chemical potential.
And the 3rd order Mott relation and WF law show that the 3rd order electric conductivity is proportional to the first (second) derivative of the  thermoelectric (thermal) conductivity with respect to the chemical potential.
It is found that the derivative on the thermoelectric and the thermal conductivity increases linearly with the nonlinear order. The derivative in the  WF law is one order higher than that of the Mott relation.  We call this structure as a "hierarchy rule".
Although we only explicitly calculate the nonlinear response up to the third order, we speculate that this ``hierarchy rule" between Mott relation and the WF law exists to higher order, revealing a deeper relationship between them. 
%
Moreover, it is discovered that the Lorentz number characterizing the relation of linear thermoelectric and thermal-thermal response applies to the nonlinear order.

An interesting and important fact is that for the second order response, the Mott relation is only proportional to the second order electric conductivity by the linear Lorentz number. Since the off-diagonal element of the second electric conductivity is just the nonlinear Hall conductivity which has been measured in experiments, the off-diagonal element of the second thermoelectric conductivity (i.e. the second order Nernst coefficient) can be obtained immediately by using the experimental data of the nonlinear Hall conductivity. We estimate that the transverse charge current density can be the order of $10^{-6}$ A/$(cm)^{2}$ for a temperature gradient of $0.01$ K/cm based on few layers WTe$_{2}$ \cite{kang2019nonlinear}. This charge current density induced by a temperature gradient can be explored in experiments.
 For the second order WF law, the electric conductivity is proportional to the first derivative of the second order thermal conductivity with respect to the chemical potential. The proportional factor is related to the Lorentz number, and the second thermal conductivity can be sizeable. Therefore, the quantities from the second order response can be measured in experiments without introducing more difficulties. We expect that our predictions can be tested in the near future experiments.

Although the derived quantum theory of nonlinear thermal response is based on a formalism for fermions, it can be utilized to boson systems. As an application, we specifically calculate the magnon thermal Hall conductivity in a strained collinear antiferromagnet model. We predict that with the combined $\mathcal{T}c_{x}$ symmetry and broken inversion symmetry, the linear magnon thermal Hall conductivity vanishes and the second order thermal Hall effect dominates.

\section{acknowledgements}

This work is supported in part by  the NSFC (Grant Nos. 11974348, 11674317, and 11834014), and the National Key R\&D Program of China (Grant No. 2018FYA0305800). It is also supported by the Fundamental Research Funds for the Central Universities, and the Strategic Priority Research Program of CAS (Grant Nos. XDB28000000, and and XDB33000000).

\clearpage
\appendix

\begin{widetext}

\section{Details of derivation for \Eq{hct}, definition of heat current, and the relation to entropy flux}\label{app11}
The definition of heat current in the presence of the  "gravitational" potential, has been presented previously \cite{PhysRevLett.99.197202}. However, since it is important to the rest of our discussion, we shall review this here.
Considering a non-interacting electron system, the energy density is written as
\begin{equation}
\hat{h}^{\Psi}(\bm{r})=[1+\Psi(\bm{r})]\left\{\frac{m}{2}[\hat{\bm{v}}\hat{\varphi} (\bm{r})]^{\dagger}\cdot [\hat{\varphi}(\bm{r})\hat{\bm{v}} ] + \hat{\varphi}^{\dagger}(\bm{r})[V(\bm{r})] \hat{\varphi}(\bm{r}) \right\}.
\end{equation}
where $\hat{\varphi}(\bm{r})(\hat{\varphi}^{\dagger}(\bm{r}))$ is the electron annihilation (creation) field operator.  The energy current operator is defined by the conservation equation
\begin{equation}\label{conserv}
\frac{\partial \hat{h}^{\Psi}(\bm{r})}{\partial t}=\frac{1}{i\hbar}[\hat{h}^{\Psi}(\bm{r}),\hat{H}^{\Psi}]=-\bm{\nabla}\cdot \bm{J}^{\Psi}_{E}(\bm{r}),
\end{equation}
where $\hat{H}^{\Psi}=\frac{m}{2}\hat{\bm{v}}[1+\Psi(\bm{r})]\hat{\bm{v}}+[1+\Psi(\bm{r})]V(\bm{r})$.
Substituting the energy density operator into the conservation equation, it yields
\begin{equation}
\frac{\partial \hat{h}^{\Psi}(\bm{r})}{\partial t}=-\bm{\nabla}\cdot \left\{\frac{1}{2}[1+{\Psi}(\bm{r})]\left([\hat{\bm{v}}\hat{\varphi} (\bm{r})]^{\dagger}[\hat{H}^{\Psi}\hat{\varphi} (\bm{r})] +[\hat{H}^{\Psi}\hat{\varphi} (\bm{r})]^{\dagger} [\hat{\bm{v}}\hat{\varphi} (\bm{r})]\right) \right\}.
\end{equation}
Therefore the energy current operator is identified as
\begin{equation}
\begin{aligned}
\bm{J}_{E}^{\Psi}(\bm{r})=&\left\{\frac{1}{2}[1+{\Psi}(\bm{r})]\left([\hat{\bm{v}}\hat{\varphi} (\bm{r})]^{\dagger}[\hat{H}^{\Psi}\hat{\varphi} (\bm{r})] +[\hat{H}^{\Psi}\hat{\varphi} (\bm{r})]^{\dagger} [\hat{\bm{v}}\hat{\varphi} (\bm{r})]\right) \right\}\\
=&\frac{1}{2}[1+{\Psi}(\bm{r})]^{2}\left\{\left([\hat{\bm{v}}\hat{\varphi} (\bm{r})]^{\dagger}[\hat{H}_{0}\hat{\varphi} (\bm{r})] +[\hat{H}_{0}\hat{\varphi} (\bm{r})]^{\dagger} [\hat{\bm{v}}\hat{\varphi} (\bm{r})]\right) \right\}+\bm{\nabla}[1+{\Psi}(\bm{r})]^{2}\times \hat{\bm{\Lambda}}   .
\end{aligned}
\end{equation}
Where $\hat{\bm{\Lambda}}=-\frac{i\hbar}{8}[\hat{\bm{v}}\hat{\varphi} (\bm{r})]^{\dagger}\times [\hat{\bm{v}}\hat{\varphi} (\bm{r})]$.
Noting that the current operator is only defined up to a curl by the equation of continuity. The form of the energy current can be determined the scaling law
\begin{equation}
\bm{J}_{E}^{\Psi}(\bm{r})=[1+\Psi(\bm{r})]^{2}\bm{J}_{E}(\bm{r}),
\end{equation}
therefore the the energy current operator becomes
\begin{equation}
\bm{J}_{E}^{\Psi}(\bm{r})\rightarrow \bm{J}_{E}^{\Psi}(\bm{r})-\bm{\nabla}[1+\Psi(\bm{r})]^{2}\times \hat{\bm{\Lambda}},
\end{equation}
\begin{equation}
\bm{J}_{E}(\bm{r})\rightarrow \bm{J}_{E}(\bm{r})+\bm{\nabla}\times \hat{\bm{\Lambda}}.
\end{equation}
The heat current is defined as $\bm{J}_{Q}(\bm{r})\equiv \bm{J}_{E}(\bm{r})-\mu \bm{J}_{N}(\bm{r})$.
In the absence of temperature gradient field, the zero-field heat current operator is given by
\begin{equation}
\bm{J}_{Q}(\bm{r})=\frac{1}{2}\left([\hat{\bm{v}}\hat{\varphi} (\bm{r})]^{\dagger}[\hat{K}_{0}\hat{\varphi} (\bm{r})] +[\hat{K}_{0}\hat{\varphi} (\bm{r})]^{\dagger} [\hat{\bm{v}}\hat{\varphi} (\bm{r})]\right) -\bm{\nabla}\times \hat{\bm{\Lambda}},
\end{equation}
where $\hat{K}_{0}=\hat{H}_{0}-\mu_{0}\hat{N}$. Noting that apart from the first term which is recognized as the usual anticommutator representation of the heat current, the second term appears is essential for satisfying the scaling law. It has been proved that in calculating the Kubo formula, the second term cancels out  \cite{PhysRevB.55.2344, PhysRevLett.99.197202}, this could be the reason why the anticommutator representation usually leads to the right results.

Alternatively, the heat current can be defined through the thermodynamics of entropy flux \cite{kadanoff2017entropy}, and it is equivalent to the definition via conservation equation. To see this we start form the Luttinger's Hamiltonian. 
The particle number conservation equation is given as
\begin{equation}\label{conservp}
\frac{\partial \hat{n}^{\Psi}(\bm{r})}{\partial t}=\frac{1}{i\hbar}[\hat{n}^{\Psi}(\bm{r}),\hat{H}^{\Psi}]=-\bm{\nabla}\cdot \bm{J}^{\Psi}_{N}(\bm{r}),
\end{equation}
Combining \Eq{conserv} and \Eq{conservp}, the conservation equation of heat is written as
\begin{equation}
\frac{\partial \hat{k}^{\Psi}(\bm{r})}{\partial t}=\frac{1}{i\hbar}[\hat{k}^{\Psi}(\bm{r}),\hat{H}^{\Psi}]=-\bm{\nabla}\cdot \bm{J}^{\Psi}_{Q}(\bm{r}),
\end{equation}
in which $\hat{k}^{\Psi}(\bm{r})= \hat{h}^{\Psi}(\bm{r})-\mu\hat{n}^{\Psi}(\bm{r})$ is the grand-canonical ensemble energy density.
 the Luttinger's Hamiltonian can be rewritten as
\begin{equation}
H_{L}(t)=\int d^{3}r\int_{-\infty}^{t}dt^\prime \bm{J}_{Q}(t^\prime)\cdot \bm{\nabla}\Psi(\bm{r},t),
\end{equation}
by converting the "gravitational" potential in form of thermal vector potential, $\partial\bm{A}_{T}(\bm{r},t)/\partial t = \bm{\nabla}\Psi (\bm{r},t)=\bm{\nabla}T/T$, the perturbation Hamiltonian is written as
\begin{equation}
H_{L}(t)=-\int d^{3}r \bm{J}_{Q}(t^\prime)\cdot \bm{A}_{T}(\bm{r},t).
\end{equation}
The rate of the change of the entropy $S$ due to 
a heat current is \cite{landau2013electrodynamics}
\begin{equation}
\frac{d S}{dt}=-\int d^{3}r\frac{1}{T}\bm{\nabla}\cdot \bm{J}_{Q}=-\int d^{3}r\frac{\nabla T}{T^{2}}\cdot \bm{J}_{Q}.
\end{equation}
And the change of entropy modifies the thermodynamic potential $E-TS-\mu N$ ($E$ is the internal energy). The perturbation Hamiltonian induced by the temperature gradient field becomes
\begin{equation}
H_{S}=\frac{1}{T}\int d^{3}r\int_{-\infty}^{t}dt^\prime\bm{J}_{Q}(t^\prime) \cdot \bm{\nabla}T.
\end{equation}
It recovers the Luttinger's Hamiltonian after the replacement $\bm{\nabla}\Psi (\bm{r},t)=\bm{\nabla}T/T$. Similar definition of the heat current can be found in \cite{sergeev2021entropy}.

\section{Expansion of the Hermitian derivatives}\label{app1}
The second order Hermitian derivative on the unperturbed Hamiltonian is expanded as
\begin{equation}
\begin{aligned}
\hat{\mathcal{K}}^{\alpha\beta}=&\hat{\mathcal{D}}^{\alpha}\hat{\mathcal{D}}^{\beta}[\hat{K}_{0}]\\
=&\frac{1}{4}(\hat{K}_{0}\hat{h}^{\alpha}\hat{h}^{\beta}+\hat{K}_{0}^{2}\hat{h}^{\alpha\beta}+2\hat{K}_{0}\hat{h}^{\alpha\beta}\hat{K}_{0}+\hat{K}_{0}\hat{h}^{\beta}\hat{h}^{\alpha}+\hat{h}^{\alpha}\hat{h}^{\beta}\hat{K}_{0}+\hat{h}^{\alpha\beta}\hat{K}_{0}^{2}+\hat{h}^{\beta}\hat{h}^{\alpha}\hat{K}_{0}),
\end{aligned}
\end{equation}
Its normal derivative is given by
\begin{equation}
\begin{aligned}
\hat{D}^{\mu}[\hat{\mathcal{K}}^{\alpha\beta}]=&\frac{1}{4}(\hat{h}^{\mu}\hat{h}^{\alpha}\hat{h}^{\beta}+\hat{K}_{0}\hat{h}^{\mu\alpha}\hat{h}^{\beta}+\hat{K}_{0}\hat{h}^{\alpha}\hat{h}^{\mu\beta}+\hat{h}^{\mu}\hat{K}_{0}\hat{h}^{\alpha\beta}+\hat{K}_{0}\hat{h}^{\mu}\hat{h}^{\alpha\beta}
+\hat{K}_{0}^{2}\hat{h}^{\mu\alpha\beta}+2\hat{h}^{\mu}\hat{h}^{\alpha\beta}\hat{K}_{0}+2\hat{K}_{0}\hat{h}^{\mu\alpha\beta}\hat{K}_{0}\\
&+2\hat{K}_{0}\hat{h}^{\alpha\beta}\hat{h}^{\mu}+\hat{h}^{\mu}\hat{h}^{\beta}\hat{h}^{\alpha}+\hat{K}_{0}\hat{h}^{\mu\beta}\hat{h}^{\alpha}+\hat{K}_{0}\hat{h}^{\beta}\hat{h}^{\mu\alpha}+\hat{h}^{\mu\alpha}\hat{h}^{\beta}\hat{K}_{0}+\hat{h}^{\alpha}\hat{h}^{\mu\beta}\hat{K}_{0}
+\hat{h}^{\alpha}\hat{h}^{\beta}\hat{h}^{\mu}+\hat{h}^{\mu\alpha\beta}\hat{K}_{0}^{2}\\
&+\hat{h}^{\alpha\beta}\hat{h}^{\mu}\hat{K}_{0}+\hat{h}^{\alpha\beta}\hat{K}_{0}\hat{h}^{\mu}+\hat{h}^{\mu\beta}\hat{h}^{\alpha}\hat{K}_{0}+\hat{h}^{\beta}\hat{h}^{\mu\alpha}\hat{K}_{0}+\hat{h}^{\beta}\hat{h}^{\alpha}\hat{h}^{\mu}).
\end{aligned}
\end{equation}

\section{Derivation of \Eq{m12f} and \Eq{m22f}}\label{appx}
Using the relation \Eq{pb1}, \Eq{pb2}, and the following identities
\begin{equation}
\int_{\beta}^{\infty}f(\varepsilon)d\lambda = \frac{\beta}{\varepsilon}\ln(1+ e^{-\beta \varepsilon})/\beta,
\end{equation}
\begin{equation}
\int_{\beta}^{\infty}\frac{\partial f(\varepsilon)}{\partial\varepsilon}d\lambda =-\frac{\beta}{\varepsilon}f(\varepsilon)+\frac{\beta}{\varepsilon^{2}}\int^{\infty}_{\varepsilon}f(\lambda)d\lambda,
\end{equation}
the first order particle magnetization response \Eq{m12f} and heat magnetization response \Eq{m22f} are obtained by integrating the auxiliary particle magnetization with respect to $\beta$.

\section{Expansion of the integral kernels used in length gauge.}\label{app4}
For the Kubo contribution of the linear thermoelectric response, the integrand is calculated as
\begin{equation}
\begin{aligned}
&{\rm{Tr}}\left\{v^{\alpha}\left(d(\omega) \circ\mathcal{D}^{\beta}\left[\rho^{(0)}\right]\right)\right\}\\
=&\sum_{p,q}\frac{1}{2}v_{pq}^\alpha d_{qp}(\omega)\left[H_{0},D^{\beta}\left[\rho^{(0)}\right]\right]_{+,qp}\\
=&\sum_{p,q}\frac{1}{2}v_{pq}^\alpha d_{qp}(\omega)\left\{\left[H_{0},\partial^{\beta}\rho^{(0)}\right]_{+,qp}-i\frac{1}{2}\left[H_{0},\left[\mathcal{A}^{\beta},\rho^{(0)}\right]_{-}\right]_{+,qp}\right\}\\
=&\sum_{p} \frac{1}{\omega}v_{p}^\alpha\varepsilon_{p}\partial^{\beta}f_{p}-\sum_{p,q}i\frac{1}{2(\omega-\varepsilon_{qp})}(\varepsilon_{p}+\varepsilon_{q})v_{pq}^{\alpha}\mathcal{A}_{qp}^{\beta}f_{pq}.\\
\end{aligned}
\end{equation}
The integrand in the $2$nd order thermoelectric response is calculated as
\begin{equation}
\begin{aligned}
&{\rm{Tr}}\left\{v^{\alpha}\left(d(\omega)\circ \mathcal{D}^{\beta}\left[d(\omega-\omega_{1}) \circ\mathcal{D}^{\gamma}[\rho^{(0)}]\right]\right)\right\} \\
=&\sum_{p,q}\frac{1}{4} v_{pq}^\alpha d_{qp}(\omega)\left[\vphantom{[H_{0}D^{\alpha_{2}} + D^{\alpha_{2}} H_{0},\rho^{(0)}]}H_{0},D^{\beta}\left[d(\omega-\omega_{1}) \circ\left[H_{0},D^{\gamma}[\rho^{(0)}]\right]_{+}\right]\right]_{+,qp}\\
=&\Pi^{(2),\beta\gamma}+\Pi^{(2),\beta}+\Pi^{(2),\gamma}+\Pi^{(2)},
\end{aligned}
\end{equation}
where
\begin{equation}
\begin{aligned}
\Pi^{(2),\beta\gamma}=& \sum_{p,q}\frac{1}{4} v_{pq}^\alpha d_{qp}(\omega)\left[\vphantom{[H_{0}D^{\alpha_{2}} + D^{\alpha_{2}} H_{0},\rho^{(0)}]}H_{0},\partial^{\beta}\left[d(\omega-\omega_{1}) \circ\left[H_{0},\partial^{\gamma}[\rho^{(0)}]\right]_{+}\right]\right]_{+,qp}\\
=&\sum_{p}v_{p}^\alpha \frac{1}{\omega}\varepsilon_{p}\frac{1}{\omega-\omega_{1}}\partial^{\beta}(\varepsilon_{p}\partial^{\gamma}f_{p}),
\end{aligned}
\end{equation}
\begin{equation}
\begin{aligned}
\Pi^{(2),\beta}=
& \sum_{p,q}\frac{1}{4} v_{pq}^\alpha d_{qp}(\omega)\left[\vphantom{[H_{0}D^{\alpha_{2}} + D^{\alpha_{2}} H_{0},\rho^{(0)}]}H_{0},\partial^{\beta}\left[d(\omega-\omega_{1}) \circ\left[H_{0},\left[\mathcal{A}^{\gamma},\rho^{(0)}\right]_{-}\right]_{+}\right]\right]_{+,qp}\\
=&\sum_{p,q}-i\frac{1}{4} v_{pq}^{\alpha}\frac{1}{\omega-\varepsilon_{qp}}(\varepsilon_{p}+\varepsilon_{q})\partial^{\beta}\left\{\frac{1}{\omega-\omega_{1}-\varepsilon_{qp}} \times (\varepsilon_{p}+\varepsilon_{q})\mathcal{A}^{\gamma}_{qp}f_{pq}\vphantom{\frac{1}{2}}\right\},
\end{aligned}
\end{equation}
\begin{equation}
\begin{aligned}
\Pi^{(2),\gamma}=
& \sum_{p,q}\frac{1}{4} v_{pq}^\alpha d_{qp}(\omega)\left[\vphantom{[H_{0}D^{\alpha_{2}} + D^{\alpha_{2}} H_{0},\rho^{(0)}]}H_{0},\left[\mathcal{A}^{\beta},\left[d(\omega-\omega_{1}) \circ\left[H_{0},\partial^{\gamma}[\rho^{(0)}]\right]_{+}\right]\right]_{-}\right]_{+,qp}\\
=&\sum_{p,q}-i\frac{1}{2}v_{pq}^{\alpha}\frac{1}{(\omega-\varepsilon_{qp})}\frac{1}{(\omega-\omega_{1})}\mathcal{A}^{\beta}_{qp}(\varepsilon_{p}\varepsilon_{q}\partial^{\gamma} f_{pq}+\varepsilon_{p}^{2}\partial^{\gamma}f_{p}-\varepsilon_{q}^{2}\partial^{\gamma}f_{q}),
\end{aligned}
\end{equation}
\begin{equation}
\begin{aligned}
\Pi^{(2)}=
& \sum_{p,q}\frac{1}{4} v_{pq}^\alpha d_{qp}(\omega)\left[\vphantom{[H_{0}D^{\alpha_{2}} + D^{\alpha_{2}} H_{0},\rho^{(0)}]}H_{0},\left[\mathcal{A}^{\beta},\left[d(\omega-\omega_{1}) \circ\left[H_{0},\left[\mathcal{A}^{\gamma},\rho^{(0)}\right]_{-}\right]_{+}\right]\right]_{-}\right]_{+,qp}\\
=&\sum_{p,q,r}-\frac{1}{4} v_{pq}^{\alpha}\frac{1}{\omega-\varepsilon_{qp}}(\varepsilon_{q}+\varepsilon_{p})\left[\frac{1}{\omega-\omega_{1}-\varepsilon_{rp}}\mathcal{A}^{\beta}_{qr}\mathcal{A}^{\gamma}_{rp}\varepsilon_{p}(\varepsilon_{r}+\varepsilon_{p})-\frac{1}{\omega-\omega_{1}-\varepsilon_{qr}}\mathcal{A}^{\gamma}_{qr}\mathcal{A}^{\beta}_{rp}\varepsilon_{p}(\varepsilon_{r}+\varepsilon_{q})\right].
\end{aligned}
\end{equation}

The integrand in the $2$nd order thermal-thermal response is calculated as
\begin{equation}
\begin{aligned}
&{\rm{Tr}}\left\{\frac{1}{2}\left[H_{0},v^{\alpha}\right]_{+}\times \left(d(\omega)\circ \mathcal{D}^{\beta}\left[d(\omega-\omega_{1}) \circ\mathcal{D}^{\gamma}[\rho^{(0)}]\right]\right)\right\} \\
=&\sum_{p,q}\frac{1}{4} v_{pq}^\alpha d_{qp}(\omega)\left[\vphantom{[H_{0}D^{\alpha_{2}} + D^{\alpha_{2}} H_{0},\rho^{(0)}]}H_{0},D^{\beta}\left[d(\omega-\omega_{1}) \circ\left[H_{0},D^{\gamma}[\rho^{(0)}]\right]_{+}\right]\right]_{+,qp}\\
=&\Xi^{(2),\beta\gamma}+\Xi^{(2),\beta}+\Xi^{(2),\gamma}+\Xi^{(2)},
\end{aligned}
\end{equation}
where
\begin{equation}
\begin{aligned}
\Xi^{(2),\beta\gamma}=& \sum_{p,q}\frac{1}{8}\left[H_{0},v^{\alpha}\right]_{+,pq} d_{qp}(\omega)\left[\vphantom{[H_{0}D^{\alpha_{2}} + D^{\alpha_{2}} H_{0},\rho^{(0)}]}H_{0},\partial^{\beta}\left[d(\omega-\omega_{1}) \circ\left[H_{0},\partial^{\gamma}[\rho^{(0)}]\right]_{+}\right]\right]_{+,qp}\\
=&\sum_{p}v_{p}^\alpha \frac{1}{\omega}\tilde{\varepsilon}_{p}^{2}\partial^{\beta}\left[\frac{1}{\omega-\omega_{1}}\tilde{\varepsilon}_{q}\partial^{\gamma}f_{p}\right],
\end{aligned}
\end{equation}
\begin{equation}
\begin{aligned}
\Xi^{(2),\beta}=& \sum_{p,q}\frac{1}{8}\left[H_{0},v^{\alpha}\right]_{+,pq} d_{qp}(\omega)\left[\vphantom{[H_{0}D^{\alpha_{2}} + D^{\alpha_{2}} H_{0},\rho^{(0)}]}H_{0},\partial^{\beta}\left[d(\omega-\omega_{1}) \circ\left[H_{0},\left[\mathcal{A}^{\gamma},\rho^{(0)}\right]_{-}\right]_{+}\right]\right]_{+,qp}\\
=&\sum_{p,q}-i\frac{1}{8}v_{pq}^{\alpha}(\tilde{\varepsilon}_{p}+\tilde{\varepsilon}_{q})^{2}\frac{1}{\omega-\varepsilon_{pq}}\partial^{\beta}\left[\frac{1}{\omega-\omega_{1}-\varepsilon_{qp}} (\tilde{\varepsilon}_{p}+\tilde{\varepsilon}_{q})\mathcal{A}^{\gamma}_{qp}f_{pq}\vphantom{\frac{1}{\omega-\omega_{1}-\varepsilon_{qp}}}\right],
\end{aligned}
\end{equation}
\begin{equation}
\begin{aligned}
\Xi^{(2),\gamma}=& \sum_{p,q}\frac{1}{8}\left[H_{0},v^{\alpha}\right]_{+,pq} d_{qp}(\omega)\left[\vphantom{[H_{0}D^{\alpha_{2}} + D^{\alpha_{2}} H_{0},\rho^{(0)}]}H_{0},\left[\mathcal{A}^{\beta},\left[d(\omega-\omega_{1}) \circ\left[H_{0},\partial^{\gamma}[\rho^{(0)}]\right]_{+}\right]\right]_{-}\right]_{+,qp}\\
=&\sum_{p,q}-i\frac{1}{8}(\tilde{\varepsilon}_{p}+\tilde{\varepsilon}_{q})^{2}v_{pq}^{\alpha}\frac{1}{(\omega-\varepsilon_{qp})}\frac{1}{(\omega-\omega_{1})}\mathcal{A}^{\beta}_{qp}(\tilde{\varepsilon}_{p}\tilde{\varepsilon}_{q}\partial^{\gamma} f_{pq}+\tilde{\varepsilon}_{p}^{2}\partial^{\gamma}f_{p}-\tilde{\varepsilon}_{q}^{2}\partial^{\gamma}f_{q}),
\end{aligned}
\end{equation}
\begin{equation}
\begin{aligned}
\Xi^{(2)}=& \sum_{p,q}\frac{1}{8}\left[H_{0},v^{\alpha}\right]_{+,pq}d_{qp}(\omega)\left[\vphantom{[H_{0}D^{\alpha_{2}} + D^{\alpha_{2}} H_{0},\rho^{(0)}]}H_{0},\left[\mathcal{A}^{\beta},\left[d(\omega-\omega_{1}) \circ\left[H_{0},\left[\mathcal{A}^{\gamma},\rho^{(0)}\right]_{-}\right]_{+}\right]\right]_{-}\right]_{+,qp}\\
=&\sum_{p,q,r}-\frac{1}{8} v_{pq}^{\alpha}\frac{1}{\omega-\varepsilon_{qp}}(\varepsilon_{q}+\varepsilon_{p})^{2}\left[\frac{1}{\omega-\omega_{1}-\varepsilon_{rp}}\mathcal{A}^{\beta}_{qr}\mathcal{A}^{\gamma}_{rp}\varepsilon_{p}(\varepsilon_{r}+\varepsilon_{p})-\frac{1}{\omega-\omega_{1}-\varepsilon_{qr}}\mathcal{A}^{\gamma}_{qr}\mathcal{A}^{\beta}_{rp}\varepsilon_{p}(\varepsilon_{r}+\varepsilon_{q})\right].
\end{aligned}
\end{equation}

\section{Expansion of the integral kernels at the $3$rd order }\label{app6}
The integrand in the $3$rd order thermoelectric response is calculated as
\begin{equation}
\begin{aligned}
&{\rm{Tr}}\left\{v^{\alpha}\left(d(\omega)\circ \mathcal{D}^{\beta}\left[d(\omega-\omega_{1}) \circ\mathcal{D}^{\gamma}\left[d(\omega-\omega_{[2]})\circ \mathcal{D}^{\delta}[\rho^{(0)}]\right]\right]\right)\right\} \\
=&\sum_{p,q}\frac{1}{8} v_{pq}^\alpha d_{qp}(\omega)\left[H_{0},D^{\beta}\left[d(\omega-\omega_{1}) \circ\left[H_{0},D^{\gamma}\left[d(\omega-\omega_{[2]})\circ \left[H_{0},D^{\delta}[\rho^{(0)}]\right]_{+}\right]\right]_{+}\right]\right]_{+,qp}\\
=&\Pi^{(3),\beta\gamma\delta}+\Pi^{(3),\beta\gamma}+\Pi^{(3),\beta\delta}+\Pi^{(3),\gamma\delta}+\Pi^{(3),\beta}+\Pi^{(3),\gamma}+\Pi^{(3),\delta}+\Pi^{(3)},
\end{aligned}
\end{equation}
where
\begin{equation}
\begin{aligned}
\Pi^{(3),\beta\gamma\delta}=& \sum_{p,q}\frac{1}{8} v_{pq}^\alpha d_{qp}(\omega)\left[\vphantom{[H_{0}D^{\alpha_{2}} + D^{\alpha_{2}} H_{0},\rho^{(0)}]}H_{0},\partial^{\beta}\left[d(\omega-\omega_{1}) \circ\left[H_{0},\partial^{\gamma}\left[d(\omega-\omega_{[2]})\circ \left[H_{0},[\partial^{\delta} [\rho^{(0)}]\right]_{+}\right]\right]_{+}\right]\right]_{+,qp}\\
=&\sum_{p}v_{p}^\alpha \frac{1}{\omega}\frac{1}{\omega-\omega_{1}}\frac{1}{\omega-\omega_{[2]}}\varepsilon_{p}\partial^{\beta}\left[\varepsilon_{p}\partial^{\gamma}(\varepsilon_{p}\partial^{\delta}f_{p})\right],
\end{aligned}
\end{equation}
\begin{equation}
\begin{aligned}
\Pi^{(3),\beta\gamma}=
& \sum_{p,q}\frac{1}{8} v_{pq}^\alpha d_{qp}(\omega)\left[H_{0},\partial^{\beta}\left[d(\omega-\omega_{1}) \circ\left[H_{0},\partial^{\gamma}\left[d(\omega-\omega_{[2]})\circ \left[H_{0}, \left[\mathcal{A}^{\delta},\rho^{(0)}\right]_{-}\right]_{+}\right]\right]_{+}\right]\right]_{+,qp}\\
=&\sum_{p,q,r}-i\frac{1}{8} v_{pq}^{\alpha}\frac{1}{\omega-\varepsilon_{qp}}(\varepsilon_{p}+\varepsilon_{q})\partial^{\beta}\left[\frac{1}{\omega-\omega_{1}-\varepsilon_{qp}}(\varepsilon_{p}+\varepsilon_{q}) \left[\partial^{\gamma}\frac{1}{\omega-\omega_{[2]}-\varepsilon_{pq}} (\varepsilon_{p}+\varepsilon_{q})\mathcal{A}^{\delta}_{qp}f_{pq}\vphantom{\frac{1}{2}}\right]\right],
\end{aligned}
\end{equation}
\begin{equation}
\begin{aligned}
\Pi^{(3),\beta\delta}=
& \sum_{p,q}\frac{1}{8} v_{pq}^\alpha d_{qp}(\omega)\left[H_{0},\partial^{\beta}\left[d(\omega-\omega_{1})\circ\left[H_{0},\left[\mathcal{A}^{\gamma},\left[d(\omega-\omega_{[2]}) \circ\left[H_{0},\partial^{\delta}[\rho^{(0)}]\right]_{+}\right]\right]_{-}\right]_{+}\right]\right]_{+,qp}\\
=&\sum_{p,q}-i\frac{1}{8}v_{pq}^{\alpha}\frac{1}{\omega-\varepsilon_{qp}}(\varepsilon_{p}+\varepsilon_{q})\frac{1}{\omega-\omega_{[2]}}\partial^{\beta}\left[\frac{1}{\omega-\omega_{1}-\varepsilon_{qp}}\mathcal{A}^{\gamma}_{qp}(\varepsilon_{p}\varepsilon_{q}\partial^{\delta} f_{pq}+\varepsilon_{p}^{2}\partial^{\delta}f_{p}-\varepsilon_{q}^{2}\partial^{\delta}f_{q})\right],
\end{aligned}
\end{equation}
\begin{equation}
\begin{aligned}
\Pi^{(3),\gamma\delta}=
& \sum_{p,q}\frac{1}{8} v_{pq}^\alpha d_{qp}(\omega)\left[H_{0},\left[\mathcal{A}^{\beta},\left[d(\omega-\omega_{1})\circ\left[H_{0},\left[\partial^{\gamma}\left[d(\omega-\omega_{[2]}) \circ\left[H_{0},\partial^{\delta}[\rho^{(0)}]\right]_{+}\right]\right]_{-}\right]_{+}\right]\right]_{-}\right]_{+,qp}\\
=&\sum_{p,q}-i\frac{1}{8}v_{pq}^{\alpha}\frac{1}{\omega-\varepsilon_{qp}}\frac{1}{\omega-\omega_{1}}\frac{1}{\omega-\omega_{[2]}}\mathcal{A}^{\beta}_{qp}\left[\varepsilon_{q}\varepsilon_{p}\partial^{\gamma}(\varepsilon_{p}\partial^{\delta} f_{p})-\varepsilon_{q}\varepsilon_{p}\partial^{\gamma}(\varepsilon_{q}\partial^{\delta} f_{q})+\varepsilon_{p}^{2}\partial^{\gamma}(\varepsilon_{p}\partial^{\delta} f_{p})-\varepsilon_{q}^{2}\partial^{\gamma}(\varepsilon_{q}\partial^{\beta} f_{q})\right],
\end{aligned}
\end{equation}
\begin{equation}
\begin{aligned}
\Pi^{(3),\delta}=
& \sum_{p,q}\frac{1}{8} v_{pq}^\alpha d_{qp}(\omega)\left[H_{0},\left[\mathcal{A}^{\beta},\left[d(\omega-\omega_{1})\circ\left[H_{0},\left[\mathcal{A}^{\gamma},\left[d(\omega-\omega_{[2]}) \circ\left[H_{0},\partial^{\delta}[\rho^{(0)}]\right]_{+}\right]\right]_{-}\right]_{+}\right]_{-}\right]_{+}\right]_{+,qp}\\
=&\sum_{p,q,r}-i\frac{1}{8}v_{pq}^{\alpha}\frac{1}{\omega-\varepsilon_{qp}}(\varepsilon_{p}+\varepsilon_{q})\left[\frac{1}{\omega-\omega_{1}-\varepsilon_{rq}}\frac{1}{\omega-\omega_{[2]}-\varepsilon_{rp}}\mathcal{A}^{\beta}_{qr}\mathcal{A}^{\gamma}_{rp}(\varepsilon_{p}\varepsilon_{r}\partial^{\delta} f_{pr}+\varepsilon_{p}^{2}\partial^{\delta}f_{p}-\varepsilon_{r}^{2}\partial^{\delta}f_{r})\right.\\
&\left.-\frac{1}{\omega-\omega_{1}-\varepsilon_{rp}}\frac{1}{\omega-\omega_{[2]}-\varepsilon_{rp}}\mathcal{A}^{\gamma}_{qr}\mathcal{A}^{\beta}_{rp}(\varepsilon_{q}\varepsilon_{r}\partial^{\delta} f_{rq}+\varepsilon_{r}^{2}\partial^{\delta}f_{r}-\varepsilon_{p}^{2}\partial^{\delta}f_{p})\right],
\end{aligned}
\end{equation}
\begin{equation}
\begin{aligned}
\Pi^{(3),\gamma}=
& \sum_{p,q}\frac{1}{8} v_{pq}^\alpha d_{qp}(\omega)\left[H_{0},\left[\mathcal{A}^{\beta},\left[d(\omega-\omega_{1})\circ\left[H_{0},\partial^{\gamma}\left[d(\omega-\omega_{[2]}) \circ\left[H_{0},[\mathcal{A}^{\delta},\rho^{(0)}]_{-}\right]_{+}\right]\right]_{+}\right]_{-}\right]_{+}\right]_{+,qp}\\
=&\sum_{p,q,r}-i\frac{1}{8}v_{pq}^{\alpha}\frac{1}{\omega-\varepsilon_{qp}}\frac{1}{\omega-\omega_{1}-\varepsilon_{rq}}(\varepsilon_{p}+\varepsilon_{q})\left[\mathcal{A}^{\beta}_{qr}(\varepsilon_{p}+\varepsilon_{r})\partial^{\gamma}\left[\frac{1}{\omega-\omega_{[2]}-\varepsilon_{rp}}\mathcal{A}_{rp}^{\delta}(\varepsilon_{r}+\varepsilon_{p})f_{pr}\right]\right.\\
&\left.+(\varepsilon_{r}+\varepsilon_{q})\partial^{\gamma}\left[\frac{1}{\omega-\omega_{[2]}-\varepsilon_{qr}}\mathcal{A}_{qr}^{\delta}(\varepsilon_{q}+\varepsilon_{r})f_{rq}\mathcal{A}^{\beta}_{rp}\right]\right],
\end{aligned}
\end{equation}
\begin{equation}
\begin{aligned}
\Pi^{(3),\beta}=
& \sum_{p,q}\frac{1}{8} v_{pq}^\alpha d_{qp}(\omega)\left[H_{0},\partial^{\beta}\left[d(\omega-\omega_{1})\circ\left[H_{0},\left[\mathcal{A}^{\gamma},\left[d(\omega-\omega_{[2]}) \circ\left[H_{0},[\mathcal{A}^{\delta},\rho^{(0)}]_{-}\right]_{+}\right]\right]_{-}\right]_{+}\right]\right]_{+,qp}\\
=&\sum_{p,q,r}-i\frac{1}{8}v_{pq}^{\alpha}\frac{1}{\omega-\varepsilon_{qp}}(\varepsilon_{p}+\varepsilon_{q})\partial^{\beta}\left[\frac{1}{\omega-\omega_{1}-\varepsilon_{rp}}\mathcal{A}^{\gamma}_{qr}\mathcal{A}^{\delta}_{rp}\varepsilon_{p}(\varepsilon_{r}+\varepsilon_{p})-\frac{1}{\omega-\omega_{1}-\varepsilon_{qr}}\mathcal{A}^{\delta}_{qr}\mathcal{A}^{\gamma}_{rp}\varepsilon_{p}(\varepsilon_{r}+\varepsilon_{q})\right],
\end{aligned}
\end{equation}
\begin{equation}
\begin{aligned}
\Pi^{(3)}=
& \sum_{p,q}\frac{1}{8} v_{pq}^\alpha d_{qp}(\omega)\left[H_{0},\left[\mathcal{A}^{\beta},\left[d(\omega-\omega_{1})\circ\left[H_{0},\left[\mathcal{A}^{\gamma},\left[d(\omega-\omega_{[2]}) \circ\left[H_{0},\left[\mathcal{A}^{\delta}, [\rho^{(0)}]\right]_{-}\right]_{+}\right]\right]_{-}\right]_{+}\right]_{-}\right]_{+}\right]_{+,qp}\\
=&\sum_{p,q,r,s}-\frac{1}{8} v_{pq}^{\alpha}\frac{1}{\omega-\varepsilon_{qp}}(\varepsilon_{q}+\varepsilon_{p})\left[\left[\frac{1}{\omega-\omega_{1}-\varepsilon_{sp}}\mathcal{A}^{\gamma}_{rs}\mathcal{A}^{\delta}_{sp}\varepsilon_{p}(\varepsilon_{s}+\varepsilon_{p})-\frac{1}{\omega-\omega_{1}-\varepsilon_{rs}}\mathcal{A}^{\delta}_{rs}\mathcal{A}^{\gamma}_{sp}\varepsilon_{p}(\varepsilon_{s}+\varepsilon_{r})^{\beta}\right]\mathcal{A}_{sp}\right.\\
&\left. +\mathcal{A}_{qr}^{\beta}\left[\frac{1}{\omega-\omega_{1}-\varepsilon_{rs}}\mathcal{A}^{\gamma}_{qr}\mathcal{A}^{\delta}_{rs}\varepsilon_{s}(\varepsilon_{r}+\varepsilon_{s})-\frac{1}{\omega-\omega_{1}-\varepsilon_{qr}}\mathcal{A}^{\delta}_{qr}\mathcal{A}^{\gamma}_{rs}\varepsilon_{s}(\varepsilon_{r}+\varepsilon_{q})\right]\right].
\end{aligned}
\end{equation}
The integrand in the $3$nd order thermal-thermal response is calculated as
\begin{equation}
\begin{aligned}
&{\rm{Tr}}\left\{\frac{1}{2}[H_{0}, v^{\alpha}]_{+}\times\left(d(\omega)\circ \mathcal{D}^{\beta}\left[d(\omega-\omega_{1}) \circ\mathcal{D}^{\gamma}\left[d(\omega-\omega_{[2]})\circ \mathcal{D}^{\delta}[\rho^{(0)}]\right]\right]\right)\right\} \\
=&\sum_{p,q}\frac{1}{16} [H_{0,}v^\alpha]_{+,pq} d_{qp}(\omega)\left[H_{0},D^{\beta}\left[d(\omega-\omega_{1}) \circ\left[H_{0},D^{\gamma}\left[d(\omega-\omega_{[2]})\circ \left[H_{0},D^{\delta}[\rho^{(0)}]\right]_{+}\right]\right]_{+}\right]\right]_{+,qp}\\
=&\Xi^{(3),\beta\gamma\delta}+\Xi^{(3),\beta\gamma}+\Xi^{(3),\beta\delta}+\Xi^{(3),\gamma\delta}+\Xi^{(3),\beta}+\Xi^{(3),\gamma}+\Xi^{(3),\delta}+\Xi^{(3)}
\end{aligned}
\end{equation}
where
\begin{equation}
\begin{aligned}
\Xi^{(3),\beta\gamma\delta}=& \sum_{p,q}\frac{1}{16} [H_{0},v^{\alpha}]_{+,pq} d_{qp}(\omega)\left[\vphantom{[H_{0}D^{\alpha_{2}} + D^{\alpha_{2}} H_{0},\rho^{(0)}]}H_{0},\partial^{\beta}\left[d(\omega-\omega_{1}) \circ\left[H_{0},\partial^{\gamma}\left[d(\omega-\omega_{[2]})\circ \left[H_{0},[\partial^{\delta} [\rho^{(0)}]\right]_{+}\right]\right]_{+}\right]\right]_{+,qp}\\
=&\sum_{p}v_{p}^\alpha \frac{1}{\omega}\frac{1}{\omega-\omega_{1}}\frac{1}{\omega-\omega_{[2]}}\varepsilon_{p}^{2}\partial^{\beta}\left[\varepsilon_{p}\partial^{\gamma}(\varepsilon_{p}\partial^{\delta}f_{p})\right],
\end{aligned}
\end{equation}
\begin{equation}
\begin{aligned}
\Xi^{(3),\beta\gamma}=
& \sum_{p,q}\frac{1}{16} [H_{0},v^{\alpha}]_{+,pq} d_{qp}(\omega)\left[H_{0},\partial^{\beta}\left[d(\omega-\omega_{1}) \circ\left[H_{0},\partial^{\gamma}\left[d(\omega-\omega_{[2]})\circ \left[H_{0}, \left[\mathcal{A}^{\delta},\rho^{(0)}\right]_{-}\right]_{+}\right]\right]_{+}\right]\right]_{+,qp}\\
=&\sum_{p,q,r}-i\frac{1}{8} v_{pq}^{\alpha}\frac{1}{\omega-\varepsilon_{qp}}(\varepsilon_{p}+\varepsilon_{q})^{2}\partial^{\beta}\left[\frac{1}{\omega-\omega_{1}-\varepsilon_{qp}}(\varepsilon_{p}+\varepsilon_{q}) \left[\partial^{\gamma}\frac{1}{\omega-\omega_{[2]}-\varepsilon_{pq}} (\varepsilon_{p}+\varepsilon_{q})\mathcal{A}^{\delta}_{qp}f_{pq}\vphantom{\frac{1}{2}}\right]\right],
\end{aligned}
\end{equation}
\begin{equation}
\begin{aligned}
\Xi^{(3),\beta\delta}=
& \sum_{p,q}\frac{1}{16} [H_{0},v^{\alpha}]_{+,pq} d_{qp}(\omega)\left[H_{0},\partial^{\beta}\left[d(\omega-\omega_{1})\circ\left[H_{0},\left[\mathcal{A}^{\gamma},\left[d(\omega-\omega_{[2]}) \circ\left[H_{0},\partial^{\delta}[\rho^{(0)}]\right]_{+}\right]\right]_{-}\right]_{+}\right]\right]_{+,qp}\\
=&\sum_{p,q}-i\frac{1}{16}v_{pq}^{\alpha}\frac{1}{\omega-\varepsilon_{qp}}(\varepsilon_{p}+\varepsilon_{q})^{2}\frac{1}{\omega-\omega_{[2]}}\partial^{\beta}\left[\frac{1}{\omega-\omega_{1}-\varepsilon_{qp}}\mathcal{A}^{\gamma}_{qp}(\varepsilon_{p}\varepsilon_{q}\partial^{\delta} f_{pq}+\varepsilon_{p}^{2}\partial^{\delta}f_{p}-\varepsilon_{q}^{2}\partial^{\delta}f_{q})\right],
\end{aligned}
\end{equation}
\begin{equation}
\begin{aligned}
\Xi^{(3),\gamma\delta}=
& \sum_{p,q}\frac{1}{16} [H_{0},v^{\alpha}]_{+,pq} d_{qp}(\omega)\left[H_{0},\left[\mathcal{A}^{\beta},\left[d(\omega-\omega_{1})\circ\left[H_{0},\left[\partial^{\gamma}\left[d(\omega-\omega_{[2]}) \circ\left[H_{0},\partial^{\delta}[\rho^{(0)}]\right]_{+}\right]\right]_{-}\right]_{+}\right]\right]_{-}\right]_{+,qp}\\
=&\sum_{p,q}-i\frac{1}{16}v_{pq}^{\alpha}(\varepsilon_{p}+\varepsilon_{q})\frac{1}{\omega-\varepsilon_{qp}}\frac{1}{\omega-\omega_{1}}\frac{1}{\omega-\omega_{[2]}}\mathcal{A}^{\beta}_{qp}\left[\varepsilon_{q}\varepsilon_{p}\partial^{\gamma}(\varepsilon_{p}\partial^{\delta} f_{p})-\varepsilon_{q}\varepsilon_{p}\partial^{\gamma}(\varepsilon_{q}\partial^{\delta} f_{q})+\varepsilon_{p}^{2}\partial^{\gamma}(\varepsilon_{p}\partial^{\delta} f_{p})\right.\\
&\left. -\varepsilon_{q}^{2}\partial^{\gamma}(\varepsilon_{q}\partial^{\beta} f_{q})\right],
\end{aligned}
\end{equation}
\begin{equation}
\begin{aligned}
\Xi^{(3),\delta}=
& \sum_{p,q}\frac{1}{16} [H_{0},v^{\alpha}]_{+,pq} d_{qp}(\omega)\left[H_{0},\left[\mathcal{A}^{\beta},\left[d(\omega-\omega_{1})\circ\left[H_{0},\left[\mathcal{A}^{\gamma},\left[d(\omega-\omega_{[2]}) \circ\left[H_{0},\partial^{\delta}[\rho^{(0)}]\right]_{+}\right]\right]_{-}\right]_{+}\right]_{-}\right]_{+}\right]_{+,qp}\\
=&\sum_{p,q,r}-i\frac{1}{16}v_{pq}^{\alpha}\frac{1}{\omega-\varepsilon_{qp}}(\varepsilon_{p}+\varepsilon_{q})^{2}\left[\frac{1}{\omega-\omega_{1}-\varepsilon_{rq}}\frac{1}{\omega-\omega_{[2]}-\varepsilon_{rp}}\mathcal{A}^{\beta}_{qr}\mathcal{A}^{\gamma}_{rp}(\varepsilon_{p}\varepsilon_{r}\partial^{\delta} f_{pr}+\varepsilon_{p}^{2}\partial^{\delta}f_{p}-\varepsilon_{r}^{2}\partial^{\delta}f_{r})\right.\\
&\left.-\frac{1}{\omega-\omega_{1}-\varepsilon_{rp}}\frac{1}{\omega-\omega_{[2]}-\varepsilon_{rp}}\mathcal{A}^{\gamma}_{qr}\mathcal{A}^{\beta}_{rp}(\varepsilon_{q}\varepsilon_{r}\partial^{\delta} f_{rq}+\varepsilon_{r}^{2}\partial^{\delta}f_{r}-\varepsilon_{p}^{2}\partial^{\delta}f_{p})\right],
\end{aligned}
\end{equation}
\begin{equation}
\begin{aligned}
\Xi^{(3),\gamma}=
& \sum_{p,q}\frac{1}{16} [H_{0},v^{\alpha}]_{+,pq} d_{qp}(\omega)\left[H_{0},\left[\mathcal{A}^{\beta},\left[d(\omega-\omega_{1})\circ\left[H_{0},\partial^{\gamma}\left[d(\omega-\omega_{[2]}) \circ\left[H_{0},[\mathcal{A}^{\delta},\rho^{(0)}]_{-}\right]_{+}\right]\right]_{+}\right]_{-}\right]_{+}\right]_{+,qp}\\
=&\sum_{p,q,r}-i\frac{1}{16}v_{pq}^{\alpha}\frac{1}{\omega-\varepsilon_{qp}}\frac{1}{\omega-\omega_{1}-\varepsilon_{rq}}(\varepsilon_{p}+\varepsilon_{q})^{2}\left[\mathcal{A}^{\beta}_{qr}(\varepsilon_{p}+\varepsilon_{r})\partial^{\gamma}\left[\frac{1}{\omega-\omega_{[2]}-\varepsilon_{rp}}\mathcal{A}_{rp}^{\delta}(\varepsilon_{r}+\varepsilon_{p})f_{pr}\right]\right.\\
&\left.+(\varepsilon_{r}+\varepsilon_{q})\partial^{\gamma}\left[\frac{1}{\omega-\omega_{[2]}-\varepsilon_{qr}}\mathcal{A}_{qr}^{\delta}(\varepsilon_{q}+\varepsilon_{r})f_{rq}\mathcal{A}^{\beta}_{rp}\right]\right],
\end{aligned}
\end{equation}
\begin{equation}
\begin{aligned}
\Xi^{(3),\beta}=
& \sum_{p,q}\frac{1}{16} [H_{0},v^{\alpha}]_{+,pq} d_{qp}(\omega)\left[H_{0},\partial^{\beta}\left[d(\omega-\omega_{1})\circ\left[H_{0},\left[\mathcal{A}^{\gamma},\left[d(\omega-\omega_{[2]}) \circ\left[H_{0},[\mathcal{A}^{\delta},\rho^{(0)}]_{-}\right]_{+}\right]\right]_{-}\right]_{+}\right]\right]_{+,qp}\\
=&\sum_{p,q,r}-i\frac{1}{16}v_{pq}^{\alpha}\frac{1}{\omega-\varepsilon_{qp}}(\varepsilon_{p}+\varepsilon_{q})^{2}\partial^{\beta}\left[\frac{1}{\omega-\omega_{1}-\varepsilon_{rp}}\mathcal{A}^{\gamma}_{qr}\mathcal{A}^{\delta}_{rp}\varepsilon_{p}(\varepsilon_{r}+\varepsilon_{p})-\frac{1}{\omega-\omega_{1}-\varepsilon_{qr}}\mathcal{A}^{\delta}_{qr}\mathcal{A}^{\gamma}_{rp}\varepsilon_{p}(\varepsilon_{r}+\varepsilon_{q})\right],
\end{aligned}
\end{equation}
\begin{equation}
\begin{aligned}
&\Xi^{(3)}\\
=& \sum_{p,q}\frac{1}{16} [H_{0},v^{\alpha}]_{+,pq} d_{qp}(\omega)\left[H_{0},\left[\mathcal{A}^{\beta},\left[d(\omega-\omega_{1})\circ\left[H_{0},\left[\mathcal{A}^{\gamma},\left[d(\omega-\omega_{[2]}) \circ\left[H_{0},\left[\mathcal{A}^{\delta}, [\rho^{(0)}]\right]_{-}\right]_{+}\right]\right]_{-}\right]_{+}\right]_{-}\right]_{+}\right]_{+,qp}\\
=&\sum_{p,q,r,s}-\frac{1}{16} v_{pq}^{\alpha}\frac{1}{\omega-\varepsilon_{qp}}(\varepsilon_{q}+\varepsilon_{p})^{2}\left[\left[\frac{1}{\omega-\omega_{1}-\varepsilon_{sp}}\mathcal{A}^{\gamma}_{rs}\mathcal{A}^{\delta}_{sp}\varepsilon_{p}(\varepsilon_{s}+\varepsilon_{p})-\frac{1}{\omega-\omega_{1}-\varepsilon_{rs}}\mathcal{A}^{\delta}_{rs}\mathcal{A}^{\gamma}_{sp}\varepsilon_{p}(\varepsilon_{s}+\varepsilon_{r})^{\beta}\right]\mathcal{A}_{sp}\right.\\
&\left. +\mathcal{A}_{qr}^{\beta}\left[\frac{1}{\omega-\omega_{1}-\varepsilon_{rs}}\mathcal{A}^{\gamma}_{qr}\mathcal{A}^{\delta}_{rs}\varepsilon_{s}(\varepsilon_{r}+\varepsilon_{s})-\frac{1}{\omega-\omega_{1}-\varepsilon_{qr}}\mathcal{A}^{\delta}_{qr}\mathcal{A}^{\gamma}_{rs}\varepsilon_{s}(\varepsilon_{r}+\varepsilon_{q})\right]\right].
\end{aligned}
\end{equation}

\section{3rd order thermal-thermal response }\label{app7}

The 3rd order thermal-thermal response is calculated in an similar way. The Kubo contribution in this case to the heat current is
\begin{equation}
\begin{aligned}
J_{h}^{{\rm{Kubo}},(3),\alpha}(\omega )
=&\int_{\bm{k}}{\rm{Tr}}\left[j_{h}^{\alpha} \rho^{(3)}\right].
\end{aligned}
\end{equation}
Following the same steps in calculating $L_{12}^{{\rm{Kubo}},\alpha\beta\delta\zeta}$, the   3rd order Kubo thermal-thermal response is rewritten as
\begin{equation}
\begin{aligned}
L_{22}^{{\rm{Kubo}},\alpha\beta\delta\zeta}(\omega ;\omega_{1},\omega_{2},\omega_{3})=\int_{\bm{k}}\left[\Xi^{(3),\beta\delta\zeta}+\Xi^{(3),\beta\delta}+\Xi^{(3),\beta\zeta}+\Xi^{(3),\delta\zeta}+\Xi^{(3),\beta}+\Pi^{(3),\delta}+\Xi^{(3),\zeta}+\Xi^{(3)}\right].
\end{aligned}
\end{equation}
The expressions of the $\Xi$s are shown in the appendix. The poles of $\Xi^{(3),...}$ are identical to those of $\Pi^{(3),...}$, with the leading term contributed by $\Xi^{(3),\beta\delta\zeta}$ and $\Xi^{(3),\delta\zeta}$. Hence the Kubo contribution in DC limit is found as
\begin{equation}
\begin{aligned}
L_{22}^{{\rm{Kubo}},\alpha\beta\delta\zeta}(\omega ;\omega_{1},\omega_{2},\omega_{3})
&=\sum_{p}\int_{\bm{k}}\left\{\frac{-i}{\omega\omega_{[2]}\omega_{3}}v_{p}^{\alpha}\tilde{\varepsilon}_{p}^{2}\partial^{\beta}
\left[\tilde{\varepsilon}_{p}\partial^{\delta}(\tilde{\varepsilon}_{p}\partial^{\zeta}f_{p})\right]+\frac{1}{4\omega_{[2]}\omega_{3}}
(\tilde{\varepsilon}_{p}+\tilde{\varepsilon}_{q})\mathcal{A}^{\beta}_{qp}\left[\tilde{\varepsilon}_{q}\tilde{\varepsilon}_{p}\partial^{\delta}(\tilde{\varepsilon_{p}}\partial^{\zeta} f_{p})\right.\right.\\
&\left.\left.-\tilde{\varepsilon}_{q}\tilde{\varepsilon}_{p}\partial^{\delta}(\tilde{\varepsilon}_{q}\partial^{\zeta} f_{q})
 +\tilde{\varepsilon}_{p}^{2}\partial^{\delta}(\tilde{\varepsilon}_{p}\partial^{\zeta} f_{p})-\tilde{\varepsilon}_{q}^{2}\partial^{\delta}(\tilde{\varepsilon}_{q}\partial^{\zeta} f_{q})\right]\vphantom{\frac{1}{2}} \right\},
\end{aligned}
\end{equation}
which can be written as
\begin{equation}\label{L22lk3}
\begin{aligned}
&L_{22,DC}^{{\rm{Kubo}},\alpha\beta\delta\zeta}(\omega ;\omega_{1},\omega_{2},\omega_{3})=\sum_{p}\int_{\bm{k}}\left\{-i\frac{1}{\omega\omega_{[2]}\omega_{3}}v_{p}^{\alpha}\tilde{\varepsilon}_{p}^{2}\partial^{\beta}\left[\tilde{\varepsilon}_{p}\partial^{\delta}(\tilde{\varepsilon}_{p}\partial^{\zeta}f_{p})\right]-\frac{1}{\omega_{[2]}\omega_{3}}\tilde{\varepsilon}_{p} w^{\gamma}_{p}\partial^{\delta}(\tilde{\varepsilon}_{p}\partial^{\zeta} f_{p})\right\}.
\end{aligned}
\end{equation}
The 3rd order heat magnetization is given as
\begin{equation}
\begin{aligned}\label{L22lleq3}
M_{Q}^{(3),\gamma}(\omega)={\rm{Tr}}\left[\int_{\bm{k}}\rho^{(2)}(\omega) w^{\gamma}-\frac{1}{e^2}\int d\varepsilon \tilde{\varepsilon} \sigma^{\gamma}(\varepsilon)\rho^{(2)}(\omega) \right].
\end{aligned}
\end{equation}
 and obtain the 3rd order thermal-thermal magnetization response
\begin{equation}\label{L22ltr3}
L_{22,DC}^{{\rm{tr}},\alpha\beta\delta\zeta}(\omega ;\omega_{1},\omega_{2},\omega_{3})
=L_{22,D}^{{\rm{tr}},\alpha\beta\delta\zeta}(\omega ;\omega_{1},\omega_{2},\omega_{3})+L_{22,A}^{{\rm{tr}},\alpha\beta\delta\zeta}(\omega ;\omega_{1},\omega_{2},\omega_{3})
\end{equation}
with
\begin{equation}
\begin{aligned}
&L_{22,D}^{{\rm{tr}},\alpha\beta\delta\zeta}(\omega ;\omega_{1},\omega_{2},\omega_{3})=\sum_{p}\int_{\bm{k}}\frac{-i}{\omega\omega_{[2]}\omega_{3}}v_{p}^{\alpha}\tilde{\varepsilon}_{p}^{2}\partial^{\beta}\left[\tilde{\varepsilon}_{p}\partial^{\delta}(\tilde{\varepsilon}_{p}\partial^{\zeta}f_{p})\right],\\
&L_{22,A}^{{\rm{tr}},\alpha\beta\delta\zeta}(\omega ;\omega_{1},\omega_{2},\omega_{3})=\sum_{p}\int d\varepsilon_p \frac{1}{\omega\omega_{2}} \left[6\tilde{\varepsilon}_{p}^{2}\frac{\partial f_{p}}{\partial \varepsilon_p}+6\tilde{\varepsilon}_{p}^{3}\frac{\partial^2 f_p}{\partial \varepsilon^2_{p}}+ \tilde{\varepsilon}_{p}^4 \frac{\partial^3 f}{\partial \varepsilon^3_{p}}\right] v^{\delta}(\varepsilon_p)v^{\zeta}(\varepsilon_p)\sigma^{\gamma}_{p}(\varepsilon).
\end{aligned}
\end{equation}

\end{widetext}



\begin{thebibliography}{50}
\bibitem [{ref()}]{ref1}%
  \BibitemOpen
  \href@noop {} {}\bibinfo {note} {It can be seen from the convergence analysis
  $\lim_{T\rightarrow 0}\frac{1}{T^{2}}\partial g_{p}/\partial k_{y}\propto
  \frac{1}{T^{3}}\frac{e^{\beta\varepsilon_{p}}}{(e^{\beta\varepsilon_{p}}-1)^{2}}\frac{\partial
  \varepsilon_{p}}{\partial k^{y}}=0$.}\BibitemShut {Stop}%
\end{thebibliography}


%

\end{document}